\documentclass[12pt]{article}
\usepackage{amsmath,amsthm,amssymb,euscript,epsf,epsfig}
\usepackage{mathtools}
\usepackage{array}
\usepackage{fancybox}
\usepackage{comment} 
\usepackage{wasysym} 

\usepackage{hyperref}
\usepackage{rotating}
 
\usepackage{tikz,pbox}

\newcommand{\be}{\begin{equation}}
\newcommand{\ee}{\end{equation}}

\newcommand{\beq}{\begin{equation}}
\newcommand{\eeq}{\end{equation}}

\newcommand{\bea}{\begin{eqnarray}\displaystyle}
\newcommand{\eea}{\end{eqnarray}}

\newcommand{\nnm}{\nonumber}

\makeatletter
\@addtoreset{equation}{section}
\makeatother

\def\one{{\hbox{ 1\kern-.8mm l}}}
\def\zero{{\hbox{ 0\kern-1.5mm 0}}}

  \def\cC{{\cal C}}
  \def\cF{{\cal F}}
  
 \def\cK{{\cal K}} \def\cL{{\cal L}}
  \def\cO{{\cal O}}
\def\cP{{\cal P}} \def\cQ{{\cal Q}}

 \def\cZ{{\cal Z}}

\newtheorem{lemma}{Lemma}
\newtheorem{corollary}{Corollary}

\newtheorem{definition}{Definition}
\newtheorem{theorem}{Theorem}
\newtheorem{proposition}{Proposition}

\newcommand{\mC}{{\mathbb C}}
\newcommand{\C}{{\mathbb C}}

\newcommand{\R}{\mathbb{R}}

\newcommand{\Z}{ \mathbb{Z} }

\newcommand{\N}{{\mathbb N}}

\def\Bch{ {\rm Bch} } 

\def\g{ \gamma} 
\def\s{ \sigma}

\def\Sym{ \hbox{Sym} } 
\def\Orb{ {\rm Orb}} 
\def\Aut{ {\rm Aut} } 
\def\Rib{ {\rm Rib}}
\newcommand{\id}{\rm id}

\newcommand{\wchiRla}{ \widehat{\chi}^{R}_\lambda }

\def\lam { \lambda }

\def\del { \delta}

\newcommand{\Tr}{{\rm Tr}}
\newcommand{\tr}{{\rm tr}}

\def\fac{ {\rm{factored}} }

\newcommand{\shP}{{\#\textsc{P}} \!}
\newcommand{\pP}{{\textsc{P}} }
\newcommand{\NP}{{\textsc{NP }}\!}

\begin{document}

\begin{flushright}
DIAS-STP-24-06\\
QMUL-PH-24-06\\
\end{flushright}

\begin{center}

 {\Large \bf  
 Counting of surfaces and computational complexity  \\ in column sums of symmetric group  character tables
 }
 \medskip

\bigskip

Joseph Ben Geloun$^{a,c,*}$ and Sanjaye Ramgoolam$^{b , d ,\dag}  $

\bigskip

$^a${\em Laboratoire Bordelais de Recherche en Informatiqye UMR CNRS 5800} \\
{\em Universit\'e de Bordeaux, 351 cours de la lib\'eration, 33522 Talence, France} \\

\medskip
$^{b}${\em Department  of Physics and Astronomy} , {\em  Centre for Theoretical Physics }\\
{\em Queen Mary University of London, London E1 4NS, United Kingdom }\\
\medskip

$^c${\em Laboratoire d'Informatique de Paris Nord UMR CNRS 7030} \\
{\em Universit\'e Paris 13, 99, avenue J.-B. Clement,
93430 Villetaneuse, France} \\

\medskip
$^{d}${\em School of Theoretical Physics } \\
{\em Dublin Institute of Theoretical Physics,
10 Burlington Road, Dublin 4, Ireland } \\
\medskip
E-mails:  $^{*}$bengeloun@lipn.univ-paris13.fr,
\quad $^{\dag}$s.ramgoolam@qmul.ac.uk

\begin{abstract}  
The character table of the symmetric group  $S_n$, of permutations of $n$ objects, is of fundamental  interest in theoretical physics, combinatorics as well as  computational complexity theory. We  investigate the implications of an identity, which has a geometrical interpretation in combinatorial topological field theories,  relating the column sum of normalised central characters of $S_n$ to a sum of  structure constants of multiplication in the centre of the group algebra of $S_n$. The identity leads to the proof   that a combinatorial computation of the column sum belongs to complexity class \shP. The sum of structure constants has an interpretation in terms of the counting of branched covers of the sphere. This allows the identification of a tractable subset of the structure constants related to genus zero covers. We use this subset to  prove 
 that  the column sum for  a conjugacy class labelled by partition $\lam$  is non-vanishing  if and only if the permutations in the conjugacy class are even. This leads to the result that  the determination of  the vanishing or otherwise of the column sum  is in complexity  class \pP\!\!.  The subset gives a positive lower bound on the column sum for any even $ \lambda$.  For any disjoint decomposition of $ \lambda$ as $ \lambda_1 \sqcup \lambda_2 $ we obtain a lower bound for the column sum at $ \lambda$ in terms of the product of the column sums for $ \lambda_1$ and  $ \lambda_2$. This can be expressed as a super-additivity property for the logarithms of column sums of normalized characters. 
\end{abstract}

\end{center}


\newpage 
\tableofcontents 

\section{Introduction}

The representation theory of symmetric groups, and their character tables  in particular, have a wide range of applications in theoretical physics and is also of interest in computational complexity theory. In this paper, we will use equations arising in  combinatorial topological string theories (CTST) as well as the connection between branched covers and symmetric groups to gain new  insights into combinatorial and complexity-theoretic aspects of character tables of symmetric groups. 

\vskip.2cm 

Low-dimensional instances of gauge-string duality, notably the formulation of the large $N$ expansion of two dimensional Yang Mills theory in terms of branched covers of surfaces \cite{Gross:1993hu}\cite{Gross:1993yt}, exploit the representation theory of symmetric groups as well as the connection between symmetric groups and branched covers (see \cite{Cordes:1994fc} for review of the results from the nineties and \cite{Aharony:2023tam,Benizri:2024mpx} for related recent developments). Combinatorial Topological String Theories (CTST) based on  Dijkgraaf-Witten two-dimensional topological field theories (abbreviated $G$-Flat TQFTs here)  of flat bundles of a finite group $G $ \cite{DW, Witten91, Freed:1991bn, Fukuma:1993hy} have recently been introduced as an avenue for employing physical constructions to inform mathematical questions about finite groups \cite{ICFDTS}. CTST employ a  basic aspect of  string theory, namely the idea of extracting useful physical information from computations across a range of two dimensional surface topologies, as a tool for  understanding structures  in the  algebraic data of finite groups. These theories are motivated, from a physics perspective,  by the idea of using two dimensional quantum gravity theories, enriched by concepts from gauge-string duality,  as a laboratory for investigating  combinatorial principles relevant to quantum gravity in general dimensions, a theme which has been of active interest in string theory, e.g. \cite{MarMax,GardMeg,ICFDTS,BanMoore,Panorama,EricDilaton}.

\vskip.2cm

It is useful to review a few points from the recent CTST literature which motivate the present paper. A classic construction due to Burnside gives an algorithm for constructing the character table of a finite group from the group multiplication table. The group multiplication data in the input  consists of the structure constants of the multiplication in the centre of the group algebra, in the natural basis for this centre labelled by conjugacy classes of the group. 
In \cite{ICFDTS},  the algorithmic construction due to Burnside  was interpreted in terms of amplitudes in $G$-Flat-TQFT associated with a genus one surface with a number of  boundaries labelled by a fixed conjugacy class. In \cite{Ramgoolam:2022xfk} the amplitudes of $G$-Flat-TQFT for surfaces of arbitrary genus having one boundary labelled by a conjugacy class were shown to lead to the construction of  certain partial sums of normalised central characters along a column of the character table of $G$.  The construction was shown to imply the integrality of these partial sums. In  \cite{Padellaro:2023oqj} similar constructive proofs of integrality of certain partial sums of entries along rows of the character table were obtained by using a different TQFT: one based on the fusion coefficients of the group. This 2D TQFT was referred to as $G$-Fusion TQFT. The analogies between the equations arising in $G$-Flat TQFTs  and $G$-Fusion TQFTs, which enable the proofs of integrality along rows and columns, were discussed in the context of a ``row-column duality'' between the two types of TQFTs \cite{Padellaro:2023oqj}. A broader goal of CTST is to develop a systematic landscape of  algorithms, with interpretations in terms of physical amplitudes in models of string theory and quantum gravity,  as a way to address questions in combinatorial representation theory (see e.g.\cite{Barcelo1997CombinatorialRT,Stanley2000}) of combinatorial constructions of representation-theoretic data.  In this paper, we focus on the perspectives of computational complexity theory in connection with these algorithms, following developments in string theory seeking to incorporate these perspectives into our understanding of the quantum physics of black holes (see e.g. \cite{Susskind,Bouland:2019pvu}) and of the lansdcape of string theory vacuua (see e.g. \cite{Halverson:2018cio,Grimm:2023lrf}). 

\vskip.2cm 

Algorithms based on string-theoretic objects  were  developed to study Kronecker coefficients of symmetric groups  in \cite{BenGeloun:2020yau}. For a triple of Young diagrams with $n$ boxes,  a system of  matrices with integer entries, acting  on a vector space spanned by bi-partite ribbon graphs with $n$ edges,  was shown to have a null space of dimension equal to the square of the Kronecker coefficient for the triple of Young diagrams. This was proposed as a way to address 
Murnaghan's long-standing question on the combinatorial interpretation of these coefficients
\cite{Murnaghan1938TheAO}. These matrix equations arise from the action of a certain number of central elements in $ \cK ( n ) $, a semi-simple associative algebra on the space of bi-partite ribbon graphs with $n$ edges. These central elements are expressed as  sums of  cyclic permutations  in the group algebra of $S_n$,  embedded  in three different ways in $ \cK ( n )$.  The algebra $ \cK (n) $ was identified in the study of these bi-partite graphs as arising in the context of counting invariant polynomials of $3$-index complex tensors \cite{BenGeloun:2013lim,Mattioli:2016eyp,BenGeloun:2017vwn}. The Wedderburn-Artin decomposition of $ \cK(n)$ involves the square of the  Kronecker coefficients in a natural way and this plays a role in \cite{BenGeloun:2020yau}. Permutation algebras have also been shown to be relevant to $O(N)$ tensor invariants   \cite{Avohou:2019qrl,BenGeloun:2020lfe}, and this demonstrates the robustness of the framework.

\vskip.2cm 

Such  systems of matrix equations exist generally in any associative algebra with combinatorial bases  and provide an approach for constructing as well as detecting Wedderburn-Artin basis elements in these algebras. Similar semi-simple associative algebras  based on permutations,  characterised as permutation centraliser algebras, have been shown to organise many aspects of the combinatorics of correlators in one-matrix or multi-matrix models \cite{BHR1,KR1,dMK1,dMK2,EHS,PCA} for general matrix size $N$.  The eigenvalue systems  were used in the context of quantum algorithms to define a general projector detection task for such associative algebras \cite{Geloun:2023zqa}, with motivations from the AdS/CFT correspondence \cite{Malda,GKP,Witten}.   

\vskip.2cm 

Character tables of symmetric groups, Littlewood-Richardson coefficients and Kronecker coefficients are the subject of active interest in  classical  and quantum computational computational complexity theory (see e.g. \cite{Narayanan06,GCTIII,Ikenmeyer2017OnVO,IPP2022}).  It is known that the determination of whether a given Littlewood-Richardson coefficient is positive or zero can be done in polynomial time \cite{Narayanan06}, while it is also known that for Kronecker coefficients this is NP-hard \cite{Ikenmeyer2017OnVO}. This means that the determination of whether a Kronecker coefficient is zero or not is as hard as the hardest problems in class NP. We refer the reader to textbooks such as \cite{AroraBoaz} for the definition of the classical complexity classes. 
 It has been shown that the verification of a non-vanishing  Kronecker coefficient can be done, using a construction of projectors associated with the Kronecker coefficient,  in a quantum version of polynomial time called bounded error quantum polynomial time (BQP), see \cite{BCGH24,IS23}. Row sums of symmetric group character tables are known to be expressible in terms of Kronecker coefficients and the verification of non-zero row sums is also shown to be in BQP \cite{BCGH24}. Given the usefulness of the notion of row-column duality in relating 2D topological quantum field theories related to character tables of finite groups, it is natural to investigate analogous complexity questions for column sums. The investigations in this paper are primarily motivated by this idea.

\vskip.2cm 
 
In this work, we are interested in a representation theoretic quantity:  
the sum of normalized characters, over the complete set of irreducible representations, evaluated on central elements of the symmetric group labelled by conjugacy classes.  These are sums along the columns of the character table.   Let $n$ be a positive integer and let $R$ denote an irreducible representation (irrep) of the symmetric group  $S_n$. Specifying $R$ involves specifying a vector space $V_R$ and  
linear operators $D^{R} ( \sigma ) : V_R \rightarrow V_R $ for all $ \sigma \in S_n$.   Let $d_R$ be the dimension of $V_R$. 
The character of a  permutation $ \sigma \in S_n$ is $ \chi^R ( \sigma ) = \tr_{ V_R} ( D^R (\sigma ) ) $ which is a class function, invariant under conjugation of $ \sigma$, by any $ \gamma \in S_n$ : $ \sigma \rightarrow \gamma \sigma \gamma^{-1} $. 
Let $\lambda $ be  partition of $n$,  and let   $\cC_\lambda $ denote the conjugacy class
labelled by $\lambda $, with cardinality $ |\cC_{ \lambda } | $.  $ \chi^R_{ \lambda } $ is defined as $\chi^R ( \sigma ) $ for any $ \sigma \in \cC_{ \lambda } $. The group algebra of $S_n$, denoted as $ \mC ( S_n) $ is realised as formal sums of group elements with complex coefficients, and the centre of the algebra is denoted $ \cZ ( \mC ( S_n ) ) $.  Sums  $T_\lambda =\sum_{ \s \in \cC_\lambda} \s $ are in the center $ \cZ ( \mC ( S_n) ) $.  We refer to the quantity 
$\chi^{R} (  T_\lambda )/d_R  =  |\cC_{ \lambda } |  \chi^{ R }_{\lambda } /d_R  $ as  {\it the normalized central character}  for irrep $R$ and conjugacy class $ \lambda$.  Our starting point in this paper is Proposition \ref{columnsum}, which is a formula for the sum $ \sum_{ R } \chi^{R} (  T_\lambda )/d_R  $ in terms of structure constants of multiplication in $ \cZ ( \mC ( S_n)) ) $. While the character sum has positive and negative integer contributions, the right hand side in the proposition is manifestly positive.  This formula is closely related to the amplitude for a genus one surface with boundary labelled by conjugacy class $ \cC_{ \lambda}  $ in 2D TQFT of flat $G$-bundles: see equation (2.7) in \cite{Padellaro:2023oqj} where this is discussed in the context of row-column duality between $G$-flat TQFT and $G$-Fusion TQFT. 

\vskip.2cm 

The main results in this paper are the following : 

\begin{enumerate} 

\item  Proposition \ref{columnsum} is used  to prove  Theorem \ref{theo:RIBn} : the function $\lam \mapsto \sum_{R  \vdash n} { \chi^{ R} (  T_\lambda ) \over d_R } $ belongs to the complexity class  \shP, i.e. the verification of positivity of the sum for a given $ \lambda $ can be done in a number of time steps which is polynomial in $n$. 
 
\item Theorem \ref{lamEven}: The sum of normalised central characters for a fixed conjugacy class $\lambda$ is strictly positive, i.e. $\sum_{ R \vdash n } { \chi^R (  T_\lambda ) \over d_R }  >  0$, if and only if the permutations in the conjugacy class labelled by  $\lam$ are even. 
 
\item Theorem \ref{coro:RIBn}:  The determination of whether $ \sum_{ R \vdash n } { \chi^R (  T_\lambda ) \over d_R } $ is positive or not is  in complexity class $ \pP$, i,e. it can be done in a number of time steps which is polynomial in the data size of  $\lambda$ . This holds whether the data of the partition $ \lambda $ is presented in binary or unary (a more detailed discussion of the significance of the presentation of  the data is included in the derivation of the theorem). This is analogous to the result that the complexity of determining the vanishing or otherwise of  Littlewood-Richardson coefficients for a triple of Young diagrams (or a triple of partitions) is in the class $ \pP$ \cite{GCTIII}.

\item  Theorem \ref{theo:LowBoundsGen}: Expressions for positive integer lower bounds on $ \sum_{ R \vdash n } { \chi^R (  T_\lambda ) \over d_R }   $ for all even $\lambda$. 

\item  We prove (Theorem \ref{theo:prodpartition}) that $ F_{ \lambda } \equiv  \sum_{ R \vdash n } { \chi^R_{\lambda}  \over d_R }$, for any $ \lambda = \lambda^{(1)} \sqcup \lambda^{(2)} $ obeys the inequality
\bea 
F_{ \lambda^{(1)} \sqcup \lambda^{(2)} } \ge F_{ \lambda^{(1)} }  F_{ \lambda^{(2)} } \nnm 
\eea
 Here $ \lambda^{(1)} $ is a partition of $n_1$, $ \lambda^{(2)} $ is a partition of $n_2$, while  $ \lambda^{(1)} \sqcup \lambda^{(2)}$ is a partition of $ n_1 + n_2$, with parts being the concatenation of the parts of $ \lambda^{(1)} $ and $ \lambda^{(2)}$. For example if $ \lambda^{(1) } = [ 3,2, 2] , \lambda^{(2)} = [ 4,3,1] $, then 
$ \lambda^{(1)} \sqcup \lambda^{(2)}  = [ 4,3,3,2,2,1] $. We may say, equivalently, that $ \log (F_{ \lambda }  ) $ is a super-additive function of partitions, where addition of partitions is defined by concatenation. 

\end{enumerate}

The proof of   Theorem \ref{theo:RIBn} involves re-expressing the algebraic formula, Proposition \ref{columnsum}, in terms of a counting  of factorisations of permutations $ \sigma_3^* $ in conjugacy class $ \lambda $ into pairs of permutations belonging to the same conjugacy class $\mu$. The proof of Theorems  \ref{lamEven} and \ref{theo:LowBoundsGen}
 relies on restriction of the sum in Proposition  \ref{columnsum} to conjugacy classes $ \mu $ of the form $ [ 2^{ k } , 1^{ n  -2k } ]$ for any $ k$, abbreviated $[2^* ,1^*]$.  The special role of this form of $ \mu$ is understood  from a geometrical  perspective, which  is based on a realisation, given in Theorem \ref{theo:RGexpans},  of the column sum as a sum over bi-partite ribbon graphs. Using the theory of combinatorial maps, the sum is equivalently over surfaces covering the sphere with three branch points on the target sphere.  For the class  $\mu = [ 2^* , 1^* ] $ all the  covering surfaces are genus zero or equivalently the associated bi-partite graphs are planar.  
 
 \vskip.2cm 

The technical computations leading to Theorems \ref{lamEven} and \ref{theo:LowBoundsGen}
obtain cues towards  the derivation from  a diagrammatic representation of the  permutation factorisations for this form of $\mu$ in terms of a combinatorial construction using chord diagrams equipped with inner and outer chords, along with a generalisation of these diagrams possessing a planarity property. Theorem \ref{coro:RIBn} is a  direct consequence of \ref{lamEven}. Theorem \ref{theo:prodpartition} is  an application of Proposition \ref{columnsum}. 
 
 \vskip.2cm 

Given our use of  permutation factorizations, it is worth noting that  permutation factorizations have a long history \cite{Hurwitz1891}. In his seminal work, Hurwitz investigates Riemann surfaces with given branch points, laying foundational principles for the study of algebraic curves and complex analysis. His research provides   insights into the enumeration and properties of these surfaces, influencing subsequent developments in various mathematical fields. In particular, cycle factorizations relate to the famous Cayley permutation tree and have a wealth of applications ranging from combinatorics to geometry, such as the enumeration of branched coverings of the sphere, intersection theory, and singularity theory; see, for instance, \cite{BousquetMlouSchaeffer2000, Irving2004CombinatorialCF, Irving2006MinimalTF} and  references therein. 

\vskip.2cm

The structure of the paper is as follows. After describing the basics of characters and the conjugacy class basis for $ \cZ ( \mC ( S_n ) ) $  for the symmetric groups in section \ref{sec:basics}, we state
 Proposition \ref{columnsum}  in section \ref{sec:columnsum}, while the proof is in Appendix \ref{app:proofcolumnsum}.  This gives a formula for the  column sum of normalized  central characters in terms of the structure constants of the algebra $ \cZ ( \mC ( S_n ) ) $ in the conjugacy class basis. Section \ref{sect:conj} is devoted to expressing Proposition \ref{columnsum}  as  a combinatorial construction, namely  the counting of the number of elements in a finite  set defined by group multiplications in $S_n$.  The column sum is further re-expressed in terms of  bi-partite ribbon graphs in section \ref{sect:refined}. This discussion is used to identify a simple set of planar ribbon graphs which contribute to the column sum. This allows us to prove that the column sum is positive for any even $ \lambda$. The fact that it is zero for any odd $ \lambda$ follows easily from Proposition  \ref{columnsum}. These facts lead to the statement, in section \ref{sec:nonvanishing},  that the decision problem of determining the vanishing or otherwise of column sums is in complexity class $\pP$. Section \ref{sect:lowbound} develops  lower bounds on the column sum of normalized characters, progressing from the simplest to the most general $\lambda$, by exploiting $ \mu $ of the form $ [ 2^*, 1^*]$.  We derive a general formula and present numerical evidence supporting our bounds.   The paper concludes with a discussion and outlook. The appendices  include detailed proofs of certain statements from the main text, as well as codes that perform several tasks demonstrated in the text.

\section{Characters and class algebras } 
\label{sec:basics}

We review key properties of the representation theory of the symmetric group $S_n$ 
on $n$ objects, the group algebra $\C ( S_n) $  and its centre $\cZ ( \C ( S_n) )$. 
The basics of group theory and the representation theory of $S_n$ can be found in standard textbooks e.g.  \cite{Hamermesh} \cite{goodman2009symmetry} \cite{macdonald1998symmetric,  FultonYoung}. The relevant formulae are collected  in Appendices of recent physics papers such as \cite{Pasukonis:2013ts,BenGeloun:2020yau}.

\subsection{Representation theory and characters of $S_n$}
\label{sub:repr} 

A representation of $S_n$ is group homomorphism $D: S_n \to GL(V)$, 
where $V$ is a finite dimensional complex vector space. 
The irreducible representations (irreps) of $S_n$ are labelled by 
partitions $R$ of $n$ that we denote $R\vdash n$ \cite{FultonYoung}, and there are  vector spaces $V_R$ and homomorphisms $D^R$ for each $R$. 
A partition  of $n$ is a  sequence  of positive integers $R = [R_1, R_2, \dots, R_\ell]$, 
such that $\sum_{i=1}^\ell R_i = n =: |R|  $ and $ R_1 \ge R_2 \ge \cdots \ge R_{ \ell}$. We write $\ell(R):= \ell$.

A partition $R  \vdash n$ can be also denoted $R = (r_1^{a_1}, r_{2}^{a_2}, 
\dots, r_{s }^{a_s})$, where now the sequence of $r_i$ is strictly { decreasing} and $a_i$ 
denotes the multiplicity of the appearance of a given $r_i$ within $R$. 
For instance, we have $[3,3,2,1,1,1] = [ 3^2, 2,1^3] \vdash 11$, $[1,1,1,1] = [1^{4}] \vdash 4$. 

A Young diagram (YD)
  of shape $R$ is an arrangement of boxes labeled by couples $(i,j)$, where rows are labeled by $i$, 
with  $1\le i \le \ell(R)$, and columns labeled by $j$, with $ 1\le j \le R_i$. 

Consider $D^{R}$ an irreducible representation of $S_n$ labeled by $R$. 
This associates to each group element in $S_n$ a  matrix of size $d_R \times d_R $, where $d_R  = n!/f(R)$, where $f(R)$ is the  product of hook-lengths
\cite{FultonYoung}. We will be interested in the character of the irreps 
$R$ which is the trace of the representation matrix of a  given element: 
\bea
\chi^{R}(\s) = \Tr(D^R (\s)) 
\eea
The characters are central class functions: $\chi^R(\g \s \g^{-1})= \chi^R(\s)$. 

The conjugacy class of an element $\s \in S_n$ is the set 
\bea
\cC (\s) = \{\s' \in S_n : \;\; 
\s' = \g \s \g^{-1},\; \forall \g \in S_n\}
\eea
It is well known that any conjugacy class is labeled by a partition of $n$.  
Thus $S_n = \cup_{\lam \vdash n } \cC_{\lam}$, where 
the elements of $\cC_\lam $ are permutations with  cycle lengths uniquely  determined
by the partition  $\lam \vdash n $. 
It is straightforward to observe that characters are constant on $\cC_\lam$: 
\bea
 \chi^{R}_\lam =  \chi^{R}(\s) \,, \quad  \forall \s \in \cC_\lam
\eea
The table of characters records these $\chi^R_{\lam}$. Conventionally, the irrep label $R$ is used for the rows and the conjugacy class label $ \lam$ for the columns. 

The 12th problem of Stanley \cite{Stanley2000} asks for a combinatorial interpretation of  the row sum of the table $ \sum_{ \lambda } (\chi^R_\lam)$, for any $R$. 
Macdonald has a formula for the column sum, i.e. $\sum_{R} \chi^R_\lam$ (see Example 11, p. 120, \cite{macdonald1998symmetric}), and these sums have been studied recently in a paper \cite{ayyer2024large} which appeared as we were finalising the write-up of the current paper. We are interested in a variant where the summands are normalised characters
\bea
\widehat{\chi}^{R}_{\lambda }  =\frac{ \chi^{R}_{ \lambda }  }{d_R} \, , 
\eea
i.e. characters of a fixed conjugacy class divided by the dimensions of the respective irreducible representations. This arises as an amplitude in two-dimensional topological field theory of flat $S_n$ connections for a genus one surface with one boundary - see eqn. (2.7) of \cite{Padellaro:2023oqj} and there are generalisations involving powers of these normalised central characters along with powers of $ d_R$ from other amplitudes (see \cite{Padellaro:2023oqj} for these more general equations and for background references on these equations). 
 One of the results in this work (Theorem \ref{theo:RGexpans}) is a combinatorial interpretation of
\bea
\sum_{ R \vdash n  } |\cC_{\lambda } | \wchiRla
\eea
 in terms of ribbon graphs. Throughout the text, we will generally refer to this expression as ``the column sum."  The result in section \ref{sec:BdDisj}  will be expressed most simply in terms of the sum with the weight $|\cC_\lam| \rightarrow n! $, i.e. 
 \bea 
n!  \sum_{ R \vdash n } \wchiRla = n! \sum_{ R \vdash n } { \chi^R_{\lambda } \over d_R } 
 \eea
 which is the sum of normalised characters, rescaled by $n!$, which we refer to as ``sum of normalised characters" for short. 
  
\subsection{The group algebra $\C ( S_n) $  and its centre $\cZ ( \C ( S_n) )$}
\label{sub:centre}

Let $\mC ( S_n )$ be the group algebra of the symmetric group $S_n$, i.e. 
the group algebra defined by  linear combinations of permutations: $\sum_{\s \in S_n} a_\s \s$, $ a_\s \in \C$. 
Representations and characters extend by linearity to $\mC(S_n)$. 

Let $\cZ ( \mC ( S_n ) ) $ be the center of $\mC ( S_n )$.  There is a basis of $\cZ ( \mC ( S_n ) ) $ 
formed by sums over group elements in conjugacy classes.  For a conjugacy class  $ \cC_{\mu}  \subset  S_n $, 
 we have a central element 
\bea \label{Tmu}
T_{ \mu} = \sum_{ \sigma \in \cC_{ \mu } } \sigma \, . 
\eea
 obeying $\g T_\mu \g^{-1}  = T_\mu$, for any $\g \in S_n$. 
 $(T_\mu)_{\mu \vdash n}$ forms a basis of the 
 centre $\cZ ( \C ( S_n) )$  \cite{Kemp:2019log}. 

The product of two general  central elements for $ \mu, \nu \vdash n $  yields 
\bea \label{tmutnu}
 T_{ \mu} T_{\nu} = \sum_{ \lam \vdash n } 
C_{ \mu \nu }^{ \;\;\; \lam } T_{ \lam} 
\eea
where the integer structure constants $C_{ \mu \nu }^{ \;\;\; \lam }$ can be  organized as a matrix $C_{\mu}$ with matrix elements : 
\bea 
( C_{ \mu } )_{ \nu}^{ \lambda } = C_{ \mu \nu}^{ \;\;\; \lambda } \, . 
\eea
It is known that  the eigenvalues of $( C_{ \mu } )_{ \nu}^{ \lambda } $ are the normalized characters $ \widehat \chi^R ( T_{ \mu}) :=  { \chi^R ( T_{ \mu} ) \over d_R } $ in the irrep labelled by the Young diagram $R\vdash n$ of dimension  $d_R$. 

\subsection{Column sum of  normalized central characters } 
\label{sec:columnsum}

The character of a given $\s \in S_n$ in an irrep labeled by $R$
is denoted by $\chi^R(\s)$. The characters  extend by linearity to $\C(S_n)$
so that  $\chi^R(T_\lam)$ makes sense. 
We will be interested in  such characters $\widehat{\chi}^R(T_\lam)= |\cC_{\lam}|\chi^R(\s)/d_R$, for any $\s$ representative 
of $\cC_\lam$.

\vskip.2cm 

\noindent 
The following statement holds: 
\begin{proposition}[Column sums in the  table of normalized central characters]
\label{columnsum}
For any $\lam \vdash n$, 
\bea \label{sumC=sumR}
 \sum_{ R  \vdash n } \; \widehat \chi^R ( T_{ \lam})  = \sum_{  \mu \vdash n} 
  C_{ \mu \lambda }^{\;\;\;    \mu  }  
\eea
\end{proposition}
\proof See Appendix \ref{app:proofcolumnsum}. 

\qed

\section{Construction and complexity of the column sums of normalised central characters }
\label{sect:conj}

In this section, we derive two key results. First, we show
that the column sums of  normalized characters of central elements associated with conjugacy classes have a combinatorial construction in terms of a counting problem involving the multiplication of permutations.  Second, we show that the computation of the column sums is in the complexity  class $\shP$. This is obtained by identifying  a  more general \shP
problem from which the column sum problem is deduced.

\subsection{Combinatorial construction from group multiplications }
\label{sect:combinconj}

Proposition \ref{columnsum} shows an  equality between the column sum of normalized central characters and the sum 
$\sum_{  \mu \vdash n}   C_{ \mu \lambda } ^{\;\;\; \mu}  $. 
We can simply exploit the definition of 
$ C_{ \mu \lambda } ^{\;\;\;  \mu} $
to provide a combinatorial construction of this sum. 

\

The integer structure constants, defined in terms of multiplication  of central elements   in 
\eqref{tmutnu},  deliver already a combinatorial 
intepretation of $C_{\nu \lam}^{ \;\;\;  \mu}$: 
\bea
C_{\nu \lam}^{\;\;\;   \mu}
\text{ is  the number of times that } T_\mu 
\text{ appears in the product } T_\nu T_\lam 
\eea
Then equating $\mu = \nu$ and summing over $\mu$, we obtain 
\bea
&&
\sum_{\mu \vdash n}
C_{\mu \lam}^{ \;\;\; \mu} \text{ enumerates the   number of times that all central elements} 
\crcr
&&
\text{ appear in their own product with } T_\lam 
\eea
 
Since the central elements $T_\mu$ are sums of elements  in  conjugacy classes
in the group algebra,  we can re-express the above as a  combinatorial interpretation in terms of sets of elements in  conjugacy classes. We provide 
that translation in the following.

Use the relation \eqref{deltaTTT} and write
\beq 
\label{Cmunulam}
 C_{  \nu \lambda } ^{\;\;\; \mu } 
 = \frac{1}{|\cC_\mu|} \delta( T_\nu T_\lam T_\mu) 
   =  \frac{1}{|\cC_\mu|} \delta(  T_\mu T_\nu T_\lam) 
 = \frac{1}{|\cC_\mu|} \sum_{\s\in \cC\mu } \delta(\s  T_\nu T_\lam) 
\eeq
where we introduce Kronecker delta function on $S_n$, $\delta: S_n \to \{0,1\}$, 
such that $\delta(\s)=1$, if $\s=\id$, and $0$ otherwise, 
and extend it to the group algebra $\C(S_n)$. 

The summand $ \delta(\s  T_\nu T_\lam)$ turns out to be constant on the conjugacy class $\cC_\mu$: 
for any pair $\s,\s'\in \cC_\mu$, related via $\s'  = \g \s \g^{-1} $, 
$ \delta(\s' T_\nu T_\lam) =  \delta(\g \s  \g^{-1} T_\nu T_\lam)
=  \delta(\s  \g^{-1} T_\nu  \g  T_\lam) =  \delta(\s  \g^{-1}  \g  T_\nu   T_\lam) =  \delta(\s  T_\nu T_\lam)  $, 
where  we use the property that $T_\nu$ and $T_\lam$ are central elements. Thus 
\bea
\label{cmunulam}
 C_{  \nu   \lambda } ^{ \;\;\;\mu} 
 =  \delta(\s_\mu ^*  T_\nu T_\lam) 
\eea 
where $\s_\mu^*$ is a fixed representative element  in $\cC_\mu$. 
Thus $  C_{  \nu   \lambda } ^{ \;\;\; \mu}  $ 
 counts the number of occurrences of terms of $T_\lam$ in the product $\s_\mu^* T_\nu$, namely: 
\bea
\label{eq:combiCol}
 &&
  \delta(\s_\mu^*  T_\nu T_\lam) 
    = 
     \sum_{\tau \in \cC_\lam} 
          \sum_{\s \in \cC_\nu} 
     \delta(\s_\mu^*  \s \tau ) \crcr
   &&
   =   \text{number of pairs } (\s,\tau) \in \cC_\nu \times \cC_\lam    \text{ such that }   \s_\mu^* \s \tau = \id
 \eea 
 Note that $\cC_\lam = \cC'_{\lam}$, where $\cC'_{\lam}$  is the conjugacy class consisting of the inverses of  
 the elements of $\cC_\lam$. This justifies that $  \s_\mu^* \s = \tau^{-1}$
 belongs to $\cC_\lam$. 
 
 Finally, equating $\mu=\nu$  and performing a sum over the partitions $\mu$  
 \eqref{cmunulam}, we obtain a combinatorial construction of the column sum 
 of normalized characters of central elements: 
\bea
 &&
 \text{Given}\;  \lam,\;  \text{the column sum of normalized characters of central elements enumerates} \crcr
 &&   \text{the number  of pairs } 
 (\s,\tau) \in \cC_\mu \times \cC_\lam    \text{ such that }   \s_\mu^* \s \tau = \id, \text{ for all } \mu \vdash n
 \; . 
 \nonumber
 \eea
There is a reduction of this counting that yields a direct result. 
Indeed, the case $\lam =[1^n]$ leads us to
 \bea
\sum_{\mu}  C_{  \mu   [1^n]} ^{ \;\;\; \mu}   
 =  \sum_{ R  \vdash n } \; \widehat \chi^R ( T_{ [1^n]}) = \sum_{ R  \vdash n } 1 = p(n)
 \eea
 where $p(n)$ denotes the number of partitions of $n$. 
Such  information may be interesting because the various 
objects that we will enumerate could be bijectively mapped 
to partitions when $\lam =[1^n]$. Observing
\eqref{eq:combiCol},  $\cC_{[1^n]}$ reduces to $\{\id\}$, therefore
$\tau = \id$, and so 
$\big| \left\{ \s\in \cC_\mu | \;\;   \s_\mu^* \s = \id \right\}\big|=1$, 
which  corresponds to $\s =( \s_\mu^*)^{-1}$. 
Per conjugacy class $\cC_\mu$, we have a single $\s$ that obeys
the condition.  Hence, counting all solutions for all $\mu$, is precisely 
counting $\mu$ partitions of $n$. 

\

\noindent{\bf Other combinatorial constructions --}
Alternatively, we list here other equivalent  combinatorial constructions of 
the column sum.  We deduce from \eqref{eq:combiCol}: 
\bea
\label{eq:sTTmu}
 &&
  \delta(\s_\mu^*  T_\nu T_\lam) 
   = \sum_{\tau \in \cC_\lam} 
     \delta(\s_\mu^*  T_\nu \tau ) \crcr
     &&
   =  \text{number of terms in } \s_\mu^*  T_\nu 
   \text{ 
   which belong to }  \cC_\lam
\crcr
&&
 = 
 \text{number of elements } \s \in \cC_\nu   \text{ such that } 
   \s_\mu^* \s  \text{ belongs to }   \cC_\lam
 \eea
 Then, we take $\mu= \nu$  and gather the countings by varing $\mu$. 
The  formula  \eqref{eq:sTTmu} is implemented using GAP, see Appendix \ref{app:GapTTT}.

Let $\mu, \nu$ and $\lam$ be three partitions of $n$, a positive integer,   
and a representative  $\s^*_\mu \in \cC_\mu$. 
Consider  $ C_{ \nu \lambda }^ {\;\;\; \mu}  $  and
 its construction given by  \eqref{eq:combiCol}, 
 we introduce the sets
\bea
&&
{\rm Fact}'(\mu ; \, \nu, \lam) = \{ (\s , \tau ) \in \cC_\nu \times \cC_\lam \, |\,  \s_\mu^* \s \, \tau = \id \} \crcr
&&
{\rm Fact}(\mu , \nu, \lam) = \{ ( \rho,  \s , \tau ) \in \cC_\mu \times \cC_\nu \times \cC_\lam \, |\,  \rho \s \, \tau =  \id \}
\eea
We recognise the relations between  cardinalities: 
 $$
 |{\rm Fact}'(\mu ; \, \nu, \lam) | =  C_{  \nu \lambda } ^{ \;\;\;\mu}  =
 \frac{C_{\mu\nu\lam}}{|\cC_\mu|}  = \frac{1}{|\cC_\mu|}  | {\rm Fact}(\mu , \nu, \lam) |  
 $$
as  shown earlier. $ {\rm Fact}(\mu , \nu, \lam)  $ is symmetric in the
indices. Note also that $ {\rm Fact}(\mu , \nu, \lam)  $
is non empty if and only if $ |{\rm Fact}'(\mu ; \, \nu, \lam) | $
is non empty. 

We introduce the set 
\bea \label{CClam}
 {\rm Fact}(\lam) =  \bigcup_{\mu}  {\rm Fact}'(\mu ; \, \mu, \lam)
\eea

From the above equations, it follows: 
\begin{theorem}
Given $\lam \vdash n$, we have 
\label{sumF}
\bea
 \sum_{ R  \vdash n } \; \widehat \chi^R ( T_{ \lam}) 
  = | {\rm Fact}(\lam) |
\eea
\end{theorem}

\subsection{\shP problems on conjugacy classes}

We recall that 
\shP is the complexity class including all counting  problems 
associated with \NP    problems, thus it is colloquially refered 
to as the enumerative analogue of \textsc{NP}.
 Formally, $\shP$ is the class of 
functions $f: \{0,1\}^{*}\to \N$, where  $\{0,1\}^{*}$ refers
to the set of binary words, such that there exists a nondeterministic polynomial Turing machine   $M$  such that for all $x\in \{0,1\}^*$, 
$f(x)$ counts the number of accepting paths  of $M$ launched on $x$ (see chap.9 \cite{AroraBoaz}). 

In other words, a \shP problem $f$ is associated with a decision problem
that is in class \NP   and which consists in determining if the function $f$ is strictly positive or not. A  decision problem is in \NP \,  if its ``yes'' instances admit a certificate of polynomial size  that can be checked in polynomial time. 

A main result is the following statement. 

\begin{theorem}
\label{theo:Cmunulam}
The map  $\textsc{Fact}_n: (\mu,\nu,\lam) \mapsto|{\rm Fact}(\mu , \nu, \lam) |$ is in \shP. 
\end{theorem}

We need preliminary lemmas to prove this theorem. 

\begin{lemma}
\label{lemclass}
Let $n$ be a positive integer, $\s \in S_n$, $\mu$
a partition of $n$, and $\cC_\mu$ the conjugacy class labeled by $\mu$.
Checking if $\s \in \cC_\mu$  or otherwise   can be achieved in a number of time steps which is polynomial in  $n$. 
\end{lemma}

\proof 

A permutation  $\s$ is  a list $[a_1,a_2,\dots, a_n]$  where
the $a_i \in \{1,2,\dots, n\}$, $i=1,2,\dots,n$, are pairwise distinct. This list is the image of $\s$, i.e.  $\s(i)= a_i$, for all $i$ (list-based-encoding).
A partition $\mu  = [\mu_1, \mu_2, \dots, \mu_l] \vdash n$
is also a list.

To check if $\s \in \cC_\mu$, we must compute the cycle structure 
of $\s$ and compare it with $\mu$. 
A simple algorithm allows one to extract 
the cycle structure $\s$: we compute the orbits of $\s$  
acting on the segment $\{1, \dots, n\}$, 
and compute their length. 
Computing the  orbits of $\s$ is 
to let $\s$  act   repeatedly on some elements starting by $1$
and excluding elements once reached.
Appendix \ref{app:basicfunctions} shows an algorithm 
that performs this task in time  bounded by
a polynomial in $n$. 
A simple modification of this algorithm
 enables one to  constructs the list the orbit sizes 
(simply by adding a counting variable, see Appendix \ref{app:basicfunctions}). This does not change the previous complexity.  

We sort the list of orbit sizes $\nu =[\nu_1,\dots, \nu_k]$ in decreasing order. 
Sorting this list can be done in $\cO( k \log k)$ by common counting sorting algorithms, when the size of elements
are negligible compared to the size of the list 
$k \le n$. Thus the sorting algorithm is surely 
polynomial in $n$. 
 
Comparing the list of orbit sizes and the list $\mu=[\mu_1,\mu_2,\dots, \mu_l]$ requires at most
$\cO(l)$ comparison tests, otherwise we stop the procedure. 
As $l$ and  $\mu_i$  are smaller than $n$, the cost of the comparison test is polynomial  in $n$. 

The overall procedure requires a number of operations
within $\cO(n^c)$, where $c$ is a constant independent of $n$ 
and is bounded above by $3$. A more detailed treatment 
can be expected to reduce this upper bound.

\qed

\begin{lemma}
\label{lemcompo}
Given two permutations $\s$ and $\g$ in $S_n$, 
composing these permutations as $\g \s$ 
can be done in polynomial time of the input size. 
\end{lemma}
\proof 

We obtain 
the resulting permutation $\g \s$ is  by listing the elements $\g \s(i)$,  $i=1,2,\dots,n$, each of which is obtained by applying $\s$ to the element $\g(i)$ 
and then reading the corresponding element in the list $\s(\g(i))$. Note that we use the convention that
composition is made from left to right. 
The cost of the procedure cannot exceed the list sizes. 
The complexity of this algorithm 
is polynomial in $n$. 

\qed

\ 

We are in position to prove our main statement. 

\noindent{\bf Proof of Theorem \ref{theo:Cmunulam}.} 
The decision problem  associated with $\textsc{Fact}_n$ 
will be  denoted  ${\rm FCT}_n$. It takes as input a triple 
of partitions $\mu$, $\nu$ and $\lam$ of $n$
and asks the question:  
 Is the set ${\rm Fact}(\mu  , \nu, \lam)$ not empty? 

We will prove that ${\rm FCT}_n$ is in class \NP. 

The decision above equivalently asks: 
Is there a triple $(\rho, \s , \tau ) \in   \cC_\mu \times \cC_\nu \times \cC_\lam \, $ such that  $\rho \s  \tau = \id$?

Assume that one delivers to the verifier the tuple $(\rho, \s, \tau, \mu,\nu, \lam)$ claiming that it 
provides  a ``yes'' answer. The different tasks in the verification are:

 -1- Check that $\rho  \in \cC_\mu$,  $\s   \in \cC_\nu$ and $ \tau   \in   \cC_\lam \,$ ; 
  
-2- Compose  $\rho \s \tau $ and check if this equals $\id$.

 Lemma \ref{lemclass} shows that point 1
 can be done in time polynomial in  $n$. 
  Lemma \ref{lemcompo} shows that
  the composition   $\rho \s \tau $
is also achieved in time polynomial in  $n$.
It simply remains to compare   $\rho \s \tau $ 
and $\id$ which are two lists of size $n$ 
and involves at most $n$ comparisons. 
The overall verification procedure therefore uses a certificate 
of size polynomial in  $n$ and is completed in a number 
of time steps which is   polynomial in $n$. 

\qed 

\noindent{\bf Remark --}
We did not yet discuss the type of encoding of the input data of this problem. 
Note that ${\rm FCT}_n$ takes as input 
three partitions $\mu,\nu, \lam \vdash n$. 
Choosing a representation of the data, e.g. a partition $\mu = [\mu_1, \mu_2,\dots,  \mu_\ell]$ presented as a weakly decreasing sequence of numbers
$\mu_i$ given in binary form, the data  size can be of order $\cO(\log n)$ 
(e.g. when the partition is simply $[n]$). This makes both 
 the certificate  $(\rho, \s, \tau, \mu,\nu, \lam) \in \cO(n\log n)$  and the cost of verification exponential in the size of
 the data input.  However, in the unary representation, 
 a partition $\mu = [\mu_1, \mu_2,\dots,  \mu_\ell]$ 
 is represented as a list of $1$ of size $\mu_1$, 
followed by a list of $1$ of size  $\mu_2$, and so on. 
 Thus,  in unary representation,  the partitions  $\mu,\nu, \lam $ 
 are of order $\cO(n)$.  In this case, both certificate and verification become
 simply polynomial of the data entry size. 
 Thus, claiming that ${\rm FCT}_n$
 is \NP must be understood in the case of unary 
 representation of the data.

\begin{theorem}
\label{theo:RIBn}
The map: 
$ \lam \mapsto \sum_{ R  \vdash n } \; \widehat \chi^R ( T_{ \lam})  $ is in \shP\!\!. 
\end{theorem}
\proof 
The result follows from two facts: 
(1) the first is 
$ \sum_{ R  \vdash n } \; \widehat \chi^R ( T_{ \lam})  
= |{\rm Fact}(\lam)|$,
by Theorem \ref{sumF},  and 
(2)  the map 
$f: \lam \mapsto \ |{\rm Fact}(\lam)|$ is in \shP. 
This remains to be proved. 

Defining ${\rm Fact}(\lam)$ as in \eqref{CClam}, 
we aim  to show  that the map  $f : \lam \mapsto| {\rm Fact}(\lam)  |= \sum_{\mu \vdash n} \frac{1}{|\cC_\mu|} |{\rm Fact}(\mu , \mu, \lam) |$ is in \shP. 

 The decision  problem associated with this enumeration is denoted 
${\rm FCT}'_n$. It takes as input a partition $\lam$ of $n$
and asks: Is the set ${\rm Fact}(\lam)$ non-empty? 

We now prove that ${\rm FCT}'_n$ is in class \NP.

A certificate consists of a tuple $(\rho, \sigma, \tau, \mu, \lambda)$ that yields an affirmative answer to ${\rm FCT}'_n$. The verification consists in the proof that  ${\rm Fact}(\mu , \mu, \lambda)$ is non-empty, that is we must prove 

-1- $\rho $ and $\sigma$ belong to $\cC_\mu$ and
$\tau$ belongs to $\cC_\lam$; 

-2- $\rho \sigma \tau$ is the identity;

Using Theorem \ref{theo:Cmunulam}, this represents a particular instance of the problem ${\rm FCT}_n$, when $\mu = \nu$.  Therefore, verifying points 1 and 2 
can be done in polynomial time in $n$.  
Thus, the verification procedure uses a certificate 
of size polynomial in $n$ and a number of time 
steps  that is also polynomial in $n$.

\qed 

We reiterate that, implicitly, the size of input data must be chosen in unary for the conclusion regarding the \shP nature of the column sum problem.

\

\section{Column sums from ribbon graphs}
\label{sect:refined}

This section gives a review of the description of bi-partite ribbon graphs with $n$ edges  and associated maps of degree $n$  between surfaces, called Belyi maps, in terms of  triples of symmetric group elements in $S_n$. A useful textbook reference is the book by Lando and Zvonkin \cite{lando2013graphs} which relates the subject to the original mathematical literature. Closely related to the considerations of this paper focused on character tables and associated representation theoretic constructions, ribbon graphs  have been used to give integer matrix algorithms for Kronecker coefficients \cite{BenGeloun:2020yau}.

In  previous sections, we have related column sums of normalised central characters to the structure constants of multiplication in the centre of the group algebra of $ S_n$, in the basis of conjugacy class sums. Here we relate these structure constants to the counting of bi-partite ribbon graphs weighted with a positive integer  associated with a symmetry of the classes involved and the symmetry of the bi-partite ribbon graph, equivalently the symmetry of the associated Belyi map.

\subsection{Ribbon graphs as permutation triples}

A bi-partite ribbon graph is a graph with black and white vertices that is equipped with a cellular  embedding on a two-dimensional surface. The bi-partite character
means that  every edge connects a black vertex to a white vertex. The cellular embedding means that if we cut the surface along the edges, we get a collection of regions that are topologically equivalent to open disks. In this paper, we will use, for short,  the term ribbon graphs to refer to bi)partite ribbon graphs. 

We can describe ribbon graphs having $n$ edges using permutations of the set $\{1, 2, \dots, n\}$, which form the symmetric group $S_n$. We label the edges of the graph with the integers from this set, and then we follow a fixed orientation on the surface to read the edge labels around each vertex. This gives us two permutations $\tau_1$ and $\tau_2$, corresponding to the cycles around the black and white vertices, respectively. If we change the edge labels by applying a permutation $\mu \in S_n$, we get a new pair of permutations $(\mu \tau_1 \mu^{-1}, \mu \tau_2 \mu^{-1})$. Pairs related by such conjugations are associated with different labellings of the same unlabelled ribbon graph.   Therefore, ribbon graphs can be identified with equivalence classes of pairs of permutations in $S_n \times S_n$, where two pairs are equivalent if they can be obtained from each other by 
simultaneous conjugation: 
\bea\label{adj}
  ( \tau_1 , \tau_2  ) \sim  ( \tau_1' , \tau_2' ) \,  ~~~
  \Leftrightarrow ~~~   \exists \mu \in S_n \,,  ~~~ ( \tau_1' , \tau_2' ) =  ( \mu \tau_1 \mu^{-1} , \mu \tau_2 \mu^{-1} ) 
\eea 
The set of permutation pairs within a fixed equivalence class forms an orbit for the diagonal conjugate action of $S_n$ on $ S_n \times S_n$ given in \eqref{adj}.  
It turns out that $\tau_3= (\tau_1 \tau_2)^{-1}$ contains relevant information
about the embedding: the cycles of $\tau_3$ relates to the faces
of the ribbon graph. The Euler characteristic 
of a (bi-partite) ribbon graph: 
\bea
2k - 2g = C_{\tau_1} + C_{\tau_2}  + C_{\tau_3} - n 
\eea
where $k$ is the number of connected components, $g$ the  genus of the ribbon graph, and $C_{\s}$  the number of cycles of
the permutation $\s\in S_n$. This is justified by  the fact that the number of 
vertices is  $V = C_{\tau_1} + C_{\tau_2} $, number of edges is $n$ and number of faces is $F=  C_{\tau_3}$. 

\

 \begin{figure}[h]
 \begin{center}
     \begin{minipage}[h]{.7\textwidth}\centering
\includegraphics[angle=0, width=4cm, height=2.5cm]{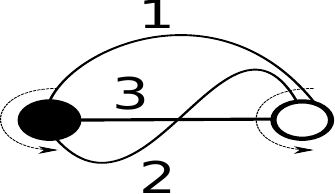}
\vspace{0.3cm}
\caption{ {\small A bi-partite ribbon graph with  $n=3$ edges
defined by a pair $(\tau_1=(123), \tau_2=(123))$}} 
\label{fig:ribb}
\end{minipage}
\end{center}
\end{figure}

 \begin{figure}[h]
 \begin{center}
     \begin{minipage}[h]{.65\textwidth}\centering
\includegraphics[angle=0, width=8cm, height=3.5cm]{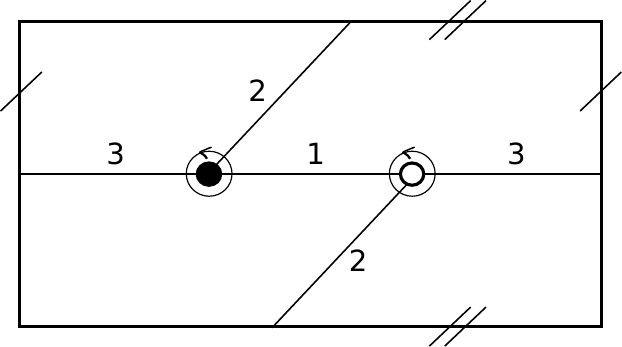}
\vspace{0.3cm}
\caption{ {\small Embedding the bi-partite graph $(\tau_1=(123), \tau_2=(123))$ on the torus, represented as a rectangle with opposite sides identified. }} 
\label{fig:torus1}
\end{minipage}
\end{center}
\end{figure}

We illustrate the above account with  an example at $n=3$. Among the possible 
ribbon graphs with $n=3$ edges, 
we focus the graph  of Figure \ref{fig:ribb}. 
The bi-partite ribbon graph is defined by the equivalence class of the couple $(\tau_1 = (123), \tau_2 =(123))$. 
We compute its number of faces through
the number of cycles of $\tau_3 = (\tau_1\tau_2)^{-1} = (132)^{-1}=(123)$. 
Thus the Euler characteristic  gives: 
$ 2-2g = 1+1+1 - 3 = 0$ and shows that the ribbon graph  has genus $g = 1$. We identify it with a torus.

\subsection{Counting surfaces and the column sum problem}

\noindent{\bf Counting ribbon graphs.}
The counting of ribbon graphs having $n$ edges, each ribbon graph counted with weight one,  is related to the counting of observables in tensor models. The formulation in terms of equivalence classes of permutation pairs leads to a simple formula in terms of a sum over symmetry factors $\Sym ~ p $ of partitions $p$ of $n$
(see section 4. \cite{BenGeloun:2013lim}):
\beq
\label{CountZn}
Z_n  = \sum_{ p \vdash n  } \Sym ~ p 
\eeq

The counting of ribbon graphs, with each ribbon graph weighted by an inverse symmetry factor of the graph and with constraints on the conjugacy class of the permutations associated with the  black or white vertices, is relevant to matrix model correlators, 
\cite{Itzykson,dMKSRBelyi,TomComplex}. 

In the following, we will be counting ribbon graphs, with specified conjugacy classes for the triple of permutations, and  with positive integer weights associated with symmetries of the conjugacy classes or of the graph itself. 

\ 

{\noindent\bf Equivalence of bi-partite  ribbon graphs, orbits and automorphisms.}
Consider the equivalence \eqref{adj} between the couples $(\tau_1,\tau_2)$ defining
bi-partite ribbon graphs that now extends to the triples $( \tau_1 , \tau_2,   \tau_3  )$, with $\tau_3= (\tau_1\tau_2)^{-1}$, in the following way
\bea\label{adj2}
( \tau_1 , \tau_2  ,   \tau_3 ) \sim   ( \mu \tau_1 \mu^{-1} , \mu \tau_2 \mu^{-1}, \mu \tau_3 \mu^{-1} )
\eea
The diagonal conjugate action of $S_n$ on the Cartesian product $S_n \times S_n \times S_n $ partitions it in orbits.
Each orbit uniquely corresponds to a bi-partite ribbon graph.
We denote by $\Orb(r)$ an orbit corresponding to $r$, the bi-partite ribbon graph. Let $\Rib(n)$ be the set of bi-partite ribbon graphs,
and thus the label $r$ runs from $r=1$ to $|\Rib(n)|$, where $|\Rib(n)|=Z_n$ \eqref{CountZn}.

 The orbit-stabilizer theorem implies that $|\Orb(r)| = |S_n|/|\Aut(r)|$, where $ | \Aut (r) | $ is the order  of an equivalence class of subgroups of $S_n$ defined by the ribbon graph $r$. 
 Given a representative triple $(\tau_1^{(r)},\tau_2^{(r)}, \tau_3^{(r)} )$ for the ribbon graph $r$, the set of $ \mu \in S_n$ which stabilises the  triple under the action \eqref{adj2} forms a   stabiliser group $  {\rm Stab}(\tau_1^{(r)},\tau_2^{(r)}, \tau_3^{(r)})$. In general, as the representative changes, the stabiliser group will change to another conjugate subgroup of $S_n$, but the order $ |\Aut (r) |$ will be invariant.

Due to the diagonal conjugate action,
equivalent triples  $( \tau_1 , \tau_2  ,   \tau_3 )$  and $( \tau_1' , \tau_2'  ,   \tau_3')$, with  $\tau_i $ and $\tau_i' = \mu \tau_i \mu^{-1} $,
$i=1,2,3$, have the same set of three conjugacy classes in $S_n$, equivalently three cycle structures. Thus, on a given orbit, the cycle structure triple is an invariant. The set of all bi-partite ribbon graphs uniquely and disjointly decomposes into subsets labelled by the triples of partitions of  $n$
\bea
\label{Ribmunulam}
\Rib(n) = \bigcup_{\mu,\nu,\lam \vdash n} \Rib_{\mu,\nu,\lam} (n) \, . 
\eea
The subset $ \Rib_{\mu,\nu,\lam}(n)$ consists of all the bi-partite ribbon graphs
with representative triples $( \tau_1 , \tau_2  ,   \tau_3 )\in  \cC_{\mu} \times \cC_{\nu}  \times \cC_{\lam}$.

\begin{definition}
A bi-partite ribbon graph  element of $\Rib_{\mu,\nu,\lam}(n)$ identified with the equivalence class of 
a triple $(\tau_1,\tau_2,\tau_3)\in  \cC_{\mu} \times \cC_{\nu}  \times \cC_{\lam} $ is said to have structure $(\mu,\nu,\lam)$.  
\end{definition}
Alternatively, we will use: ``a (bi-partite) ribbon graph with structure  $(\mu,\nu,\lam)$'' or ``is of structure $(\mu,\nu,\lam)$'' to express the same content of 
the previous definition. 

\ 

{\noindent\bf Enumerating edge labelled bi-partite ribbon graphs and unlabelled graphs with inverse automorphisms. }
Here we show that enumerating edge-labelled bi-partite ribbon graphs, i.e. counting triples $(\tau_1,\tau_2,\tau_3)$  obeying the constraint $\tau_1 \tau_2 \tau_3 = \id$ is equivalent to counting unlabelled ribbon graphs $r$  with weight $ n! \over |\Aut (r)| $. Indeed defining 
\bea 
Z'_n \equiv
 \sum_{(\tau_1,\tau_2, \tau _3 ) \in S_{n}^{\times 3} }
   \delta(\tau_1 \tau_2 \tau_3 )  = (n!)^2 
\eea
we have 
\bea
\label{Zprimn}
Z'_n 
&=&
    \sum_{r}  \sum_{(\tau_1,\tau_2, \tau _3 )\in \Orb(r) }
   \delta(\tau_1 \tau_2 \tau_3 ) \crcr
   &=&
    \sum_{r}  \sum_{(\tau_1,\tau_2, \tau _3 )\in \Orb(r) }
1 = 
 \sum_{r}|\Orb(r)|   = 
 \sum_{r}\frac{| S_n|}{|\Aut(r)|} 
\crcr
& =&  n! \sum_{r}\frac{1}{|\Aut(r)|} 
\eea
where the sum over $r$ runs from $1$ to $|\Rib(n)|$,
and we have used in an intermediate step the orbit-stabilizer theorem:
$|\Orb(r)|   = | S_n| / |\Aut(r)| $.

Using  the above equation,  $Z'_{n}$  \eqref{Zprimn} can be interpreted in any of the following three ways 
\begin{itemize} 
\item   the counting of labelled bi-partite ribbon graphs, each with 
weight 1, 
\item   the counting of unlabelled bi-partite ribbon graphs, each with weight
 its  orbit size $|\Orb(r)|$ 
 \item  the counting of unlabelled bi-partite ribbon graphs, each with weight the inverse of its automorphism group order, $|\Aut(r)|$ (up to a $n!$ factor). 
\end{itemize} 

We can compare \eqref{Zprimn} and \eqref{CountZn}: 
$Z_n = \sum_r 1$  counts (unlabelled)
bi-partite ribbon graphs with weight 1 \cite{BenGeloun:2013lim}. 

\

{\noindent\bf Enumerating constrained bi-partite ribbon graphs.}
Consider three partitions $\mu, \nu$, and $\lam $ and their respective
conjugacy classes, $\cC_\mu, \cC_\nu$ and $\cC_\lam$. 
Our interest lies in the interpretation of the following sum 
\bea
\label{chimucc}
\sum_{R \vdash n } \widehat \chi^R ( T_{ \lam})  = \sum_{\mu \vdash n} \frac{1}{|\cC_\mu|} 
\delta(T_\mu T_\mu T_\lam )
\eea
We expand
\bea
 \delta(T_\mu T_\nu T_\lam ) := 
\sum_{\tau_1 \in \cC_\mu,\tau_2\in \cC_\nu , \tau_3 \in \cC_\lam}
   \delta(\tau_1 \tau_2 \tau_3 )
\eea
The sum \eqref{chimucc} recasts using the orbit/ribbon graph decomposition as 
\bea
\sum_{R \vdash n } \widehat \chi^R ( T_{ \lam})  
    &=&  \sum_{\mu  \vdash n} \frac{1}{|\cC_\mu|} 
     \sum_{r} \sum_{r  \in \Rib_{\mu,\mu,\lam}(n)}  
    | \Orb(r)|
 \crcr
    &     =  &
     \sum_{\mu  \vdash n} \frac{1}{|\cC_\mu|} 
    \sum_{r   \in \Rib_{\mu,\mu,\lam}(n)  } \frac{n!}{|\Aut(r)|}
    \label{cZnlam}
\eea

This leads to the following interpretations of  $\sum_{R \vdash n } \widehat \chi^R ( T_{ \lam})  $: 
\begin{itemize} 
\item  $\sum_{R \vdash n } \widehat \chi^R ( T_{ \lam})  $ counts the number of edge labelled bi-partite ribbon graphs with structure $(\mu, \mu, \lam)$, for a fixed $\lam$ and all $\mu$, each with weight $1/|\cC_\mu|$; 
\item $\sum_{R \vdash n } \widehat \chi^R ( T_{ \lam})  $ counts the number of (unlabelled) bi-partite ribbon graphs with structure $(\mu, \mu, \lam)$ for a fixed $\lam$ and all $\mu$, 
each with   weight its corresponding orbit size  divided by $|\cC_\mu|$; 
(this weight translates also in term of the order of the automorphism group). 
\end{itemize} 

We use now the partial equivalence among the vertices of a given color, say $\tau_1 \sim \g \tau_1 \g^{-1}$, to write $|\cC_\mu| = |S_n |/ \Aut(\mu)$, where $\Aut(\mu)$is the automorphism keeping the label of vertices unchanged. 
Thus, we can write \eqref{cZnlam}  in the form: 
 \bea
\sum_{R \vdash n } \widehat \chi^R ( T_{ \lam})  
      &=&  
     \sum_{\mu  \vdash n} \frac{|\Aut(\mu)|}{n!} 
    \sum_{r   \in \Rib_{\mu,\mu,\lam}(n)  }  \; \frac{n!}{|\Aut(r)|} \crcr
    &   = &     \sum_{\mu  \vdash n}  \; 
       \sum_{r   \in \Rib_{\mu,\mu,\lam}(n)  } \;  \frac{|\Aut(\mu)|}{|\Aut(r)|} 
    \label{cZnlam21} 
\eea
Since $\Aut(r)$ is a subgroup of $\Aut(\mu)$,
 we guarantee that $|\Aut(\mu)| / |\Aut(r)|$ is an integer. 
 
 We have another interpretation of the sum:  
 
 \begin{itemize} 
\item  $\sum_{R \vdash n } \widehat \chi^R ( T_{ \lam})     $ counts the number of  
 bi-partite ribbon graphs with structure  $(\mu, \mu, \lam)$ for a fixed $\lam$ and all $\mu$, each with weight $|\Aut(\mu)|/ |\Aut(r)|$. 
 \end{itemize}

\begin{theorem}
\label{theo:RGexpans}
Let $\lam$ be a partition of $n$. The column sum of normalized central characters for conjugacy class $ \lam $ is a sum over ribbon graphs weighted by integer sizes of cosets associated with the ribbon graph: 
\bea
\sum_{ R  \vdash n } \; \widehat \chi^R ( T_{ \lam}) 
    =      \sum_{\mu  \vdash n}  \; 
       \sum_{r   \in \Rib_{\mu,\mu,\lam}(n)  } \;  \frac{|\Aut(\mu)|}{|\Aut(r)|} 
\eea
\end{theorem}
The counting of ribbon graphs can be refined according to their genus. 
Taking this into account, we write the previous counting as: 
\bea
\sum_{ R  \vdash n } \; \widehat \chi^R ( T_{ \lam}) 
    =    \sum_{ g=0}^{ g_{\max} }  \sum_{\mu  \vdash n}  \; 
         \sum_{  r   \in \Rib^g_{\mu,\mu,\lam}(n)   } \;  \frac{|\Aut(\mu)|}{|\Aut(r)|} 
\eea
where the sum over  $g$ is performed over all possible 
  genera of the surface $r$, $g_{\max}$ is an upper bound (according to the graph constraint
  or could be let to infinity), 
  and $\Rib^g_{\mu,\nu,\lam}(n)  $ is simply the set of 
  ribbon graphs with representative lying in the class  $(\mu,\nu,\lam)$ with
  genus $g$.  To understand if this genus decomposition could lead to relevant information on the sum, for instance, the properties of dominant terms, is left for future investigations. In the following, more remarks will be given on the implications of the genus of ribbon graphs.

\subsection{ Ribbon graphs and genus restrictions  }\label{genrest}

The coefficients  $ C_{ \mu \mu \lambda  }$ which enter the formula for column sums of normalized central characters for $S_n$ have an interpretation as the counting of b-ipartite ribbon graphs embedded on surfaces, and also a related interpretation in terms of 
surfaces equipped with holomorphic maps to a sphere branched over three points. This geometrical interpretation in terms of surfaces helps identify easily computable families of $C_{ \mu \mu \lambda } $ and families which are likely to be more complex.

\

\noindent{\bf Genus restrictions --}
When a permutation triple $ \sigma_1 , \sigma_2 , \sigma $ with $\sigma_1 , \sigma_2 \in \cC_{ \mu} $ and $\sigma \in \cC_{ \lambda } $ generates a transitive subgroup of $S_n$, i,e. the subgroup can map any integer in $ \{ 1, \cdots , n \}$ to any other, then surface is connected. 
 The Euler characteristic of a connected surface is given by 
\bea 
2h -2  = -2n + 2 \Bch ( \mu )   + \Bch ( \lambda )
\eea
where $h$ is the genus of the surface  and, 
for a partition $\nu $ of $n$, $\Bch ( \nu ) $ is the branching of $\nu$:  
\bea 
\Bch ( \nu ) = n - C ( \nu ) 
\eea
where $ C ( \nu )$ is the number of cycles in $\nu$. Let $\mu = [ 1^{ n - 2k_2 - 3k_3 - \cdots -  L k_L } , 2^{ k_2} , 3^{ k_3} , \cdots , L^{k_L} ] $ be a partition of $n$, then
\bea 
 2h -2  = -2n + 2 ( \sum_{ l = 2}^{ L } ( l -1) k_l )  + \Bch ( \lambda ) 
 \eea 
$ C ( \lambda ) \ge 1 $, $ \Bch ( \lambda ) \le ( n-1) $. 
This implies that 
\bea 
&&  2h \le 2 -2n + 2 ( \sum_{ l = 2}^{ L } ( l -1) k_l )   +  (n-1) \cr
&& 2h \le -n +1 + 2 ( \sum_{ l = 2}^{ L } ( l -1) k_l ) \cr 
&& 2h \le 1 - \sum_{l=1}^L  l k_l + 2 ( \sum_{ l = 2}^{ L } ( l -1) k_l ) \cr 
&& 2h \le 1 - k_1 + \sum_{ l=3}^L ( l - 2 ) k_l 
\eea
A remarkable corollary is that if $ k_3, k_4 \cdots $ are all zero, i.e. $ \mu = [ 2^* , 1^* ] $, then 
\bea 
2h \le 1- k_1 
\eea
Since $ h \ge 0$, this means that $k_1$ can be $0$ or $1$. From the above inequality, we also learn that the genus  of a connected surface associated with factorisations of permutations in $ \cC_{ \lambda}$, when the partition $ \mu $ of the 
factor permutations is of the form $ [ 2^* , 1^* ]$, cannot exceed $0$. 

 It is interesting to contrast this result with a partition $\mu$ which is close enough but 
departs from  $[ 2^* , 1^* ] $. 
Consider the case $\mu = [ 3^{ k_3} , 2^{ k_2 } , 1^{ k_1} ] $, and $\lambda $ corresponds to  a single cycle of length $n = 3 k_3 + 2k_2 + k_1$.  With this, we have 
\bea 
&& \Bch ( \mu ) = 2k_3 + k_2 \cr 
&& \Bch ( \lambda ) = ( 3k_3 + 2k_2 + k_1)  -1  = n-1
\eea
The number of handles is given by 
\bea 
 2h -2  & =&  -2 n + 2 \Bch ( \mu ) + \Bch ( \lambda ) \cr 
& =& -2 n  + 2 ( 2k_3 + k_2 ) + n -1 \cr 
&  =& - ( 3 k_3 + 2k_2 + k_1 ) + 4k_3 + 2k_2 -1 \cr 
& =& k_3 - k_1 - 1
\eea
The $h \ge  0 $ condition yields the bound 
\bea 
k_3  \ge k_1 -1 
\eea
which is independent of $k_2$,  
while $k_3$ is constrained by $k_1$. The genus $ h = ( k_3 -k_1 +1)/2$ of the connected surface grows with the difference $ (k_3 - k_1)$. These are higher order terms from the point of view of string world-sheet perturbation theory and can be expected to correspond to higher computational complexity. In the context of gauge-string duality, the higher genus surfaces correspond to higher powers of $1/N$ from the point of view of the gauge theory.

\

\section{Vanishing/non-vanishing of column sums of $S_n$  and a subset of ribbon graphs with $n$ edges} 
\label{sec:nonvanishing}

In the last section, we found that the column sum  for conjugacy class $ \lambda $, previously expressed as a sum of the structure constants $ C_{ \mu \lambda }^{ ~~ \mu }$,  can be written as a sum over ribbon graphs which are associated with surfaces equipped with Belyi maps. The ribbon graphs are represented by triples of permutations $ ( \sigma_1 , \sigma_2 , \sigma) $ in conjugacy classes  $ ( \mu , \mu , \lambda )$ and there is a sum over all partitions $\mu$. The partitions $ \mu $ of the form $[ 2^{ k } , 1^{ n-2k } ] $, abbreviated $ [ 2^* , 1^* ]$, lead to genus zero ribbon graphs. In this section, we will show that for any even $ \lambda $, there is a non-zero $ C_{ \mu \lambda }^{ ~~~ \mu }$ for $ \mu $ of this form.  

\subsection{The  column sum is non-vanishing for all  even $\lambda  $}

For simplicity, 
if we deal with a single cycle, we  assume that any odd length cycle  can be written as  $(1,2, \dots ,2k+1) $, 
or in the case of an even length cycle, as  $(1,2, \dots,2k)$. Indeed,  the following reasoning bijectively 
extends to any cycle $(a_1,a_2,\dots, a_l)$,
with $a_i \in \{1,\dots, n\}$ pairwise distinct,  $l=2k, 2k+1$. Any such cycle can be written as a product of $ (l-1)$ simple transpositions $ ( a_l , a_{ l-1} ) ( a_{ l-1} , a_{ l-2} ) \cdots(  a_2 , a_1) $. This an even length cycle is expressed as a product of an odd number of simple transpositions, and thus is an odd-parity permutation while an odd length cycle is expressed as a product of an even number of simple transpositions and is this an even-parity permutation.

We have the following statement:

\begin{lemma}
\label{lem:evenfactor}
An cyclic permutation which has even parity, i,e. has an odd-length, factors into 
2 permutations of identical cycle structure.  
\end{lemma}

\proof We first inspect a given cycle of a permutation and draw the conclusion from this point.

\begin{figure}
\centering
\includegraphics[scale=0.7]{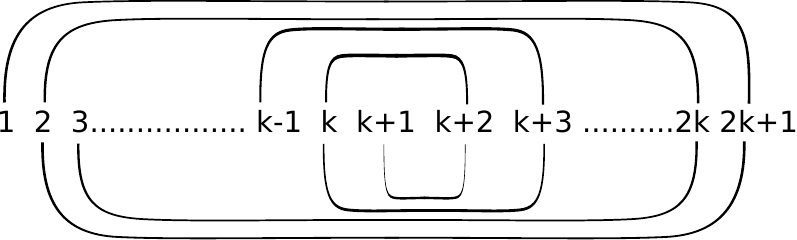}
\caption{Factorizing a cycle $(1,2,\dots, 2k+1)$ of odd length in a product of permutations.}
\label{factorodd} 
\end{figure}

A  cycle $(1,2, \dots ,2k+1) $ can be split as
\bea 
&& \sigma_1 = (  1, 2k+1)( 2, 2k) , ... (k-1,k+3)( k, k+2)  (k+1) \cr 
&& \sigma_2 =  (1)   ( 2, 2k+1) (3, 2k ) ...  ( k , k+3) ( k+1 , k+2)  \cr 
&& (1,2,\dots,  2k+1) = \sigma_1 \sigma_2 
\eea
and the factor cycle structures are $[ \sigma_1 ] = [ \sigma_2 ] = \mu =  [2^k,1] $. 
We emphasize the fact that, here and in the following, all factorizations uses  
transpositions that are disjoint from one another. 
A diagrammatic understanding is given  in Figure \ref{factorodd}:
the composition of $ \sigma_1 $ followed by $\sigma_2 $
is performed by using arcs above and then
below, alternately (in the picture $\s_1$ is above and 
$\s_2$ below). If we reach a fixed point of one permutation, there is no arc
to use for that permutation. The composition uses  the sole available arc incident on that point. 
Thus a cycle of length $2k+1$ factors into a product
of permutations such that $ [\s_1]= [2^k,1]$ and $[\s_2]  =[2^k,1]$. 
Note that  an odd length cycle decomposes into a product of
 an even number ($k+k$) of transpositions.

\begin{figure}
\centering
\includegraphics[scale=0.7]{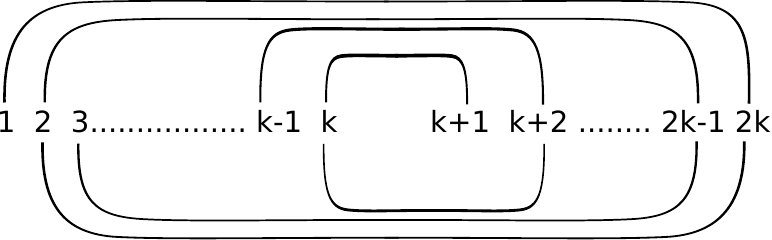}
\caption{A factorization of the cycle $(1,2,\dots, 2k)$  in a product of permutations.}
\label{factoreven1} 
\end{figure}

Meanwhile, an even length cycle $ ( 1,2, \dots, 2k)$ can be split as 
\bea 
\label{decomeven1}
&& \sigma_1 = ( 1, 2k ) (2, 2k-1) \cdots (  k -1 , k+2 ) (k , k+1)  \cr
&& \sigma_2 = ( 2, 2k )  (3, 2k-1) \cdots ( k-1 , k+3 ) (k, k+2) ( k+1) \cr 
&&  ( 1,2, \dots, 2k) =  \sigma_1 \sigma_2
\eea
In this case, the cycle of even length splits in two 
permutations with structure $ [ \sigma_1 ] = [2^k ]$ and 
$\sigma_2 = [ 2^{ k-1},1^2 ]$. 
This is explained diagrammatically in Figure \ref{factoreven1}.

\begin{figure}
\centering
\includegraphics[scale=0.7]{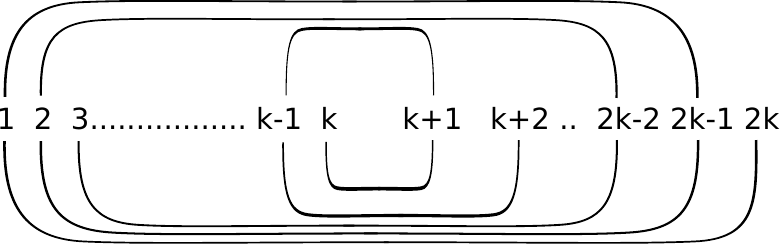}
\caption{A second factorization of  $(1,2,\dots, 2k)$  in a product of permutations.}
\label{factoreven2} 
\end{figure}

There is another possible splitting of the same even cycle: 
\bea 
\label{decomeven2}
&& \sigma_1 = ( 1, 2k-1 ) (2, 2k-2) \cdots (  k -1 , k+1 ) (k )  (2k)\cr
&& \sigma_2 =  ( 1, 2k ) (2, 2k-1) \cdots (  k -1 , k+2 ) (k , k+1) \cr 
&&    (1,2, \dots, 2k)   =  \sigma_1  \sigma_2
\eea
with two permutation factors such that 
 $\sigma_1 = [2^{k-1}, 1^2]   , \sigma_2 = [ 2^{ k} ] $. 
 The diagram associated with this 
 decomposition is given in Figure \ref{factoreven2}. 
Both decompositions \eqref{decomeven1}
 and \eqref{decomeven2} show that 
  a cycle of even length can be always decomposed
 into a product of an odd number ($k+k-1$) of transpositions.

 In summary, we have the following notation 
 for cycle splittings  according to their length: 
{ \bea
 &&
 [2k+1 ]\to [2^{k},1] \circ  [2^{k},1] \crcr
 &&
 [2k] \to [ 2^{k}] \circ [2^{k-1},1^2] \crcr
&&
 [2k] \to   [2^{k-1},1^2]  \circ [ 2^{k}] 
 \eea
 }

For a pair of even length cycles of length $2k_1$ and $2k_2$, we can use the 
$[2^{ k_1 }] \circ [2^{ k_1-1},1^2]$ splitting for the first cycle and the  $[2^{ k_2-1},1^2] \circ [2^{ k_2}]$  splitting for second cycle. This gives a $[2^{ k_1 + k_2 -1 } ,1^2]\circ 
 [2^{ k_1 + k_2 - 1},1^2]$ splitting
 in transpositions (concatenation of separated transpositions) for a pair of cycles of lengths $2k_1 $ and $ 2k_2$.

An even permutation contains some number of odd length cycles along with an even number of even length cycles
(that is the only way to guarantee an even number of transpositions). 
We can split the odd-length cycle as above. And we also pair up the even length cycles in an arbitrary way and use the second procedure to end the proof. 

\qed

We ought to mention that
Lemma \ref{lem:evenfactor} possesses generalizations. 
We will come back to this point later in the paper, but at this moment, we do not require a stronger result to achieve the following statement.

\begin{theorem}\label{lamEven}
If $\lambda$ corresponds to the cycle structure of an even 
(resp. odd) permutation, then there exists a $\mu$, such that
$C_{\mu\lam}^{\;\;\;\mu} >0$ 
and  $\sum_{ R  \vdash n } \; \widehat \chi^R ( T_{ \lam})  >0$ (resp. for all $\mu \vdash n$, $C_{\mu\lam}^{\;\;\; \mu} =0$ and $\sum_{ R  \vdash n } \; \widehat \chi^R ( T_{ \lam})  =0$). 
\end{theorem}

\proof 
Lemma \ref{lem:evenfactor} proves that any even permutation  factors
into two permutations of same cycle structure. 
So an even permutation with given cycle structure 
 $\lambda $ admits a  factorization in terms of permutation
 with  $\mu$ structure (we have constructed such a $\mu$). 
 This simply  means that, for that $\mu$, 
 $ C_{ \mu \mu }^{ \;\;\; \lambda }$ is not zero
(keeping in mind that this integer counts 
the occurances of $T_\lam$ in the product of two 
permutations of cycle type $\mu$). 
On the other hand, we have 
$C_{ \mu \mu }^{ \;\;\; \lambda } = 
C_{ \mu  \lambda}^{ \;\;\;  \mu }  \frac{ |\cC_\lam| }{   |\cC_\nu| }$, 
see \eqref{craising}. Thus,  if
 $C_{ \mu \mu }^{ \;\;\; \lambda } $
is not 0,  so is $C_{ \mu  \lambda}^{ \;\;\;\mu  } $.

 The result  when $\lam$ corresponds to
 a  odd permutation holds because there is no 
 such a factorization. The consequences 
 on the column sum are immediate. 

\qed

\subsection{Deciding the positivity of the column sum  is in \pP }

The above propositions enable  us to formulate  
a refined statement regarding the complexity 
class of the decision problem: 
``Is $\sum_{ R  \vdash n } \; \widehat \chi^R ( T_{ \lam})  >0$?''. 

 \begin{theorem}
\label{coro:RIBn}
The problem ``Is $\sum_{ R  \vdash n } \; \widehat \chi^R ( T_{ \lam})  >0$ ?'' is  in \pP\!\!. 
\end{theorem}

\proof  

To determine whether $\sum_{ R  \vdash n } \; \widehat \chi^R ( T_{ \lam})  >0$ or equals 0,  our task is to construct a ``yes" or ``no" answer to this problem via a polynomial-time algorithm.

It is convenient to represent the partition $ \lambda $ as a list $ [ c_1^{ k_1} , c_2^{ k_2} , \cdots , c_L^{ k_L } ] $ of positive  parts $c_i$  and positive multiplicities $k_i$, with $ n = \sum_{ i=1}^{L} c_i k_i $. In a binary representation of $ c_i , k_i$,  the data size  is $  D = \cO ( \sum_i ( \log k_i + \log c_i )  )$. Testing the parity of the permutations in conjugacy class $ \lambda $  involves checking that the sum of $k_i$ for even cycle lengths $c_i$ is even. This can be done as follows: \\
1. Determine the parity  $p (c_i)  \in \{ 0 , 1 \} $ of the cycle-length $c_i$, which is  the last digit in the binary representation of $c_i$.  \\ 
2. Determine the parity $ p(k_i) \in \{ 0 , 1 \}$ of each $ k_i$ which is the last digit of $k_i$.\\
3.  Calculate the parity of permutations in conjugacy class given by  the partition $ \lambda $ as
$ (-1)^{\sum_{ i=1 }^L p( k_i )  (p (c_i)  +1)} = \prod_{ i =1}^L (-1)^{ p( k_i) } (-1)^{ p (c_i) +1}  $, where the addition $p(c_i) +1$ can be  done modulo two. 
If the parity in  point 3 is $1$, the column sum is non-zero, and if it is $(-1)$ then the column sum is zero. The three steps can be completed within  $ \cO ( L  )$ operations. Since $D$ is at least order $(L \times { \rm min }( \log k_i + \log c_i )) $, this is no more than polynomial in $D$.

If the numbers $ c_i , k_i $ are given in unary, then the data size is $ D \sim  \cO ( n )$. Calculating the parities of $c_i , k_i$ takes order $c_i , k_i $ steps respectively, finding 
$  (-1)^{ p( k_i) } (-1)^{ p (c_i) +1} $ takes order $1$ steps for each $i$, and the total number of steps is $ \cO ( \sum_i c_i k_i ) = \cO ( n ) $. This is polynomial in the data size $D$. 
We therefore conclude that in either binary or unary presentation of the partition data, the determination of the vanishing or otherwise of the column sum is in complexity class $P$.

\qed

Since the polynomial time property holds in either presentation of the partition data, this is a stronger polynomiality result than the one obtained in the earlier derivation of the \shP property (Theorem \ref{theo:RIBn}) which is derived simply from the existence of the formula in Proposition \ref{columnsum} and holds only in the unary presentation of the data. Here we are using Theorem \ref{lamEven} which exploited properties of the structure constants for the special class of $\mu = [ 2^* , 1^* ] $ contributing to the equation in Theorem \ref{columnsum}, and this results in a stronger result on the complexity  of the column sums.

\vskip.2cm

 \noindent{\bf Scaling property --}
 Theorem \ref{coro:RIBn}   also enables us to discuss scaling properties  of the column sum.
  Let us consider 
 $\alpha$ a positive integer and $\lam =[c_1^{ k_1} , c_2^{ k_2} ,\dots, c_{L }^{ k_L} ] \vdash n$. We define the scaled partition 
 $\alpha \lam$  of $\alpha n$ by the partition with 
 multiplicity list
 $[\alpha k_1, \alpha k_2, \dots , \alpha k_L ]$.

 \begin{corollary}
 \label{coro:Scaled}
 Let $\lam$ be a partition of $n$ and $\alpha$ a positive integer. \\
-  If  $\sum_{ R  \vdash n } \; \widehat \chi^R ( T_{ \alpha \lam}) = 0$, then $\sum_{ R  \vdash n } \; \widehat \chi^R ( T_{ \lam})= 0$; \\
-  If  $\sum_{ R  \vdash n } \; \widehat \chi^R ( T_{ \alpha \lam})>0$ and if $\alpha$ is odd,  then $\sum_{ R  \vdash n } \; \widehat \chi^R ( T_{  \lam})>0$. 
 \end{corollary}
\proof 
The condition on the parity 
 of the sum of even parts of $\alpha \lam$ is particularly easy to handle. This sum  can be put in the form $\sum_{i : c_i \, ~\rm{ even} } (\alpha k_{i})= \alpha S_e$,  where $S_e= \sum_{i : c_i ~\rm{even} } k_i $  is the sum of even-length  parts of $\lam$. 
 
 Use Theorem \ref{coro:RIBn} and infer that 
 if $\sum_{ R  \vdash n } \; \widehat \chi^R ( T_{ \alpha \lam}) = 0$, then $\alpha S_e$ is odd. 
Both $\alpha$  and $S_e$ must be odd, 
therefore $\sum_{ R  \vdash n } \; \widehat \chi^R ( T_{ \lam}) = 0$. 

Now let us assume that   $\sum_{ R  \vdash n } \; \widehat \chi^R ( T_{ \alpha \lam})>0$. This implies, via Theorem \ref{coro:RIBn}, that $\alpha S_e$ is even. 
Thus knowing that $\alpha$ is odd  entails that $S_e$ is  even and therefore  $\sum_{ R  \vdash n } \; \widehat \chi^R ( T_{ \lam}) > 0$. 

\qed

\section{Lower bounds: Counting contributions of $\mu = [2^*, 1^*] $}
\label{sect:lowbound}

We recall  the following notation: a partition $\mu = [2^*, 1^*] $ of a positive integer $n$
is composed only of parts of length $2$ and $1$. A permutation that has cycle structure $\mu = [2^*, 1^*] $ is simply a product of disjoint transpositions, possibly with some fixed points; such a permutation is an involution and, conversely, any involution on a finite set decomposes into a finite product of disjoint 2- and 1-cycles.

This section focuses on the case $\mu = [2^*, 1^*]$
to extract lower bounds on the column sum of characters.

\subsection{Cycle factorizations} 

We determine here the sufficient conditions for a partition $\mu$ to contribute
to the column sum of normalized characters. 
From this data, one can find lower bounds on the
column sum.

\begin{lemma}
\label{facto21}
Let $p>1$ be a positive integer, 
and $\tau $ a mapping such that
$\tau(i)= i+1$, for $ i \in \{1,2,\dots, p-1\}$.   
Then $\tau$ factorizes in a unique way as $\tau= \s_1\s_2$, 
where $\s_1$ and $\s_2$ are involutions, 
$\s_1$
acting on  $\{1,2,\dots, p\}$, 
and $\s_2$ acting on  $\{2,\dots, p\}$, 
 and where $\s_1(1) = p$.  
\end{lemma}
\proof 
The mapping  $\tau$ is denoted  $[1 \to 2 \to \dots \to p]$. 
 We  start with 
\bea 
&& \s_1 \s_2 = \tau =[1 \to 2 \to \dots \to p] \cr 
&& \s_1 \supset ( 1 , p ) 
\eea 
where the notation $\s_1 \supset (1,p)$ 
means that the 2-cycle $(1,p)$ appears in $\s_1$.  More generally, we will write
 $\s_i \supset \prod_{l}(a_l, a'_l)$, with 
$i=1,2$, and $a_l, a'_l \in \{1,2,\dots, p\}$, and that will mean that the finite product of cycles appears in $\s_i$.

To ensure $\s_1 \cdot \s_2 ( 1)= 2 = \tau (1)$  implies 
\bea 
\s_2 \supset ( p , 2 ) 
\eea
Recall that the composition rule is from left to right.

Now let $\s_1 \supset ( 2, x ) $, for some 
$x\in \{3,4,\dots, p-1\}$. This implies $ \s_1 \cdot \s_2 ( x ) = p
= \tau (x) $. Hence, there is no other choice but to fix $x = p-1$. 
Therefore 
\bea 
&& \s_1 \supset ( 1, p) ( 2, p-1 ) \cr 
&& \s_2 \supset ( p , 2 ) 
\eea 
To get the correct image $ \tau ( 2 ) = 3 $, we need 
\bea 
&& \s_1 \supset ( 1, p) ( 2, p-1 ) \cr 
&& \s_2 \supset ( p , 2  ) ( p-1 , 3 ) 
\eea
For a new unknow $x\in \{4,5,\dots, p-2\}$, consider 
\bea 
&& \s_1 \supset ( 1, p ) ( 2, p-1 ) (3, x )  \cr 
&& \s_2 \supset ( p , 2  ) ( p-1 , 3 ) 
\eea 
We arrive at $ \s_1 \cdot \s_2 ( x ) = p-1 = \tau(x)$, which means $ x = p-2$. So we learn 
\bea 
&& \s_1 \supset  ( 1, p ) ( 2, p-1 ) (3, p-2 )  \cr 
&& \s_2 \supset ( p , 2  ) ( p-1 , 3 ) 
\eea
Now to ensure $\s_1 \cdot \s_2 ( 3 ) = 4 =\tau (3)$, we must have 
\bea 
&& \s_1 \supset( 1, p ) ( 2, p-1 ) (3, p-2 )   \cr 
&& \s_2 \supset ( p , 2  ) ( p-1 , 3 ) ( p-2 , 4 ) 
\eea
The iteration problem boils down to the following: 
 for    $k\leq p$,  find $x_k$ fulfilling: 
\bea
&&
\s_1(1) = p  = x_1\crcr
&& 
\s_1(k) = x_k  \; , \qquad     \s_1 \cdot \s_2(k) = \s_2(x_k)= \tau (k) = k+1 
\eea
This yields
\bea
&&
k=1 \qquad 
\s_1(1) = x_1 =  p \qquad  \s_2(p)= 2 \crcr
&&
k=2 \qquad 
\s_1(2) = x_2 \qquad  \s_2(x_2)= 3 \crcr
&&
k=3 \qquad 
\s_1(3) = x_3 \qquad  \s_2(x_3)= 4 \crcr
&&
\vdots \crcr
&&
k \qquad \quad    \;\;\;
\s_1(k) =x _{k}\qquad  \s_2(x _{k})= k+1
\eea
Note that the iteration will stop at a point depending 
on whether $p$ is odd or even. We will come back to that
later on. 
The next step determines $x_k$. Indeed, since $\s_1$ 
and $\s_2$ 	are involutions: 
\bea
\s_1(x_k) = k \;, \qquad 
\s_2(k) = x_{k-1} \;, \qquad 
\s_1 \cdot \s_2 (x_k) = \s_2(k) = x_{k-1} =\tau (x_k)  
\eea
From $ x_{k-1} =\tau (x_k)  $, 
we infer that  the following recursive equation solves the problem: 
\bea
\left\{
\begin{array}{l}
x_k= x_{k-1} -1 \\
 x_1= p
 \end{array}
 \right. 
\eea
Expanding this, we obtain 
\bea
&&
k=2  \;, \qquad 
x_2 = x_1 - 1 = p-1  \;, 
\crcr
&&
k=3  \;, \qquad 
x_3 = x_2 - 1 = p-2  \;, 
\crcr
&&
\vdots \crcr
&&
k   \;, \qquad  \; \; \; \; \;\; 
x_{k} = x_{k-1} - 1 = p- k+1 \;, 
\eea

\begin{figure}
\centering
\includegraphics[scale=0.6]{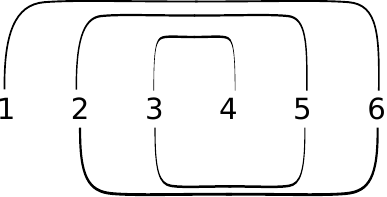}
\qquad \qquad  \qquad 
\includegraphics[scale=0.6]{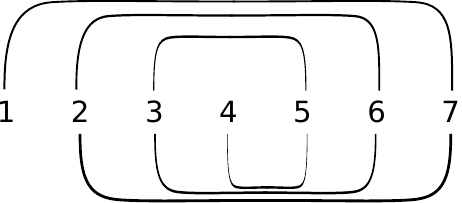}
\caption{(Left) Factorizing the mapping $[1 \to 2 \to \dots \to 6]$ as a product of involutions: 
$\s_1 = (1,6)(2,5)(3,4)$ (top arcs) and $\s_2 = (2,6)(3,5)(4)$ (bottom arcs). 
(Right) Factorizing the mapping $[1\to 2 \to \dots \to7]$ as a product of involutions: 
$\s_1 = (1,7)(2,6)(3,5)(4)$ (top arcs) and $\s_2 = (2,7)(3,6)(4,5)$ (bottom arcs). }
\label{Segment} 
\end{figure}

Thus, we   iterate this   $k$ times to learn that 
\bea 
&& \s_1 \supset ( 1, p) ( 2, p-1 ) \cdots ( k , p -k +1 )   \cr 
&& \s_2 \supset ( p , 2  ) ( p-1 , 3 )  \cdots ( p-k+1 , k+1 ) 
\label{s1s2kSeg}
\eea
This iteration will stop when $k$ reaches $p/2 $ or $(p+1)/2$. The detailed formula for the end point will depend on whether $p$ is even or odd. 
It is useful to draw 	pictures to understand the difference. See,  Figure \ref{Segment} that delivers the results for $ p = 6 $ and $7$.

\ 

\noindent{\bf $p$ is even --}
Suppose $p$ is even, then, when   $k$ reaches $p/2 $, 
\eqref{s1s2kSeg} becomes 
\bea 
\label{s1s2FactEven}
&& \s_1 = ( 1, p ) ( 2, p-1 ) \cdots ( k , p -k +1 ) \cdots ( p/2 , p/2 +1 )   \cr 
&& \s_2 =  ( p , 2  ) ( p-1 , 3 )  \cdots ( p-k+1 ,  k+1  ) \cdots (   p/2 + 2, p/2  ) (p/2+1) 
\eea
where $\s_2 (p/2+1)= p/2+1$. 
In this case, the decomposition of $\tau = \s_1\s_2$ gives 
$\s_1$ an involution of $\{1,2,\dots, p\}$, with $\s_1(1)= p$, 
and 
$\s_2$  an involution of $\{2,\dots, p\}$.

\

\noindent{\bf $p$ is odd --}
We inspect the case when $p$ is odd: 
 \eqref{s1s2kSeg} takes the form  
\bea 
\label{s1s2FactOdd}
&& \s_1  = ( 1, p ) ( 2, p-1 ) \cdots ( k , p -k +1 ) \cdots ( (p-1)/2 , (p +3)/2 )  
 (( p +1)/2  )   \cr 
&&
\hspace{-3mm} \s_2 = ( p , 2  ) ( p-1 , 3 )  \cdots ( p-k+1 ,  k+1  ) 
\cdots  (  (p +5)/2 ,   (p-1)/2  )  (  (p +3)/2 ,   (p+1)/2 ) \crcr
&&
\eea
where $\s_1 (( p +1)/2)=(p +1)/2$. 
We conclude that $\s$ decomposes again as  $\s_1\s_2$, 
with the required properties of $\s_1$ and $\s_2$.

\qed 

Lemma \ref{facto21} entails that there is unique factorization 
of the mapping $[1\to 2\to 3 \to \dots \to p]$ in terms of involutions 
with only 2- and 1-cycles, for a given 
positive integer $p$. This property will be crucial to 
determine the factorizations of  odd-cycles.  
One also observes that $\s_1$ is a permutation
of $S_p$ meanwhile, $\s_2$ is not (unless we extend 
it by imposing $\s_2(1)=1$). 
From   the proof of Lemma \ref{facto21}, in particular \eqref{s1s2FactEven}
and \eqref{s1s2FactOdd}, the following 
enumeration statement is straightforward: 
\begin{proposition}\label{enumFacto21}
Let $\tau$ be the mapping $[ 1 \to 2 \to \dots \to p]$, $p>1$. 
Let  $\s_1$ and $\s_2$ the unique involutions
that factor $\tau $ as  $\tau = \s_1\s_2$, with $\s_1(1) = p$. 
Then 
\begin{itemize}
\item If $p$ is  even, then 
$\s_1$ is of cycle structure $[2^{p/2}]$ and 
$\s_2$ of cycle structure  $[2^{p/2-1},1]$. 

\item If $p$ is odd, then $\s_1$ is of cycle structure $[2^{(p-1)/2},1]$ and 
$\s_2$ of cycle structure  $[2^{(p-1)/2},1]$. 

\end{itemize}
\end{proposition}

We are in position to state crucial factorization properties
of cycles. 
\begin{theorem}
\label{factoP21}
Let $n$ be an odd positive integer, 
and $\sigma $ a $n$-cycle permutation of $S_n$.  
There are $n$ splittings $\s = \s_1\s_2$ 
where $\s_1$ and $\s_2$ are permutations of $S_n$
that have identical cycle structure $[2^{(n-1)/2},1]$. 
\end{theorem}

\begin{figure}
\centering
 \includegraphics[scale=0.6]{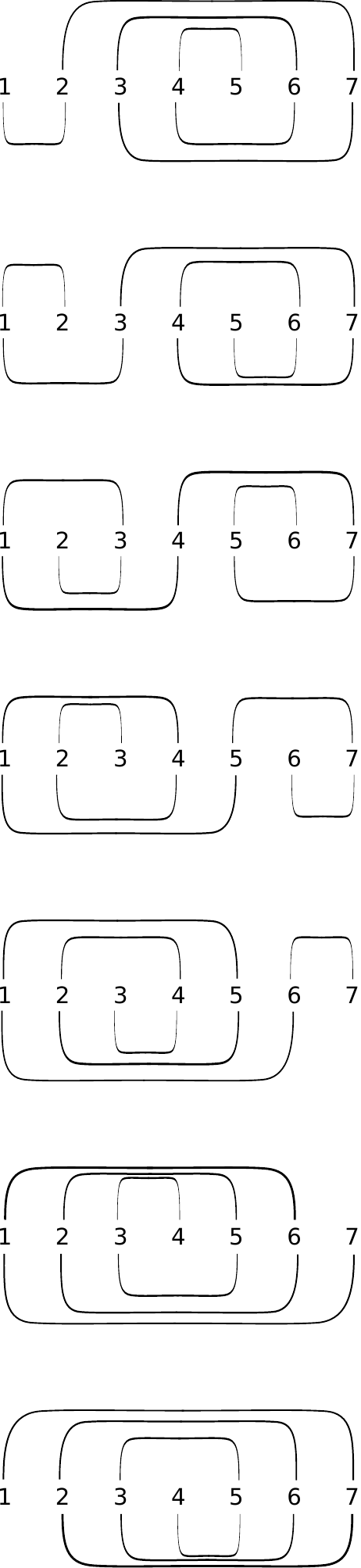} 
   \put(15,560){$\s_1=(2,7)(3,6)(4,5)(1)$}
    \put(15,530){$\s_2=(1,2)(3,7)(4,6)(5)$}
   \put(15,470){$\s_1=(1,2)(3,7)(4,6)(5)$}
    \put(15,440){$\s_2=(1,3)(4,7)(5,6)(2)$}
   \put(15,390){$\s_1=(1,3)(4,7)(5,6)(2)$}
    \put(15,360){$\s_2=(1,4)(2,3)(5,7)(6)$}
   \put(15,310){$\s_1=(1,4)(2,3)(5,7)(6)$}
    \put(15,280){$\s_2=(1,5)(2,4)(6,7)(3)$}
   \put(15,220){$\s_1=(1,5)(2,3)(6,7)(3)$}
    \put(15,190){$\s_2=(1,6)(2,5)(3,4)(7)$}
   \put(15,130){$\s_1=(1,6)(2,5)(3,4)(7)$}
    \put(15,100){$\s_2=(1,7)(2,6)(3,5)(4)$}
   \put(15,40){$\s_1=(1,7)(2,6)(3,5)(4)$}
    \put(15,10){$\s_2=(2,7)(3,6)(4,5)(1)$}
 \caption{The 7 splittings of a 7-cycle $(1,2,3,4,5,6,7)$.}
 \label{figsplitoddmax1} 
 \end{figure}
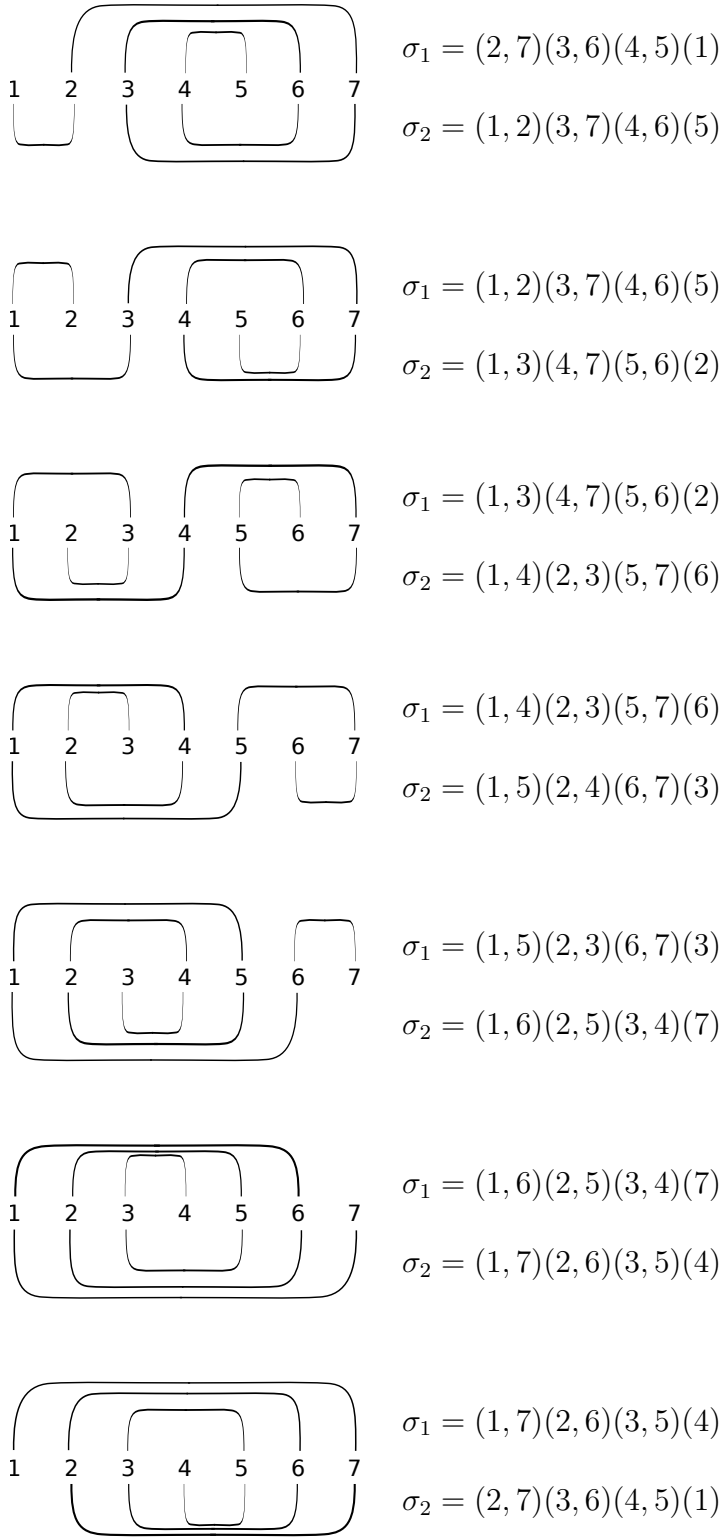

Figure \ref{figsplitoddmax1} illustrates that, for a $7$-cycle, there are $7$ such splittings. 
 We introduce the following terminology: looking each 
 diagram, we observe that the set $\{1,2,\dots, n\}$ is partitioned
 in  two subsets: 
 the first denoted by $I = \{1,\dots, \s_1(1)\}$ and 
 $\bar I = \{\s_1(1)+1, \dots, n\}$. Note that $\bar I$ may be empty
 when $\s_1(1)+1 = n+1$. 
 In a loose way, we will call {\it spiral} the compositions of transpositions staying within  $I$,  or within $\bar I$. Pictorially, these sectors are precisely the spirals observed in Figure \ref{figsplitoddmax1}.

\ 

\noindent{\bf Proof of Theorem \ref{factoP21}.} 
Take $\s = ( 1, 2, \cdots , n  ) $. We want to find $ \s = \s_1 \s_2 $ with  $\s_1 , \s_2$ in class $\mu=[2^*,1^*] $ constrained as above. We will show that once we fix $\s_1 (1) = i $ for any $ i \in \{ 1, \cdots , n \} $, there is a unique pair $\s_1$ and $\s_2 $ satisfying the equation. So we are starting with 
\bea \label{initProblem}
&& \s_1 \s_2 = \s = ( 1 , 2, \cdots, n ) \cr 
&& \s_1 \supset ( 1 , i ) 
\eea 

We are led to the next sub-problem: 
given $i$, find a decomposition of $\s' = [1  \to 2 \to \dots \to  i]$ into a
product $\s'_1 \s'_2 $, where $\s'_1$ and   $ \s'_2 $  are  involutions, and 
$\s'_1(1) = i$. 
Depending on the parity of $i$ (even or odd),  this problem admits a unique solution  according to Lemma \ref{facto21}: 

\ 

\noindent{\bf $i$ is even --}
When   $k$ reaches $i/2 $, we obtain (see \eqref{s1s2FactEven})
\bea 
&& \s'_1  =  ( 1, i ) ( 2, i-1 ) \cdots ( k , i -k +1 ) \cdots ( i/2 , i/2 +1 )   \cr 
&& \s'_2 =  ( i , 2  ) ( i-1 , 3 )  \cdots ( i-k+1 ,  k+1  ) \cdots (   i/2 + 2, i/2  ) (i/2+1) 
\eea
where $\s'_2 (i/2+1)= i/2+1$. Note that we have 
reached all points within $I=\{1,2,\dots, i\}$, but not 
points beyond $i+1$, i.e. within the set $\bar I = \{i+1, \dots, n\}$. 
Now to get $ \s ( i ) = i+1$, we require 
\bea
&&\s_1 \supset \s'_1 \crcr
&& \s_2 \supset \s'_2    \; ( 1 , i+1 ) 
\eea

Suppose  $ \s_1 ( i+1 ) = x$, then $\s_1 ( x )  =  i+1 $ 
(as $\s_1^2=id$) and $ \s_1 \cdot \s_2 ( x ) = 1$. But we know that $\s ( n )  = 1 $. So it follows that $\s_1 ( i+1 ) = n $.  Hence 
\bea \label{partials}
&& \s_1 \supset \s'_1 \ ;  \;  ( i+1, n )   \crcr
&& \s_2 \supset \s'_2\;  ( 1 , i+1) 
\eea
where the semi-colon separates the cycles that are related to the right 
and left spirals in the diagrams. 
Once again, our factorization problem is similar to the previous one: 
we wish to factorize $\s'' = [i+1 \to i+2 \to \dots \to n]$ into 
two involutions $\s''_1$ and $\s''_2$  (but acting on 
the interval $ \{i+1,i+2, \dots, n\}$ and $\{i+2,1+3,\dots, n\}$, respectively),
 such that $\s''_1(i+1)= n$. 
 Following  Lemma \ref{facto21}, there is a unique solution of this problem. 
Note that $n$ is odd and $i$ even, so $n-i$ is   odd. 
We obtain the the factorization of $\s''$ as 
 \bea 
\label{s''1s''2}
\s''_1  &=& ( i+1, n ) ( i+2, n-1 ) \cdots ( i+k , n -k +1 ) \cdots \crcr
&&  ( (n+i-1)/2 , (n+i +3)/2 )  
 (( n +i+1)/2  )   \cr\cr
 \s''_2 &=& ( n , i+2  ) ( n-1 , i+3 )  \cdots ( n-k+1 ,  i+k+2  ) 
\cdots  \crcr
&&  (   (n+i+5)/2,  (n+i -1)/2  )  (   (n+i+3)/2,  (n+i +1)/2  ) 
\eea
We easily conclude that $\s$ factorizes as $\s_1 \s_2$ with 
\bea 
\label{partial2}
&& \s_1  =  \s'_1  \; ;   \;  \s''_1 \crcr
&& \s_2 = \s'_2\;  ( 1 , i+1) \;  ;\,  \s''_2 
\eea
It is immediate that $\s_1$ and $\s_2$ become permutations of $S_n$. 
Proposition \ref{enumFacto21} yields the cycle structure of $\s_1$ and $\s_2$ as: 
\bea
&&
[\s_1]= [\s'_1][\s''_1]  = [2^{i/2}] \cup [2^{(n-i-1)/2},1] =  [2^{(n-1)/2},1]
\cr\cr
&&
[\s_2]= [\s'_2][(1,i+1)][\s''_2]  = [2^{i/2-1},1]\cup[2] \cup [2^{(n-i-1)/2}] =  [2^{(n-1)/2},1]
\eea
Thus $\s_1$ and $\s_2$ have identical cycle structure. 

\

\noindent{\bf $i$ is odd --}
The case of $i$ odd is analogous to the above case, with a few differences
that we will highlight   in a streamlined analysis. 
We call twice Lemma \ref{facto21} to be able factorize both
the partial mappings  $[1  \to 2 \to \dots \to  i]$ and $[i+1 \to i+2 \to \dots \to n]$ corresponding to the left and right spirals.
We will use the same notation altough the variable may refer to 
different quantities.

 Use Lemma \ref{facto21}  (see \eqref{s1s2FactOdd}) and write the 
 unique solution of the  problem  of factorizing $[1 \to 2 \to \dots \to i]$, when $i$ odd,  as
\bea 
\label{s1s2odd1}
&&
\hspace{-5mm} \s'_1  = ( 1,  i) ( 2, i-1 ) \cdots ( k , i -k +1 ) \cdots ( (i-1)/2 , (i +3)/2 )  
 (( i +1)/2  )   \cr 
&&
\hspace{-5mm} \s'_2 = ( i , 2  ) ( i-1 , 3 )  \cdots ( i-k+1 ,  k+1  ) 
\cdots  (  (i +5)/2 ,   (i-1)/2  )  (  (i +3)/2 ,   (i+1)/2 ) \crcr
&&
\eea
where $\s'_1(1)= i$  and $\s'_1 (( i +1)/2)=(i +1)/2$. 
 
Under the same constraints,  $\s_1(i+1)= x$, and $\s_1 \cdot \s_2(x)= 1$,
impose that $\s_1(i+1) = n$, and thus we learn that 
 \eqref{partials} applies in the present case as well. 
 
 We are left with the factorization of $[i+1 \to i+ 2 \to \dots \to n]$ 
 in two permutations made of 2- and 1-cycles with 
 initial condition $\s''_1(i+1)= n$. Lemma \ref{facto21}
 delivers once more the unique solution which
 is 
\bea 
\label{s1s2odd2}
\s''_1  &=& (  i+1,n) ( i+2, n-1 ) \cdots ( i+k , n -k +1 ) 
\crcr
&& \cdots  ((n+i)/2 -1 , (n+ i)/2  + 2 )   
 ((n+i)/2, (n+ i)/2 +1  )   
 \cr\cr
\s''_2 &=& ( n, i +2  )(n-1, i+3)  \cdots (   n-k,  i+k+2  ) 
\crcr
&& 
\cdots (( (n+ i)/2 +2, (n+i)/2  )((n+ i)/2 +1  )    
\eea
The solution is of  the same form as \eqref{partial2}
and the cycle structure of the factors $\s_1$ and $\s_2$
permutations of $S_n$ are delivered by Proposition \ref{enumFacto21}: 
\bea
&&
[\s_1]= [\s'_1][\s''_1]  = [2^{(i-1)/2}, 1] \cup [2^{(n-i)/2}] =  [2^{(n-1)/2},1]
\cr\cr
&&
[\s_2]= [\s'_2][(1,i+1)][\s''_2]  = [2^{(i-1)/2}]\cup[2] \cup [2^{(n-i)/2 -1},1] =  [2^{(n-1)/2},1]
\eea
Hence, $[\s_1]=[\s_2]$ as expected. 

In conclusion, for each choice of $i \in \{1,2,\dots, n\}$, $\s$ factorizes in 
$\s_1 \s_2$ with identical cycle structure made only 
of 2- and 1-cycles. There are $n$ such choices. 

\qed 

There is an analogous theorem for even-length cycle. 

\begin{theorem}
\label{factoP21Even}
Let $n$ be an even positive integer 
and $\sigma $ a $n$-cycle permutation of $S_n$.  
There are $n$ splittings $\s = \s_1\s_2$ 
where $\s_1$ and $\s_2$ are permutations of $S_n$,
with cycle structure $[2^*,1^*]$.
There are $n/2$ splittings determined by  
$[\s_1] = [2^{n/2}]$ and $[\s_2] =[2^{n/2-1},1^2]$, 
and, vice-versa, there are $n/2$ splittings determined 
by $[\s_1] =  [2^{n/2-1},1^2]$ and $[\s_2] = [2^{n/2}]$. 
\end{theorem}
\proof The proof of this theorem is similar to that
of Theorem \ref{factoP21} with some differences 
that we highlight.
We want to solve $\s = ( 1, 2, \cdots , n  ) = \s_1 \s_2 $ with  $\s_1 , \s_2$ in class $\mu=[2^*,1^*] $ constrained as above. Fixing $\s_1 (1) = i $ for any $ i \in \{ 1, \cdots , n \} $
leads to unique pair $\s_1$ and $\s_2 $. 

The initial problem is the same as \eqref{initProblem}. 
Depending on the parity of $i$ (even or odd),  
 Lemma \ref{facto21} delivers the unique solution  factorization 
of the following sub-problems: 

 - given $i$, find a decomposition of $\s' = [1 \to 2 \to \dots \to i]$ into a
product $\s'_1 \s'_2 $, where $\s'_1$ and   $ \s'_2 $  are  involutions, and 
$\s'_1(1) = i$.

- given $n-i$, find a  decomposition of $\s' = [i+1, \to i+ 2 \to \dots \to n]$ into a
product $\s''_1 \s''_2 $, where $\s''_1$ and   $ \s''_2 $  are  involutions, and 
$\s''_1(i+1) = n$.

\noindent{\bf $i$ is even --}
We obtain from  Lemma \ref{facto21}  (see \eqref{s1s2FactEven})
\bea 
&& \s'_1  =  ( 1, i ) ( 2, i-1 ) \cdots ( k , i -k +1 ) \cdots ( i/2 , i/2 +1 )   \cr 
&& \s'_2 =  ( i , 2  ) ( i-1 , 3 )  \cdots ( i-k+1 ,  k+1  ) \cdots (   i/2 + 2, i/2  ) (i/2+1) 
\eea
where $\s'_2 (i/2+1)= i/2+1$. Now to get $ \s ( i ) = i+1$ and  $\s ( n )  = 1 $, we require 
\bea \label{partialsEven}
&& \s_1 \supset \s'_1 \ ;  \;  ( i+1, n )   \crcr
&& \s_2 \supset \s'_2\;  ( 1 , i+1) 
\eea
At this point, we wish to factorize $\s'' = [i+1 \to i+2 \to \dots \to n]$ into 
two involutions $\s''_1$ and $\s''_2$, such that $\s''_1(i+1)= n$. 
 Following  Lemma \ref{facto21}, there is a unique solution of this problem. 
Using the fact that $n$ and $i$ both even, 
we know that $n-i$ is  even and  obtain the  factorization of $\s''$ as 
 \bea 
\label{s''1s''2Even}
\s''_1  &=& ( i+1, n ) ( i+2, n-1 ) \cdots ( i+k , n -k +1 ) \cdots \crcr
&&  ( (n+i)/2 , (n+i)/2 +1)  
   \cr\cr
 \s''_2 &=& ( n , i+2  ) ( n-1 , i+3 )  \cdots ( n-k+1 ,  i+k+2  ) 
\cdots  \crcr
&&  (   (n+i)/2+3,  (n+i )/2-1  )  (     (n+i)/2+2 , (n+i)/2) 
( (n+i)/2 + 1) 
\eea
Hence $\s$ factorizes as $\s_1 \s_2$ with 
$\s_1  =  \s'_1  \; ;   \;  \s''_1$ and $ \s_2 = \s'_2\;  ( 1 , i+1) \;  ;\,  \s''_2 $
where the semi-colon separates the right and left spirals. 
One agrees with that $\s_1$ and $\s_2$ are permutations of $S_n$. 
Proposition \ref{enumFacto21} yields the cycle structure of $\s_1$ and $\s_2$ as: 
\bea
&&
[\s_1]= [\s'_1][\s''_1]  = [2^{i/2}] \cup [2^{(n-i)/2}] =  [2^{n/2}]
\cr\cr
&&
[\s_2]= [\s'_2][(1,i+1)][\s''_2]  = [2^{i/2-1},1]\cup[2] \cup [2^{(n-i)/2-1},1] =  [2^{n/2-1},1^2]
\eea
 There are $n/2$ possibilities for such factorizations
 entirely determined by $\s_1(1)$ even. This proves one part of the claim.  

\

\noindent{\bf $i$ is odd --}
 By Lemma \ref{facto21} \eqref{s1s2FactOdd}, we infer 
\bea 
&&
\hspace{-5mm} \s'_1  = ( 1,  i) ( 2, i-1 ) \cdots ( k , i -k +1 ) \cdots ( (i-1)/2 , (i +3)/2 )  
 (( i +1)/2  )   \cr 
&&
\hspace{-5mm} \s'_2 = ( i , 2  ) ( i-1 , 3 )  \cdots ( i-k+1 ,  k+1  ) 
\cdots  (  (i +5)/2 ,   (i-1)/2  )  (  (i +3)/2 ,   (i+1)/2 ) \crcr
&&
\eea
where $\s'_1(1)= i$  and $\s'_1 (( i +1)/2)=(i +1)/2$. 
 Requiring both $\s(i)= i+1$ and $\s(n)= 1$ leads us 
 to \eqref{partialsEven}. 
The factorization of  $\s'' = [i+1 \to i+2 \to \dots \to n]$ 
 in two involutions $\s''_1$ and $\s''_2$ with  initial condition $\s''_1(i+1)= n$
 can be treated  by Lemma \ref{facto21}. 
 Keeping in mind that  $n-i$ is now odd,  it delivers: 
\bea 
\label{s1s2odd2}
\s''_1  &=& (  i+1,n) ( i+2, n-1 ) \cdots ( i+k , n -k +1 ) 
\crcr
&& \cdots  ((n+i-1)/2, (n+ i+3)/2  )    ((n+ i+1)/2) 
 \cr\cr
\s''_2 &=& ( n,  i +2  )(n-1, i+3)  \cdots (   n-k,   i+k+2  ) 
\crcr
&& 
\cdots ((n+ i+3)/2  , (n+i+1)/2  )
\eea
The solution is of  the same form as \eqref{partial2}
and the cycle structure of the factors $\s_1$ and $\s_2$
permutations of $S_n$ are delivered by Proposition \ref{enumFacto21}: 
\bea
&&
[\s_1]= [\s'_1][\s''_1]  = [2^{(i-1)/2}, 1] \cup [2^{(n-i-1)/2},1] =  [2^{n/2-1},1^2]
\cr\cr
&&
[\s_2]= [\s'_2][(1,i+1)][\s''_2]  = [2^{(i-1)/2}]\cup[2] \cup [2^{(n-i-1)/2}] =  [2^{n/2}]
\eea
which was expected.  One observes that there are $n/2$ possibilities for such factorizations uniquely determined by $\s_1(1)$ odd. This proves the second part of the claim.

\qed

\subsection{Disconnected and pairwise connected factorisations of cycles} 

\begin{figure}[h]
\centering
\includegraphics[scale=0.6]{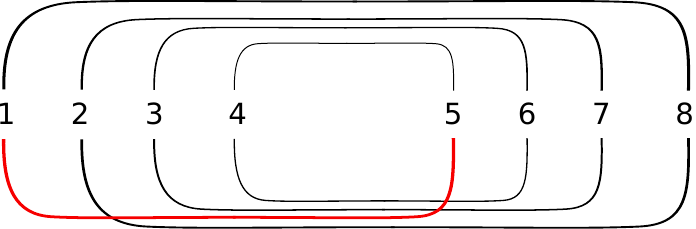}
\caption{Factorizing two cycles $(1,2,3,4)(5,6,7,8)$ as a product of permutations: 
$\s_1 = (1,8)(2,7)(3,6)(4,5)$ (top arcs) and $\s_2 = (2,8)(3,7)(4,6)(5,1)$ (bottom arcs). }
\label{Factor2cycles} 
\end{figure}

Considering a generic permutation $\sigma$ with an arbitrary number of cycles, we now consider  factorizations into two  permutations  $ \sigma_1 , \sigma_2$ of cycle structure $[2^*, 1^*]$, seeking generalisations of
 Theorem \ref{factoP21} or by \ref{factoP21Even}). Of course, we can apply those theorems to each $n$-cycle and obtain, based on the parity of $n$,  a factorization that we will call ``disconnected''. Nonetheless, the factorization may also entail maps that send elements accross different cycles. As an illustration, Figure \ref{Factor2cycles} depicts a factorization of two $4$-cycles that is distinct from the factorization obtained from individual cycles. This section elucidates factorizations of this nature. A key point will be that such factorisations will involve, beyond the ``disconnected'' factorisations just one new type of factorisation : those factorisations which pair cycles of  $ \sigma$, i.e. where  $ \sigma_1 , \sigma_2$ map the numbers involved in one cycle of $ \sigma$ with exactly one other cycle of $ \sigma$. This is a key element of Theorem  \ref{2cyclefacto}. 

\begin{figure}[h]
\centering
\includegraphics[scale=0.6]{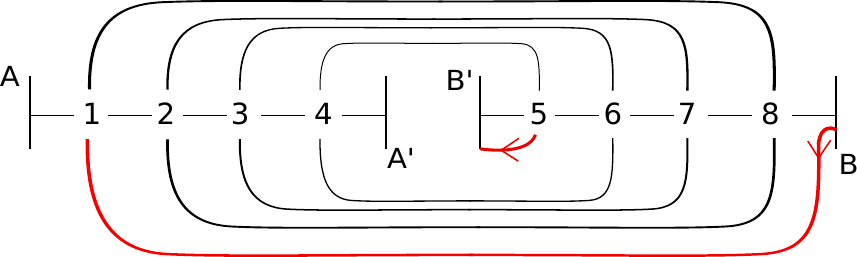}
\caption{Another way of representing the factorization 
of   $(1,2,3,4)(5,6,7,8)$ given in Figure \ref{Factor2cycles}. }
\label{Factor2cyclesNCross} 
\end{figure}

This key new element is illustrated in Figure \ref{Factor2cycles}. Given the planarity property of factorisations into $ [2^* , 1^*]$ discussed in section \ref{genrest}, it is useful to introduce an  additional diagram to illustrate the factorization, which does not  have  intersecting arcs.  Figure \ref{Factor2cyclesNCross} gives such a diagram. 
Each cycle $(1,2,\dots, n)$ is symbolized  by
a segment $[1,2,\dots, n]$ with endpoints $A$ and $A'$ which are identified.  Consequently, in the illustration,
the red arc  $(5,1)$, which connects $5$ to $1$ in Figure \ref{Factor2cycles}, 
is represented as follows:  the arc passes through $B'$
which is identified with $B$, and from $B$, and then  joins
$1$ without intersecting any other arc.

\begin{theorem}\label{2cyclefacto}
Let $n$ and $n'$ be two positive integers, and a permutation $\s =(1,2,\dots, n)(n+1,n+2,  \dots, n+ n')$. If $n=n'$ and $ \s_1 (1)  \in \{ n+1 , n+2,\cdots , n+ n' \}$, then
there are $n$ splittings $\s = \s_1 \s_2$, where $\s_1$ and $\s_2$ permutation of $S_{2n}$ with identical cycle structure 
$[2^{n}]$. The $n$ possibilities come from the choice of $n$ possible images  $ \sigma_1 (1)$. 
If  $n\ne n'$ and  $ \s_1 (1) \in \{ n+1, n+ 2,\cdots ,n+ n' \}$, there is no such a splitting. 
\end{theorem}

\

\noindent
{\bf Proof of Theorem \ref{2cyclefacto}.} 
The proof relies on the same ideas that prove Lemma \ref{facto21}. 
We want to find $\s = (1,2,\dots, n)(n+1,n+2, \dots, n+ n')= \s_1 \s_2 $ with  $\s_1 , \s_2$ in class $\mu=[2^*,1^*] $ constrained as above. 

Without loss of generality, let us assume that
$n'>n$. 
Fixing  $\s_1 (1) = n+i $ for any $ i \in \{ 1, 2,\cdots , n' \} $,
we can cycle around the points 
of the cycle $(n+1,n+2, \dots, n+i, n+i+1, \dots, n+ n')$
and map it to 
\bea
(n+i+1, n+i+2, \dots, n+n'-1, n+ n', n+1,n+2, \dots, n+i)
\eea
Then we use the obvious bijection between the labels of this cycle and the initial one
\bea
&& 
n+i+1 \to n+1 \qquad 
 n+i+2 \to  n+2  \qquad 
  \dots \qquad n+i \to n+n'
\eea
in order to study only the case $i=n'$. 
Indeed, the proof relies on the  difference between 
the cycle lengths. 
 
We show that there is a unique pair of involutions, $\s_1$ and $\s_2 $, satisfying the equation if and only if $n=n'$. 
We start with the problem: 
\bea \label{Probdisc}
&& \s_1 \s_2 = \s =(1,2,\dots, n)(n+1,n+2,n+3, \dots, n+ n') \cr 
&& \s_1 \supset ( 1 , n+n' ) 
\eea 
The construction of $\s_1$ and  $\s_2$ follows the same routine  as for Lemma \ref{facto21} and they are unique
if they exist. 
We build successively the 2-cycles of $\s_1$ and $\s_2$. 
The  difference in the present case is that 
$\s_1$ and $\s_2$ can only pair 
elements of the cycle $(1,2,\dots, n)$ with elements of 
$(n+1,n+2,n+3, \dots, n+ n')$. 
A moment of thought leads us to the $k$th step
of the iteration process 
\bea
&&
\s_1 \supset (1,n+n') (2,n+n'-1) \dots (k, n+ n'-k+1)  \crcr
&&
\s_2 \supset  (n+n',2)(n+n'-1, 3) \dots (n+n'-k+1, k+1)
\eea
Without any issue, the process can go down from $k=0$ up to $k=n<n'$. 
To solve the  constraint $\s(n)=1$  
requires 
\bea
&&
\s_1 \supset  (1,n+n') (2,n+n'-1) \dots (k, n+ n'-k+1) 
\dots  (n-1, n'+2) (n, n'+1) \crcr
&&
\s_2 \supset  (n+n',2)(n+n'-1, 3) \dots (n+n'-k+1, k+1)
 \dots (n'+2,n) (n'+1, 1) \crcr
 && 
 \label{i>n}
\eea
With this, all points of the cycle $(1,2,\dots, n)$ have 
a two arcs. From our assumption, note that $n<n' \le  n'+n $
 so it must have an image by $\s$. 

Let us call $\s_1(n')= x$. 
 Then 
$\s(n') = n'+1 =\s_2(\s_1(n') ) = \s_2(x) $. 
Because $\s_2$ is an involution, 
 according to \eqref{i>n}, $x=1$. 
 So $\s_1(n') = x = 1 $ but $\s_1$ 
 is also an involution and we already know
 that $\s_1(n+n') = 1$. This makes sense only 
 if $n+n' = n'$ implying that $n=0$
 which contradicts our hypothesis. 

In summary, a $[2^*,1^*]$-decomposition  of $\s  = \s_1 \s_2$
will be not possible when $n'>n$ as some
elements within the cycle $(n+1, \dots, n+n')$ 
(of exceeding length compared to $(1,\dots, n)$)
cannot be paired in a consistent way
during the composition $\s_1 \s_2$. 

The proof of the statement is achieved once
we agree with the fact that we choose $\s_1(1)=n+i$,  for $i\in \{1,2,\dots, n\}$,  and observe that $\s_1$ and $\s_2$ 
have no fixed points, in which case $[\s_1] = \s_2 = [2^n]$. 

\qed 

A complete expression for $\s_1$ and $\s_2$ 
can be easily withdrawn by completing the iteration: 

- if $i<n $
\bea
&&
\s_1  = (1,n+i) (2,n+i-1) \dots (k, n+ i-k+1) 
\dots  \crcr
&& \qquad 
(i, n+1) (i+1, 2n) (i+2, 2n-1) \dots (n, n+i +1)  \cr\cr
&&
\s_2 =  (n+i,2)(n+i-1, 3) \dots (n+i-k+1, k+1) 
 \dots 
   \crcr
&& \qquad 
(n+1, i+1)  (2n, i+2) (2n-1, i+3) \dots (n+i +1, 1)  
 \label{i<n2}
\eea 

- if $i=n$
\bea
&&
\s_1 = (1,2n) (2,2n-1) \dots (k, 2n-k+1) 
\dots 
(n, n+1)   \cr\cr
&&
\s_2 =  (2n,2)(2n-1, 3) \dots (2n-k+1, k+1) 
 \dots  
(n+1,  1)  
 \label{i=n2}
\eea 

\ 

\noindent{\bf Terminology.} 
For use in subsequent sections, we introduce the following terminology. 
We call {\bf disconnected splitting} a factorization 
of a permutation $\s$
as $\s =\s_1 \s_2$ where $\s_1$  and $\s_2$ map each number within a cycle of $\s$ to a number  within the same cycle.  
For $\s$ composed of two cycles 
$c_1 c_2$, we refer to the type of splitting
described by Theorem \ref{2cyclefacto} as 
 {\bf paired connected splitting}. This involves mappings that
 connect elements from one cycle to another. 
 A general splitting of a permutation $\s$
 will be a collection of paired connected 
 and disconnected splittings.

\subsection{A lower bound result for the column sum}

We can know derive lower bounds for the column sum problem. Let us keep in mind that 
\bea 
\sum_{ R} {  \chi^R ( T_{\lambda } ) \over d_R } = \sum_{ \mu } { 1 \over |\cC_\mu  | } \delta ( T_{ \mu } T_{ \mu} T_{ \lambda } ) 
 =  \sum_{ \mu } { |\cC_\lam | \over |\cC_\mu  | }  
 \delta ( T_{\mu} T_{ \mu} \sigma_3^* )
\eea 
Hence generally, we can write 
\bea
 \sum_{ R} {  \chi^R ( T_{\lambda } ) \over d_R }  = 
  \sum_{ \mu } { |\cC_\lam | \over |\cC_\mu  | }   
  \times \Big|  \{ (\s_1, \s_2) \in \cC_\mu \times \cC_\mu | \s_1 \s_2 \s_3^* = \id \}  \Big|
\eea
In the following analysis, we will confine the summation over  $\mu$ to the subset of classes $[2^*,1^*]$. 
This restriction naturally leads to a lower bound 
of the column sum. 
It appears that the above set cardinality which  represents the number of pairs $(\s_1, \s_2) \in \cC_\mu \times \cC_\mu$ obeying  $\s_1 \s_2 \s_3^* = \id$, 
or equivalently, the pairs that factor $(\s_3^*)^{-1} = \s_1\s_2$ can be achieved through some combinatorics. 

\subsubsection{Odd-length  cycles} 

Let us see concretely how Theorem \ref{factoP21} leads to a lower bound on the sum of normalized central characters. 

\

\noindent{\bf Single cycle $ \lambda = [ n ] $, with $n$ odd. }
Here we are finding for $ \lambda = [ n ] $,
with the restriction to $\mu = [2^* , 1^* ] $ that
\bea 
\delta ( T_{\mu} T_{ \mu} \sigma_3^* ) = n 
\eea
where $\sigma_3^* $ is a fixed permutation in the specified conjugacy class. 
Indeed, Theorem \ref{factoP21} tells us there are $n$ ways to  factorize 
 $(\sigma_3^*)^{-1}$  in $\s^{1}_{\mu} \s^{2}_{ \mu}$ 
where $\s^{1}_{\mu}$ and $ \s^{2}_{ \mu}$ are in the same
 class  $\mu =[ 2^{ (n-1)/2} , 1 ]$.

Here 
$|\cC_{ \lambda } | = ( n -1) ! $ and $ \mu = [ 2^{ (n-1)/2} , 1 ]$  which gives
\bea 
|\cC_{ \mu}  | = { n! \over 2^{ (n-1)/2} (( n-1)/2)! } 
\eea
We evaluate 
\bea 
{ 1 \over |\cC_{ \mu}  | } \delta ( T_{\mu} T_{ \mu} T_{ \lambda } ) = n \;  ( n-1)! \;  {  2^{ (n-1)/2} (( n-1)/2)!\over n!  }  =  2^{ (n-1)/2} (( n-1)/2)!
\eea
Given $\lambda= [n]$, 
this allows us to write  a formula for  $C_{\mu\lam}^{\;\;\;\mu}$ with $ \mu = [ 2^{ (n-1)/2} , 1 ]$: 
\bea
C_{\mu\lam}^{\;\;\;\mu} = 2^{ (n-1)/2} (( n-1)/2)!
\eea
So we have the following lower bound 
\beq 
\sum_{ R \vdash n } {  \chi^R ( T_{\lambda = [ n ]  } ) \over d_R }  
= \sum_{\mu  \vdash n} C_{\mu\lam}^{\;\;\;\mu}  \ge
\sum_{\mu \in [2^*, 1^*]} C_{\mu\lam}^{\;\;\;\mu} 
 = C_{[2^{ (n-1)/2} , 1 ]\lam}^{\;\;\;[2^{ (n-1)/2} , 1 ]}  = 
  2^{ (n-1)/2} (( n-1)/2)!
  \label{BSOddCyc}
\eeq

Varying the value of $n$ we obtain   Table \ref{tab:singlecycl} 
using  Sage programming. Appendix \ref{app:Bounds} provides the codes that generate this data.

\begin{table}[ht]
\centering
\begin{tabular}{l|l|l|l}
$\lambda(n)$ & $\sum_{ R \vdash n } {  \chi^R ( T_{\lambda = [ n ]  } ) \over d_R }  $  & Lower bound $=  C_{[2^{ (n-1)/2} , 1 ]\lam}^{\;\;\;[2^{ (n-1)/2} , 1 ]} $ & Relative error \\
\hline\hline
$[3]$ & 3     & 2    &    0.333 \\
$[5]$ & 40     & 8    &    0.8 \\ 
$[7] $ & 1260 & 48 & 0.961  \\
$[9] $ & 72576& 384& 0.994 \\
$[11] $ & 6652800& 3840& 0.999 \\
$[13] $ & 889574400& 46080& 0.999 \\
$[15] $ & 163459296000& 645120& 0.999 \\
$[17] $ & 39520825344000& 10321920& 0.999 \\
$[19] $ & 12164510040883200& 185794560& 0.999 \\
\hline
\end{tabular}
\caption{The column sum evaluation and its lower bound for  
$n=3,5,\dots, 19$ odd.}
\label{tab:singlecycl}
\end{table}

\ 
\noindent{\bf{Multiple cycles of one odd length: $ \lambda = [ a^p ] $,  $p$ cycles of fixed odd length $a$}:}

Take the case where $\lambda = [ a^{ p} ] $ with $a$ odd. Pick a fixed $ \sigma_3 \in \cC_{ \lambda }$.
We can  pair a subset of the  $p$ cycles to produce 
paired connected splittings and leave the remaining ones
 to produce  disconnected splittings.
Suppose we choose $2k$ elements of the $p$ cycles, where $ k \in \{ 1,2, \cdots , \lfloor { p \over 2 }  \rfloor \}$.
The number of possible pairings is $(2k-1)!! := (2k-1) \cdot (2k-3) \cdot \dots 3 \cdot 1$.
For each $k$, the total number of splittings is therefore
\bea
\label{k-decomp}
 \binom{p}{2k}  (2k-1)!!   \;  a^k \, a^{ p - 2k }  
 = { p! \over (2 k)! ( p - 2k) ! }  \;   \frac{ (2k)!  }{ (2k)!!} \, a^{ p - k } = 
{ p! \over 2^k k! ( p - 2k) ! }  \,  a^{ p - k }   
\eea
Above, Theorem \ref{factoP21} gives for the $k$ pairs,
 $a^{k}$ possible paired connected splittings. For the $p-2k$ remaining cycles, Theorem \ref{factoP21}
delivers the number of disconnected splittings as $a^{p-2k}$.

We seek for a lower bound on $\sum_{\mu  } C_{\mu\lam}^{\;\;\;\mu}$. First note that $\mu= \mu_k$ depends on $k$ in \eqref{k-decomp}. The $a$-length cycle splits into
$[2^{ a -1 \over 2 } ,1 ]$ (see Theorem \ref{factoP21}). A pair of $a$-length cycles splits into $ [ 2^{ a} ] $, with no $1$-cycles involved  (see Theorem \ref{2cyclefacto}). For the paired connected splittings of $[a^{ p} ] $, we therefore have permutations of cycle structure $[ 2^{ { (  p - 2 k ) (a -1 ) \over 2 }} , 1^{ p - 2k} ]  $ from the $ ( p - 2k ) $ individual splittings, while we have permutations of cycle structure $ [ 2^{ k a}  ] $ from the paired splittings. Thus for fixed $k$, we have 
\bea
&& 
\mu_k = [ 2^{ {p (  a -1 ) \over 2 }  +k  } , 1^{ p - 2k } ] 
\crcr
&&
|\cC_{ \mu_k } | = \frac{(a p ) ! }{
2^{ \frac{p (  a -1 )}{2 }  + k  } 
( \frac{p (  a -1 )}{2 }  + k ) ! \times  (p - 2k)! } 
\eea 
and 
\bea
C_{\mu_k\lam}^{\;\;\;\mu_k} =  \frac{|\cC_\lambda|}{|\cC_{\mu_k}|} \delta(T_{\mu_k} T_{\mu_k} \s_3^* ) 
=  \frac{|\cC_\lambda|}{|\cC_{\mu_k}|} f_{a,p} (k)
\eea
where the number of possible splittings of $(\s_3^*)^{-1} \in \lam = [a^{p}]$ 
in permutations belonging to  $\cC_{\mu_k}$ 
is of course 
\bea
\delta(T_{\mu_k} T_{\mu_k} \s_3^* )
= { p! \over 2^k k! ( p - 2k) ! }  \,  a^{ p - k }   = f_{a,p} (k) 
\eea 
As $\sum_{ k = 0 }^{ \lfloor { p  \over 2 } \rfloor }
C_{\mu_k\lam}^{\;\;\;\mu_k} \le 
\sum_{\mu }C_{\mu\lam}^{\;\;\;\mu}  $, 
we obtain the lower bound 
\bea
&&
 \sum_{ R \vdash n } {  \chi^R ( T_{\lambda = [ a^{p} ]  } ) \over d_R }  \ge 
 \sum_{ k = 0 }^{ \lfloor { p  \over 2 } \rfloor }  
 \frac{|\cC_\lambda|}{|\cC_{\mu_k}|} f_{a,p} (k) \crcr
 &&
   \sum_{ k = 0 }^{ \lfloor { p  \over 2 } \rfloor }  
 \frac{|\cC_\lambda|}{|\cC_{\mu_k}|} f_{a,p} (k)  = 
  \frac{(ap) ! }{a^{p} p ! }
  \sum_{ k = 0 }^{ \lfloor { p  \over 2 } \rfloor }  
\frac{
2^{ \frac{p (  a -1 )}{2 }  + k  } 
( \frac{p (  a -1 )}{2 }  + k ) !    (p - 2k)!   \;  p!   }{(a p ) ! \;  2^k k! ( p - 2k) ! } 
   a^{ p -k }\crcr
  &&
  = 2^{ \frac{p (  a -1 )}{2 }   } 
  \sum_{ k = 0 }^{ \lfloor { p  \over 2 } \rfloor }  
( \frac{p (  a -1 )}{2 }  + k ) ! 
  { 1 \over a^{k} k!  }  
  \label{BPowOddCyc}
\eea
One may ask about a reason for the integrality 
of this bound. It is enough to look at a part of  the summand: 
\bea
( \frac{p (  a -1 )}{2 }  + k ) ! 
  { 1 \over a^{k} k!  }  
  \label{interme}
\eea
Let us show that $\frac{p (  a -1 )}{2 }  + k  \ge a k $ 
in such a way ${ ( \frac{p (  a -1 )}{2 }  + k ) ! \over a^{k} k! }
 = I \cdot  { (ak)! \over   a^{k} k! }$  is an integer as
  $I$ will be necessarily an integer.  
The previous  claim is true since  $\frac{p (  a -1 )}{2 }  + k  -  a k  = \frac{p (  a -1 )}{2 }  - k(  a - 1) 
  $ $ = (a- 1) (\frac{p}{2} - k) \ge 0  $. 

We illustrate this bound in Table \ref{tab:singleOdd}. 
See Appendix \ref{app:Bounds} for a code 
that computes this table. 

\begin{table}[ht]
\centering
\begin{tabular}{l|l|l|l}
$\lambda = [a^p]$ & $\sum_{ R \vdash n } {  \chi^R ( T_{\lambda = [ a^p ]  } ) \over d_R }  $  & Lower bound & Relative error \\
\hline\hline
$[3^2]$ & 99& 16& 0.838 \\
$[3^3]$ & 4520& 112& 0.975  \\
$[3^4]$ & 504012& 1664 & 0.996 \\ 
$[5^2] $ & 139535& 768 & 0.994  \\
$[5^3] $ & 3276085815& 110592 & 0.999 \\
$[5^4] $ &309182846345600& 47480832 & 0.999  \\
$[7^2] $ & 1669223276& 92160 & 0.999 \\
$[7^3] $ & 147396002215287089& 451215360  & 0.999 \\
$[7^4] $ & 10217174199249113268200217& 9249384038400 & 0.999 \\
\hline
\end{tabular}
\caption{The column sum evaluation and its lower bound for  
$\lambda= [a^p]$, $a=3,5,7$ odd, and $p=2,3,4$.}
\label{tab:singleOdd}
\end{table}
\ 

\noindent{\bf Multiple odd-length cycles, each with multiplicity :  $ \lambda = [ a_1^{ p_1} , a_2^{ p_2} , \cdots , a_K^{ p_K } ] $, the $a_i$ are odd and multiplicities $p_i$ general. } 
Now take the partition $\lambda=[ a_1^{ p_1} , a_2^{ p_2} , \cdots , a_K^{ p_K } ] \vdash n $, where the $a_i$'s are distinct and odd.
Theorem \ref{2cyclefacto} says that there is no possible
connected splitting between cycles of different lengths.
For each $i$, we choose $2k_i$ cycles among the $p_i$
and perform pair connected splittings, and the remaining
$p_i- 2k_i $ cycles decomposed in  disconnected splittings. 
Thus, the number of splittings is computed by
considering any $p_i$ cycle of length $a_i$ and take
the product: 
\bea \label{a-odd}
F _{\{a_i\},\{p_i\} } (\{k_i\}) &=& 
\prod_{ i=1}^K f_{ a_i, p_i } (k_i )  
 = \prod_{ i=1}^K   { p_i! \over 2^{k_i} k_i! ( p_i - 2k_i) ! }  \,  a_i^{ p_i - k_i }    \crcr
 &  = & {1 \over 2 ^{\cK}}\prod_{ i=1}^K  { p_i! \over   k_i! ( p_i - 2k_i) ! }
   a_i^{ p_i - k_i }   
\eea
where  $\cK =  \sum_{ i =1}^K  k_i$. 

Note that the precise form of $ \mu $ depends again on 
 the choice of splittings in \eqref{k-decomp}. 
Taking into account all the $a_i$'s, and at fixed  $(k_1,k_2, \dots, k_K)$, $\mu$ is of the form: 
\bea 
\label{muOdd}
\mu = [ 2^{ \sum_{ i =1}^K {p_i (  a_i  -1 ) \over 2 }  +k_i  } , 1^{ \sum_{ i } (  p_i - 2k_i )}] 
 =  [ 2^{  \frac{1}{2}(n - \cP) + \cK }  , 1^{ \cP   \, - 2 \cK   } ]
\eea
with $\cP =  \sum_{ i =1}^K  p_i$, and   $n =  \sum_{ i =1}^K a_i p_i$. 
We denote this partition $\mu = \mu_{\cK}$ and 
identify 
\bea
C_{\mu_{\cK}\;  \lambda}^{\;\;\;\;\; \mu_{\cK}}  
 = { |\cC_\lambda| \over  |\cC_{\mu_{\cK}}|} \delta(T_{\mu_{\cK}} T_{\mu_{\cK}} \s_3^*)
 =  { |\cC_\lambda| \over  |\cC_{\mu_{\cK}}|} 
 \sum_{ \substack{k_i \in\{ 0,1, \dots , \lfloor { p_i  \over 2 } \rfloor\} \\   \sum_{i} k_i = {\cK} } }  F _{\{a_i\},\{p_i\} } (\{k_i\}) 
\eea
Because of this dependency in $k_i$, 
we have 
for $ \lambda = [ a_1^{ p_1} , a_2^{ p_2} , \cdots , a_K^{ p_K}  ] $, with $a_i$ odd the bound on the column sum: 
\bea 
&&
 \sum_{ R \vdash n } \frac{\chi^R ( T_{ \lam})}{d_R}  \ge 
 \sum_{{\cK} = 0}^{\widetilde \cP  }
 \; 
C_{\mu_{{\cK}}\;  \lambda}^{\;\;\;\;\; \mu_{\cK}} 
 \cr\cr
 &&
  \sum_{{\cK} = 0}^{\widetilde \cP }
 \; 
C_{\mu_{{\cK}} \; \lambda}^{\;\;\;\;\; \mu_{\cK}} 
 = 
 |\cC_\lambda|
 \sum_{{\cK} = 0}^{ \widetilde \cP  }
 \; 
 \frac{1}{|\cC_{\mu_{{\cK}}}|} 
 \sum_{ \substack{k_i \in\{ 0,1, \dots , \lfloor { p_i  \over 2 } \rfloor\} \\   \sum_{i} k_i = {\cK} } }  
 \;  
F _{\{a_i\},\{p_i\} } (\{k_i\}) 
\crcr
 &&
 = 
  {1 \over \prod_{ i =1}^K a_i^{p_i} p_i! } 
\sum_{{\cK} = 0}^{\widetilde \cP  }
 \; 
2^{  \frac{1}{2}(n - \cP) + {\cK} }(    \frac{1}{2}(n - \cP) + {\cK})! ( \cP   \, - 2 {\cK} )!  {1 \over 2 ^{\cK}}  \crcr
&& \qquad \qquad  \times 
 \sum_{ \substack{k_i \in\{ 0,1, \dots , \lfloor { p_i  \over 2 } \rfloor\} \\   \sum_{i} k_i = {\cK} } }  
 \prod_{i=1}^K   { p_i! \over   k_i! ( p_i - 2k_i) ! }   a_i^{ p_i - k_i }   
 \crcr
 &&
 = 
\sum_{{\cK} = 0}^{\widetilde \cP  }
 \; 
  2^{  \frac{1}{2}(n - \cP)  }    (    \frac{1}{2}(n - \cP) + {\cK})! ( \cP   \, - 2 {\cK} )!   
 \sum_{ \substack{k_i \in\{ 0,1, \dots , \lfloor { p_i  \over 2 } \rfloor\} \\   \sum_{i} k_i = {\cK} } }  
 \prod_{i=1}^K   { 1 \over  a_i^{k_i}  k_i! ( p_i - 2k_i) ! }
 \crcr
 &&
   \label{eq:GenBoundOdd}
\eea
where $\widetilde \cP = \sum_{i} \lfloor { p_i  \over 2 } \rfloor $. We will give numerical evaluations of this bound when we will deal 
with the more general case.

\subsubsection{Even-length cycles} 

We can extend the above result to include even-length cycles.

\

\noindent{\bf{Multiple cycles of one even length: $ \lambda = [ b^q ] $,  $q$ cycles of fixed even length $b$}.}\\

First consider a single cycle of even length : $ \lambda = [ b ] $. This has $b/2$   splittings into $ [2^{b/2 -1 }  ,1^2 ] \circ [ 2^{ b/2} ]   $  and $b/2$ splittings into $[ 2^{b /2} ]  \circ  [ 2^{ b/2 -1}, 1^2] $ (see Theorem \ref{factoP21Even}). 
There is no possible splitting involving the same partition 
$\mu$. This is as expected since a permutation which is an even-length cycle is an odd permutation. 

Next, consider the  fully disconnected splittings of $ \lambda = [ b^q ]$. To get a  splitting in two permutations of the same cycle structure $\mu$, we need $q$ to be even. 

For $q/2$ of the $a$ length cycles, we take splittings of type $ [2^{b/2 -1 }  ,1^2  ] \circ [ 2^{ b/2} ]   $  and for the remaining $q/2$ we take splittings of type  $[ 2^{ b/2} ] \circ [ 2^{ b/2 -1}, 1^2] $. This produces a splitting of $\lambda$ with $\mu$
of the form $ [ 2^{(b-1)\frac{q}{2}} , 1^q ] \circ [  2^{(b-1)\frac{q}{2}} , 1^q] $. The number of these is 
\bea 
{ q \choose q/2 } \,  ({ b \over 2  })^{ q/2} \,  ({ b \over 2  })^{ q/2}  = { q \choose q/2 } ({ b \over 2  })^{ q} 
\label{p:2even}
\eea
Applying Theorem \ref{factoP21Even}, per 
cycle of even length $b$, we have $b/2$ possible splittings.
 This yields the above result involving only disconnected splittings. 

Next consider the partially connected splittings. Of the $q$ cycles of even length $b$, let us have $k$ pairs connected to give connected splittings. Each pair gives $b$ splittings 
into $ [ 2^b ] $ (Theorem \ref{2cyclefacto}). So we have 
\bea 
{ q \choose 2k  } \;  (2k-1) !! \; b^k 
 = 
 { q \choose 2k  } 
{ (2k)! \over 2^k k! } \; b^k =
 { q! \over 2^k k! ( q - 2k) ! } \;  b^k 
\eea 
contributing $[ 2^ { bk } ] $ to the cycle structure of $\mu$.  The remaining $ ( q - 2k ) $ cycles of length $b$ split individually. Using the above combinatorics \eqref{p:2even},  we have 
\bea 
{ (q -2 k)  \choose (q-2k) /2 } . ({ b \over 2  })^{ (q -2k)} 
\eea
This produces a contribution to the cycles of $\mu $ of the form:  
 $[ 2^{\frac12 (b-1) ( q -2k) } , 1^{ (  q - 2k) }]$. 
 
The full cycle structure of $\mu $ from the paired-splittings and the individual splittings is 
\bea
 \mu_k  &=& [  2^{ \frac12(b-1) ( q -2k) + bk } , 1^{ (  q - 2k) }  ] =  [  2^{ \frac12b ( q -2k) -\frac12 ( q -2k)  + bk } , 1^{ (  q - 2k) }  ]
 \crcr
 & =&  [  2^{ \frac12  (b  -  1) q +  k   } , 1^{ (  q - 2k) }  ]
 \eea
  Multiplying the relevant factors we get 
\bea
{ q! \over 2^k k! ( q - 2k) ! }  . { (q -2 k)  \choose (q-2k) /2 }  . b^k . ({ b \over 2  })^{ (q -2k)} 
  =  g_{b,q}(k)  
\eea
Then we have 
\bea
C_{\mu_k \lambda }^{\;\;\; \mu_k} =   \frac{|\cC_{\lambda}|}{ |\cC_{\mu_k}| }\delta(T_{\mu_k} T_{\mu_k} \s_3^*)= \frac{|\cC_{\lambda}|}{ |\cC_{\mu_k}| } g_{b,q}(k) 
\eea
 and thus the bound 
\bea
 \sum_{ R \vdash n } \frac{ \chi^R ( T_{ \lam})}{d_R}  &\ge& 
 \sum_{k = 0}^{q/2} C_{\mu_k \lambda }^{\;\;\; \mu_k} 
  =  \sum_{k = 0}^{q/2}  \frac{|\cC_{\lambda}|}{ |\cC_{\mu_k}| } g_{b,q}(k)
  \crcr
   \sum_{k = 0}^{q/2}  \frac{|\cC_{\lambda}|}{ |\cC_{\mu_k}| }  g_{b,q}(k)
& =&     \frac{(bq)!}{b^q q!}  \sum_{k = 0}^{q/2} \frac{   2^{ \frac12(b-1)  q  +  k }  ( \frac12(b-1) q  + k) ! 
    (  q - 2k ) !    }{(bq) ! } \crcr
    & \times& 
  { q! \over 2^k k! ( q - 2k) ! }  . { (q -2 k)  \choose (q-2k) /2 }  . b^k . ({ b \over 2  })^{ (q -2k)} 
  \crcr
  &=  &
     \frac{1}{b^q}  \sum_{k = 0}^{q/2}    2^{ \frac12(b-1)  q  +  k }  ( \frac12(b-1) q  + k) !  
      \crcr
    & \times &
  { 1 \over 2^k k!  }  . { (q -2 k)  \choose (q-2k) /2 }  . b^k . ({ b \over 2  })^{ (q -2k)} 
  \cr\cr
   &   = &
      \sum_{k = 0}^{q/2} 
       2^{ \frac12(b-3)  q  + 2k } \frac{(\frac12(b-1)  q  +  k) ! }{b^k k!}
   { (q -2 k)  \choose (q-2k) /2 }  
\eea
Recalling that $b$ is even, 
if $b\ge 4$, then 
the summand becomes an integer because $(\frac12 (b-1)  q  + k) !  / (b^k k!)$  is a positive integer (using a previous argument in \eqref{interme}). 
If $b=2$,  then we can prove that the summand  
$t_{k} =  2^{ - \frac12 q + k }   { ( \frac12  q  + k) !          
\over   k!  }    { (q -2 k)  \choose (q-2k) /2 }  
  $ is again an integer: 
\bea
t_{k} &=&    { ( \frac12  q  + k) !    \over k ! }      
  {1 \over  2^{\frac{q}{2} -k}   }  . { (q -2k)  \choose (q-2k) /2 } 
   =    { ( \ell  + k) !    \over k ! }        
  {1 \over  2^{\ell  -k}   }  . { 2(\ell -k)  \choose  \ell -k  } \crcr
  &   =&   { ( \ell  + k) !    \over k ! }          
  {1 \over  2^{\ell  -k}   }  . { (2(\ell -k))!  \over  (\ell -k)!^2  } 
  = \frac{ (\ell + k) !   }{ k! (\ell -k)!}   \; . \;     
{ (2(\ell -k))!! \over   2^{\ell  -k}   (\ell -k)!  } \; .\;   (2(\ell -k)-1)!! 
\crcr
  &= &\frac{ (\ell + k) !   }{k! (\ell -k)!}    (2(\ell -k)-1)!!
\eea
which achieves the proof that $t_k$ is a positive
integer.

\ 

\noindent{\bf Multiple even-length cycles, each with multiplicity :  $ \lambda = [ b_1^{ q_1} , b_2^{ q_2} , \cdots , b_L^{ q_L } ] $, the $b_i$ are even and multiplicities $q_i$ general. }
 
Consider the case where we have  multiple even cycles $ [ b_1^{q_1} , b_2^{ q_2} , \cdots , b_L^{ q_L } ] $. We need $ q_1   + q_2 + \cdots + q_L $ to be even, otherwise we have an odd permutation and there is no splitting of equal cycle structure $\mu$. Let us assume we have $l_1, l_2 , \cdots , l_L$ paired
connected splittings, where we pair   $2l_1 $ cycles from the $q_1$, $2l_2$ cycles from the $q_2$,  etc. And with the remaining $ ( q_1 + q_2 + \cdots + q_L ) - ( 2l_1 + 2l_2 + \cdots + 2l_L ) $ cycles, we have to pick half the splittings to be of type where $ \sigma_1$ has the $1$-cycles and the other half to be of type where $\sigma_2$ has the 1-cycles. 
Putting this together leads us to: 
\bea \label{b-even}
G _{ \{b_i\}, \{q_i\}} ( \{l_i\})&=& 
\left ( \prod_{ i =1}^L { q_i! \over 2^{l_i} l_i! ( q_i - 2l_i) ! }  b_i^{ l_i } \right ) 
 { ( \cQ - 2\cL )\choose ( \cQ - 2\cL)/2  } \prod_{ i =1 }^L  ({ b_i \over 2  })^{ (q_i -2l_i)} \crcr
 &= &
 {1 \over 2^{\cQ - \cL}} { ( \cQ - 2\cL )\choose ( \cQ - 2\cL)/2  } 
 \prod_{ i =1}^L { q_i! \over  l_i! ( q_i - 2l_i) ! }   
b_i^{ (q_i -l_i)}  
\eea
where we define 
$\cQ = \sum_{i=1}^L  q_i $ and 
$ \cL = \sum_{i=1}^L l_i $. 

We must pick $2l_i$ cycles from $q_i$, therefore $l_i \le \lfloor q_i/2 \rfloor$, 
and 
 $  \cL  \le   \sum_{i=1}^{L} \lfloor q_i/2 \rfloor $. We then  define
$\widetilde \cQ  =  \sum_{i=1}^{L} \lfloor q_i/2 \rfloor $.

Summing over $l_i$'s, 
$ \sum_{ l_1 , l_2 \cdots , l_L } G _{ \{b_i\}, \{q_i\}} ( \{l_i\})$
 delivers the number of such possible splittings. 
For a given family $\{l_i\}$, the cycle structure of the splitting is therefore 
 \bea
 \label{muEven}
 \mu_{\cL} 
& =&  [  2^{  \sum_ {i=1}^L ( \frac{b_i}{2} ( q_i -2l_i) + b_il_i ) -\frac12 (\cQ - 2\cL)  } , 1^{ (\cQ -2 \cL)}  ]
 \crcr
 &  = &
    [  2^{  \sum_ {i=1}^L  \frac{b_i}{2} q_i  -\frac12 (\cQ - 2\cL)  } , 1^{ (\cQ - 2\cL)}  ] = [  2^{ \frac{1}{2} (n  -\cQ)   + \cL } , 1^{ (\cQ - 2\cL)}  ] 
 \eea
 where $\sum_ {i=1}^L   b_i q_i  =n   $. One observes that this formula
 is similar to \eqref{muOdd}. 
 We obtain the coefficient 
\bea
C_{\mu_{\cL} \lambda }^{\;\;\; \mu_{\cL}} 
 = { |\cC_\lambda| \over  |\cC_{\mu_{\cL}}|} \delta(T_{\mu_{\cL}} T_{\mu_{\cL}} \s_3^*) = 
 \frac{|\cC_\lambda|}{|\cC_{\mu_{\cL}}|} 
  \sum_{ \substack{l_i \in \{0,1,\dots, \lfloor q_i/2 \rfloor \} \\  \sum_{i} l_i  = \cL} } G_{ \{b_i\}, \{q_i\}} ( \{l_i\})
\eea
and write a bound on the column sum as follows: 
\bea
\sum_{ R \vdash n } {  \chi^R ( T_{\lambda   } ) \over d_R }  
&\ge&  
 \sum_{\cL = 0}^{ \widetilde \cQ  } C_{\mu_{\cL} \lambda }^{\;\;\; \mu_{\cL}}  
 \crcr
 \sum_{\cL = 0}^{ \widetilde \cQ  } C_{\mu_{\cL} \lambda }^{\;\;\; \mu_{\cL}}  
 &=& 
 \frac{ n ! }{\prod_{i=1}^L b_i^{q_i} q_i !  }
  \sum_{\cL = 0}^{ \widetilde \cQ } 
  \frac{ 2^{\frac{1}{2}  (n- \cQ) + \cL } (\frac{1}{2}  (n- \cQ) + \cL ) !   (\cQ - 2\cL)! }{n!} \crcr
  &\times &
  \sum_{ \substack{l_i \in \{0,1,\dots,   \lfloor q_i/2 \rfloor \} \\  \sum_{i} l_i  = \cL} } 
G _{ \{b_i\}, \{q_i\}} ( \{l_i\})
  \crcr
  &=& 
  \sum_{\cL = 0}^{ \widetilde \cQ } 
 {2^{\frac{1}{2}  (n- \cQ) + \cL }  \over 2^{\cQ - \cL}}
 (\frac{1}{2}  (n- \cQ) + \cL ) !   (\cQ - 2\cL)! 
  { ( \cQ - 2\cL )\choose ( \cQ - 2\cL)/2  } \crcr
  & \times &
  \sum_{ \substack{l_i \in \{0,1,\dots,   \lfloor q_i/2 \rfloor \} \\  \sum_{i} l_i  = \cL} } 
 \prod_{ i =1}^L {1\over  b_i^{ l_i}   l_i! ( q_i - 2l_i) ! }   
  \crcr
    &= & 
  \sum_{\cL = 0}^{ \widetilde \cQ  } 
 {2^{\frac{1}{2}  (n- 3\cQ)  +  2\cL}}
 (\frac{1}{2}  (n- \cQ) + \cL ) !   (\cQ - 2\cL)! 
  { ( \cQ - 2\cL )\choose ( \cQ - 2\cL)/2  } 
  \crcr
  &  \times &
  \sum_{ \substack{l_i \in \{0,1,\dots,{ \lfloor q_i/2 \rfloor } \} \\  \sum_{i} l_i  = \cL} } 
 \prod_{ i =1}^L {1\over  b_i^{ l_i}   l_i! ( q_i - 2l_i) ! }   
  \crcr
  &&
  \label{eq:GenBoundEven}
\eea
We can combine this result with the one obtained for an arbitrary number 
of odd-length cycles to establish a bound in the general case. 
This will be the purpose of the next section.

\subsubsection{The general case}

Consider a partition $\lambda \vdash n $ that, 
depending on the computation and context, can be expressed in two forms:  
$\lambda =  [c^{m_1}_1, c^{m_2}_2,\dots, c^{m_{L+N}}_{L+N}]$ 
and 
$\lambda = [ a_1^{ p_1} , a_2^{ p_2} , \ldots , a_L^{ p_L }, b_1^{q_1},  b_2^{q_2},  $ $\ldots, b_N^{q_N} ] $, 
where we 
equate $c_i = a_i$,  $m_i = p_i$, for $i=1,2,\dots, L$, and similarly, 
 $c_{i} = b_{i}$, $m_{i}  = q_i$, for $i= L+1,L+2,\dots, L+N$.  
In the latter notation, 
 $a_i$ and $b_i$ are pairwise distinct, with all $a_i$ being odd, and all $b_i$ being even. 
 Additionally, the sum $\sum_{l=1}^N q_l$ must be even; otherwise, the permutation associated with $\lambda$ will be odd, precluding any splitting with an equal cycle structure $\mu$.

In forming connected splittings, we must decide which cycles to pair. According to Theorem \ref{2cyclefacto}, odd and even cycles cannot be connected if we aim for a splitting with an identical cycle structure. 
So we must choose, on the one hand,
among the even cycles which one to  connect, 
and on the other hand, among the odd cycles which one 
to connect. This process will result in taking a product of the corresponding number of paired connected splittings. For the remaining cycles, the rules for pure odd or pure even splittings must  apply individually, resulting in disconnected splittings. This approach yields a product of contributions, as implemented subsequently.

For all $i = 1, 2, \ldots, L$, we choose to pair $2k_i$ cycles among the $p_i$ cycles of size $a_i$, where $k_i \in \{1, 2, \ldots, \lfloor \frac{p_i}{2}\rfloor \}$. For the remaining $p_i - 2k_i$ cycles, we apply disconnected splittings.

On the other hand, for each $i = 1, 2, \ldots, N$, we select $2l_i$ cycles from the $q_i$ cycles of size $b_i$ for paired connected splittings, where 
 $l_i \in \{0, 1, 2, \ldots, \lfloor \frac{q_i}{2} \rfloor \}$. Subsequently, we take half of the $\sum_{i=1}^{N} (q_i - 2l_i)$ cycles to split them such that $\sigma_1$ contains the 1-cycles, and the other half are split in the opposite manner, so that $\sigma_2$ contains the 1-cycles. We deduce that the general case is resolved by the product of the formulae \eqref{a-odd} and \eqref{b-even}.
We have 
\bea
\cF  
=    F_{ \{a_i\}, \{p_i\}} ( \{k_i\})   G_{ \{b_i\}, \{q_i\}} ( \{l_i\}) 
\eea
We introduce the notation: 
$\cP = \sum_{i=1}^L  p_i $, $\cK =  \sum_{i=1}^L k_i$, 
$n_o = \sum_{i=1}^L p_i k_i$, 
 $ \cQ = \sum_{i=1}^N q_i$,  $ \cL  = \sum_{i=1}^N l_i$, 
  $n_e = \sum_{i=1}^N q_i l_i$, $n_o + n_e = n$, 
  $\sum_{i=1}^L \lfloor { p_i  \over 2 } \rfloor = \widetilde \cP$
 and   $\sum_{i=1}^N \lfloor { q_i  \over 2 } \rfloor = \widetilde \cQ$ 
  and recast the above as  
  \bea
\cF
&= &
 {1 \over 2 ^{\cK}}\prod_{ i=1}^L  { p_i! \over   k_i! ( p_i - 2k_i) ! }
   a_i^{ p_i - k_i }
 {1 \over 2^{\cQ - \cL}} { ( \cQ - 2\cL )\choose ( \cQ - 2\cL)/2  } 
 \prod_{ i =1}^N { q_i! \over  l_i! ( q_i - 2l_i) ! }   
b_i^{ q_i -l_i}   \crcr
 &=&   {1 \over 2^{\cK+ \cQ - \cL}} { ( \cQ - 2\cL )\choose ( \cQ - 2\cL)/2  } 
  \prod_{ i =1}^{L+N} { m_i! \over  k_i! ( m_i - 2k_i) ! }   
c_i^{ m_i -k_i} 
\eea
where we extend the values of
$k_i$ beyond $i=L$, requiring that  $k_{i+L}= l_i$, for $i=1,2,\dots, N$. 
We then switch to $\lam =[ c^{m_1}_1, c^{m_2}_2,\dots, c^{m_{L+N}}_{L+N}]$
and therefore rename $\cF  = \cF_{\{c_i\}, \{m_i\} }(\{k_i\})$.

 The decomposition of a permutation from the class $\lam$ into two permutations with an identical cycle structure $\mu$ is achieved through a concatenation of the processes described in \eqref{muOdd} and \eqref{muEven}:
\bea
\mu  &=& [2^{\frac12(n_o - \cP) + \cK }, 1^{\cP - 2\cK}]\circ 
[2^{\frac12(n_e - \cQ) + \cL }, 1^{\cQ - 2\cL}] \crcr
&=&
[2^{\frac12(n - (\cP + \cQ)) + \cK  + \cL}, 1^{\cP+\cQ - 2(\cK+ \cL) }]
\eea
 which is denoted $\mu_{\cK+ \cL}$. 

The structure coefficient that we seek is of the form: 
\bea
C_{\mu_{\ell  } \lambda }^{\;\; \;    \mu_{\ell}} 
 &=&  { |\cC_\lambda| \over  |\cC_{\mu_{\ell }}|} \delta(T_{\mu_{\ell}} T_{\mu_{\ell}} \s_3^*) \crcr
 &=& 
  \frac{|\cC_\lambda|}{|\cC_{\mu_{ \ell }}|}  
  \sum_{\cK = 0}^{ \widetilde \cP  }
    \sum_{\cL   = 0}^{ \widetilde \cQ  }
    \delta_{\cK + \cL,  \ell}
   \sum_{ \substack{k_i \in \{0,1,\dots, \lfloor { m_i  \over 2 } \rfloor \}, \;\;  i\le L \\
                    k_{i} \in \{0,1,\dots,  \lfloor  \frac{m_{i}}{2}  \rfloor \} , \;\;  i>  L \\  
             \sum_{i \le L }   k_i = \cK \\ 
               \sum_{i>L}  k_i  = \cL  } } 
  \cF_{\{c_i\}, \{m_i\} } (\{k_i\}) 
\eea
The following statement holds
\begin{theorem}\label{theo:LowBoundsGen}
Let $\lam$ be a partition of $n$, the normalized sum of
central characters of the column labeled by $\lam$ admits the following bound: 
\bea
&& 
\sum_{ R \vdash n } \widehat \chi^R ( T_{ \lam}) 
\ge   
   \sum_{\ell = 0}^{   \widetilde  \cP + \widetilde \cQ   }  
2^{ \frac12(n - ( \cP + 3\cQ))  +2 \ell    }( \frac12(n - ( \cP + \cQ)) + \ell   ) ! \; ((\cP+ \cQ)- 2\ell) !
\cr\cr
&&\times 
\sum_{\cK = 0}^{ \widetilde \cP  } \; 
   \sum_{\cL   = 0}^{ \widetilde \cQ  }
  { \delta_{\cK + \cL,  \ell} \over 2^{2\cK  }} { ( \cQ - 2 \cL )\choose ( \cQ - 2\cL )/2  } 
  \sum_{ \substack{k_i \in \{0,1,\dots, \lfloor { p_i  \over 2 } \rfloor \}, \;\;  i\le L \\
                    k_{i} \in \{0,1,\dots, \lfloor  \frac{q_i}{2} \rfloor \} , \;\;  i>  L \\  
             \sum_{i\le L}  k_i = \cK \\ 
               \sum_{i > L}  k_i  = \cL} } 
  \prod_{ i =1}^{L+N} {1 \over  c_i^{k_i} k_i! ( m_i - 2k_i) ! }  \crcr
  &&
  \label{eq:MostGenBound}
  \eea
\end{theorem}
\proof  As in the previous situation, 
we use  the following bound 
\bea
\sum_{ R \vdash n } \widehat \chi^R ( T_{ \lam}) 
\ge  
 \sum_{\ell = 0}^{   \widetilde  \cP +\widetilde  \cQ  }  
 C_{\mu_{\ell } \lambda }^{\;\; \;  \mu_{\ell}}  
 \eea
 where
 \bea
 &&
C_{\mu_{\ell } \lambda }^{\;\; \;  \mu_{\ell}} = 
    \frac{|\cC_\lambda|}{|\cC_{\mu_{ \ell }}|}   
\sum_{\cK = 0}^{ \widetilde \cP  } \; 
 \sum_{\cL   = 0}^{ \widetilde \cQ  }
    \delta_{\cK + \cL,  \ell}
   \sum_{ \substack{k_i \in \{0,1,\dots, \lfloor { m_i  \over 2 } \rfloor \}, \;\;  i\le L \\
                    k_{i} \in \{0,1,\dots,  \lfloor  \frac{m_{i}}{2}  \rfloor\} , \;\;  i>  L \\  
             \sum_{i \le L }   k_i = \cK \\ 
               \sum_{i>L}  k_i  =  \cL } } 
  \cF_{\{c_i\}, \{m_i\} } (\{k_i\}) 
                \crcr
  && = 
  \frac{1}{\prod_{i=1}^{L+N} c_i ^{m_i} m_i!} 
2^{ \frac12(n - ( \cP + \cQ)) + \ell    }( \frac12(n - ( \cP + \cQ)) + \ell   ) ! \; ((\cP+ \cQ)- 2\ell) !
\crcr
&&\times 
\sum_{\cK = 0}^{ \widetilde \cP  } \; 
 \sum_{\cL   = 0}^{\widetilde \cQ  }
    \delta_{\cK + \cL,  \ell}
  {1 \over 2^{\cK+ \cQ -\cL }} { ( \cQ - 2\cL )\choose ( \cQ - 2\cL))/2  } 
  \crcr
  &&
  \times 
  \sum_{ \substack{k_i \in \{0,1,\dots, \lfloor { p_i  \over 2 } \rfloor \}, \;\;  i\le L \\
                    k_{i} \in \{0,1,\dots,  \lfloor  \frac{q_i}{2} \rfloor \} , \;\;  i>  L \\  
             \sum_{i\le L}  k_i = \cK \\ 
               \sum_{i > L}  k_i  = \cL} } 
  \prod_{ i =1}^{L+N} { m_i! \over  k_i! ( m_i - 2k_i) ! } c_i^{ m_i -k_i}  
   \cr\cr
  &&
  = 
2^{ \frac12(n - ( \cP + 3\cQ))  +2 \ell    }( \frac12(n - ( \cP + \cQ)) + \ell   ) ! \; ((\cP+ \cQ)- 2\ell) !
\cr\cr
&&\times 
\sum_{\cK = 0}^{ \widetilde \cP  } \; 
   \sum_{\cL   = 0}^{ \widetilde \cQ  }
  { \delta_{\cK + \cL,  \ell} \over 2^{2\cK  }} { ( \cQ - 2 \cL )\choose ( \cQ - 2\cL )/2  } 
  \sum_{ \substack{k_i \in \{0,1,\dots, \lfloor { p_i  \over 2 } \rfloor \}, \;\;  i\le L \\
                    k_{i} \in \{0,1,\dots, \lfloor  \frac{q_i}{2} \rfloor \} , \;\;  i>  L \\  
             \sum_{i\le L}  k_i = \cK \\ 
               \sum_{i > L}  k_i  = \cL} } 
  \prod_{ i =1}^{L+N} {1 \over  c_i^{k_i} k_i! ( m_i - 2k_i) ! }  \crcr
  &&
  \label{Cmu_ell}
\eea
summing over $\ell$ from $0$ to $  \widetilde  \cP + \widetilde \cQ$, we obtain the   expected result. 

\qed

This is the lower bound for the general case.
We illustrate this bound by computing  for all $n \in \{1,2,\dots,9 \}$
the partition $\lam = [3^*, 2^{*}, 1^{*}]$, for only even number
of parts of size 2. The results are recorded in Table \ref{tab:gen} 
in Appendix \ref{app:Bounds}, which gives the code that directly performs the sum in \eqref{eq:MostGenBound}. This somewhat complicated-looking formula was derived by restricting to $ \mu = [2^* , 1^* ] $. As a check, we have coded in GAP the same restriction and verified that the same results are produced by the GAP code.  The results have been verified match for all partitions $\lam =[3^*, 2^{*}, 1^{*}]$ of $n \le 9$. Note
that the GAP code, given in Appendix \ref{sec:AppGapBds}, 
allows us to compute the $[2^* , 1^* ]$-restricted column sum for all partitions $\lam$, not only those of the type $ [3^*, 2^{*}, 1^{*}]$. 

An interesting problem is to determine, given $n$ as an input,  the maximal computational complexity of using the formula 
\eqref{Cmu_ell} to obtain the lower bound in the large $n$ limit.  This involves the computation of the elementary functions, powers, factorials in the formula. The computational complexities of these operations are available in standard sources (e.g. \cite{WikiCompOp} \cite{Borweins}). A key step in the formula is the sum over $k_i$. This complexity question suggests  two interesting mathematical problems about partitions. \\
\vskip.2cm
{\bf Problem 1:} Given the sequence of positive integers $n$, consider the partitions $p$ of $n$ in the presentation as distinct positive  parts with their positive multiplicities $ p = [ c_1^{ m_1} , c_2^{ m_2} \cdots c_S^{m_S} ] $, where $ \sum_{ i =1}^{S} c_i m_i = n$, the function $f( p , n  ) = \prod_{ i=1}^{ S} m_i  $, and the function $\tilde{ f}( n) = {\rm Max}_{ p \vdash n }   ( f ( p , n ) ) $. Describe the asymptotic growth of $\tilde{ f} (n)$ at large $n$. \\ 

\vskip,2cm 
{\bf Problem 2:}  Given the sequence of positive integers $n$, consider the partitions $p$ of $n$ in the presentation as distinct positive  parts with their positive multiplicities $ p = [ c_1^{ m_1} , c_2^{ m_2} \cdots c_S^{m_S} ] $, where $ \sum_{ i =1}^{S} c_i m_i = n$, the function $g( p , n  ) = \prod_{ i=1}^{ S} \lfloor {m_i /2 } \rfloor  $, and the function $\tilde{g} ( n) = {\rm Max}_{ p \vdash n }  ( g ( p , n ) ) $. Describe the asymptotic growth of $\tilde{g}(n)$ at large $n$. \\

\vskip.2cm 
The second problem is more directly relevant to the calculation of the complexity desired, but the the first is closely related and plausibly has a similar asymptotic behaviour at large $n$. 
These sequences are easily explored using Mathematica for values of $n$ up to $60$. Generating functions for the sums over partitions of the products of multiplicities have been considered by MacMahon \cite{McMa} and  may perhaps be adapted to give the Max functions involved in the above problems. 

\subsection{Counting $ C_{  \mu \lam } ^{\;\;\; \mu } $}

The previous developments demonstrates that we can 
access  $ C_{  \mu \lam } ^{\;\;\; \mu } $ for  general $\lam$ and some $\mu$, 
independently of knowing their sum over $\mu$. 
 It  is, indeed,  a problem on its own to evaluate $ C_{  \mu \lam } ^{\;\;\; \mu } $. 
 This problem has applications such as counting ribbon graphs with certain properties. 

For an integer $n$, a generic partition $\lam = [ c^{m_1}_1, c^{m_2}_2,\dots, c^{m_{L+N}}_{L+N}]  \vdash n$, and 
$\mu_\ell \vdash n$ such that 
\bea
\label{mu_ell}
\mu_\ell =  [2^{\frac12(n - (\cP + \cQ)) +\ell}, 1^{\cP+\cQ - 2\ell  }]
\eea
using previous notation, we have given 
 in \eqref{Cmu_ell} an expression for 
 $ C_{  \mu _\ell \lam } ^{\;\;\; \mu_\ell  } $ 
 for $\ell \in [\![0,   \widetilde  \cP + \widetilde \cQ]\!]$. 
We can thus provide a table of the values of $ C_{  \mu _\ell \lam } ^{\;\;\; \mu_\ell  } $.
For a few values of $n \in \{1,2,\dots, 6\}$,  this is recorded in Table \ref{tab:cmulam}. 
One  checks that the  sum of  $ C_{  \mu _\ell \lam } ^{\;\;\; \mu_\ell  } $  over  the $\ell$'s within a sector corresponding to the same $\lam$ returns the right 
output. This has to be compared with  the corresponding row labeled by $\lam$ in  Table \ref{tab:gen}.

\begin{table}[h]
\centering
\begin{tabular}{l|l|l|l|}
$n$ & $\lambda $ &  $\mu$  & $C_{  \mu  \lam } ^{\;\;\; \mu  }$   \\
\hline\hline
1 & $[1]$ & $[1 ]$  & 1  \\ \hline
2 &$[1^2]$ & $[ 1^2 ]$ &  1 \\ 
 &  &  $[ 2 ] $  & 1 \\ \hline
 3 &$[3]$ &$[2,1]$      & 2  \\
 &$[1^3]$ &  $ [1^3] $   & 1  \\  
  &$[1^3]$ & $[2,1]$     & 1  \\ \hline
4 & $[3,1]$ & $ [2,1^2]$ & 4  \\
& $[2^2]$ & $ [2,1^2]$  & 1  \\
& $[2^2]$ & $ [2^2]$  & 2 \\
& $[1^4]$ & $ [1^4]$& 1  \\ 
& $[1^4]$ & $ [2,1^2]$ & 1  \\ 
& $[1^4]$ &  $ [2^2]$& 1  \\ \hline
 5& $[3,1^2]$ &$ [2,1^3] $& 6  \\
  & $[3,1^2]$ & $ [2^2,1] $ & 4  \\
& $[2^2 , 1]$ &  $ [2,1^3] $ & 3     \\
& $[2^2 , 1]$ &  $ [2^2,1] $ & 2     \\
& $[1^5]$ &  $ [1^5]$  & 1     \\
& $[1^5]$ & $ [2,1^3]$  & 1     \\ 
& $[1^5]$ & $ [2^2,1] $  & 1     \\ \hline
6& $[3^2]$ & $[2^2,1^2]$ & 8  \\
 & $[3^2]$ & $[2^3]$ & 8  \\
& $[3 , 1^3]$ &  $[2, 1^4]$ & 8    \\
& $[3 , 1^3]$ &  $[2^2, 1^2]$ & 8    \\
& $[2^2 , 1^2]$ &  $[2, 1^4]$ & 6    \\
& $[2^2 , 1^2]$ &  $[2^2, 1^2]$ &  4    \\
& $[2^2 , 1^2]$ &  $[2^3]$ & 6   \\
& $[1^6]$ & $[ 1^6]$  & 1   \\
& $[1^6]$ & $[2, 1^4]$ &   1 \\ 
& $[1^6]$ & $[2^2, 1^2]$  & 1   \\ 
& $[1^6]$ & $[2^3]$ & 1   \\ \hline
\hline
\end{tabular}
\caption{Evaluation of $ C_{  \mu  \lam } ^{\;\;\; \mu   }$ for  $\lambda= [3^*,2^*,1^*] \vdash n$, $n\in  \{1,2,\dots,6 \}$.}
\label{tab:cmulam}
\end{table}

\section{Bounds for disjoint unions of partitions}\label{sec:BdDisj}  

Consider a partition $ \lambda $ of $n$, which takes the form $ \lambda^{(1)} \sqcup \lambda^{(2)} $ where $ \lambda^{(1)} \vdash n_1 , \lambda^{(2)} \vdash n_2 $ and $ n_1 + n_2 = n$. 
For example if $ \lambda = [ 4,3,2,2] \vdash n= 11 $ we can write 
\bea 
&& \lambda = [ 4] \circ [ 3,2,2] \;, \; \; \qquad  {  [4] \vdash n_1= 4 ; \;\; [3,2,2] \vdash n_2 = 7 }\cr 
&& \lambda = [ 4,3 ] \circ [ 2,2 ] \;, \; \; \qquad  
  { 
 [ 4,3 ]\vdash n'_1= 7 ; \;\; [ 2,2 ]  \vdash n'_2 = 4} 
 \cr 
&&
{  \lambda = [ 4,2 ] \circ [ 3,2 ] \;, \; \; \qquad   
 [ 4,2]\vdash n''_1= 6 ; \;\; [3,2 ]  \vdash n''_2 = 5} 
\eea
These are just three of several possibilities. 

The sum of normalised characters is
\bea 
 G_{ \lambda } \equiv \sum_{ R }  { \chi^{ R} ( T_{\lambda }  ) \over d_{ R } } =  \sum_{ \mu \vdash  n }  { \; 
 { | \cC_{ \lambda } | \over |\cC_{ \mu } | } }  \delta ( T_{ \mu } T_{ \mu } \sigma_3^* )  
\eea
where $ \sigma_3^* $ is any chosen permutation in the class $ \lambda$. 
Picking a decomposition $ \lambda = \lambda^{(1)} \sqcup \lambda^{(2)} $, 
with  $ \lambda^{(1)} \vdash n_1 , \lambda^{(2)} \vdash n_2 $, we also pick a homomorphism 
 $  S_{n_1} \times S_{n_2} \rightarrow S_{ n_1 + n_2}$ by letting $ S_{ n_1} $ act on 
 $\{ 1, \cdots , n_1 \}$ while $S_{ n_2} $ acts on $ \{ n_1+1 , \cdots , n_1 + n_2 \}$. 
 We then  pick a permutation in the conjugacy class $\lambda $ of the form 
$  \sigma_3^* = \sigma_3^{*(1)} \circ \sigma_3^{*(2)}$, 
{ where $\sigma_3^{*(1)}$ belongs to conjugacy class $\cC_{\lambda^{(1)} }$  
and  $\sigma_3^{*(2)}$ belongs to $\cC_{\lambda^{(2)} }$.

 The quantity 
$ \delta ( T_{ \mu } T_{ \mu } \sigma_3^* ) $ will contain contributions where $ \mu$ splits 
into the form $ \mu = \mu^{(1)} \sqcup \mu^{(2)} $ with $ \mu^{(1)} \vdash n_1 , \mu^{(2)}  \vdash n_2$. Hence, 
\bea 
 G_{ \lambda } \geq
\sum_{ \mu^{(1)} \vdash n_1 } \sum_{ \mu^{(2)} \vdash n_2 } 
{ 
{ | \cC_{ \lambda^{(1)} \circ \lambda^{(2)} } | \over  |\cC_{ \mu^{(1)} \circ \mu^{(2)}} |  }  }
\delta ( T_{ \mu^{(1)} \circ \mu^{(2)}  } T_{ \mu^{(1)} \circ \mu^{(2)}  } \sigma_3^* ) 
\equiv G_{ \fac} 
\eea
{ In fact, as $\cC_{ \mu^{(1)} \circ \mu^{(2)} } \ne  
\cC_{ \mu^{(1)}   } \circ \cC_{   \mu^{(2)} } \subset S_{n_1} \times S_{n_2}$, we have 
 $T_{ \mu^{(1)} \circ \mu^{(2)}  } \ne T_{ \mu^{(1)}  } \circ T_{ \mu^{(2)}  }$, where we extend $ \circ$ to the group algebra. 
We have instead the following equality: 
\bea
\delta ( T_{ \mu^{(1)} \circ \mu^{(2)}  } T_{ \mu^{(1)} \circ \mu^{(2)}  } \sigma_3^* ) 
= 
\delta ( T_{\mu^{(1)}} T_{ \mu^{(1)}} \sigma_3^{*(1)} ) 
\delta ( T_{ \mu^{(2)}} T_{ \mu^{(2)}  } \sigma_3^{*(2)}  ) 
\eea
Let us assume that some $\s_1 , \s_2\in  \cC_{ \mu^{(1)} \circ \mu^{(2)}  }$ obeys $\s_1 \s_2 \sigma_3^* = \id$. Then, we have 
$\s_1 \s_2 \sigma_3^*$ $=\s_1 \s_2 (\sigma_3^{*(1)} \circ \sigma_3^{*(2)} )= \id$, and moreover there exist 
some $\s_i^{*(1)} \circ \s_i^{*(2)}   = \s_i $, $i=1,2$, 
where $\s_i^{*(1)}  \in  \cC_{ \mu^{(1)} }$ and 
$\s_i^{*(2)}  \in  \cC_{ \mu^{(2)} }$. Thus
\bea
(\s_1^{*(1)} \circ \s_1^{*(2)})(\s_2^{*(1)} \circ \s_2^{*(2)}) 
(\sigma_3^{*(1)} \circ \sigma_3^{*(2)} )
= (\s_1^{*(1)} \s_2^{*(1)}\sigma_3^{*(1)}  ) \circ  
 (\s_1^{*(2)} \s_2^{*(2)}\sigma_3^{*(2)}  )  = \id 
\eea
which requires each factor to be the identity on each subgroup
 $S_{n_1}$ or $S_{n_2}$. 

The converse is clearly true: 
if the product $\delta ( T_{\mu^{(1)}} T_{ \mu^{(1)}} \sigma_3^{*(1)} ) 
\delta ( T_{ \mu^{(2)}} T_{ \mu^{(2)}  } \sigma_3^{*(2)}  ) =1$,
then we can easily construct two 
$\s_1 , \s_2\in  \cC_{ \mu^{(1)} \circ \mu^{(2)}  }$ which obeys $\s_1 \s_2 \sigma_3^* = \id$, by collecting two permutations in 
the class $\cC_{ \mu^{(1)} }$ and two permutations
in $\cC_{ \mu^{(2)} }$ and pair them to produce $\s_1$ and $\s_2$. 
} 

Now note that 
\bea 
&& G_{ \fac  } = \sum_{ \mu^{(1)} \vdash n_1 } \sum_{ \mu^{(2)} \vdash n_2 } 
{ 
{ | \cC_{ \lambda^{(1)} \circ \lambda^{(2)} } | \over  |\cC_{ \mu^{(1)} \circ \mu^{(2)}} |  }  }
\delta ( T_{\mu^{(1)}} T_{ \mu^{(1)}} \sigma_3^{*(1)} ) 
\delta ( T_{ \mu^{(2)}} T_{ \mu^{(2)}  } \sigma_3^{*(2)}  )  \cr\cr 
&& =  \sum_{ \mu^{(1)} \vdash n_1 } \sum_{ \mu^{(2)} \vdash n_2 } 
{ 
{ | \cC_{ \lambda^{(1)} \circ \lambda^{(2)} } | \over  |\cC_{ \mu^{(1)} \circ \mu^{(2)}} |  }  
{|\cC_{ \mu^{(1)} } |  |  \cC_{ \mu^{(2)} } |  \over  |\cC_{ \lambda^{(1)} } | |\cC_{ \lambda^{(2)} } |  }  
}
\cr \cr
&& 
\qquad \qquad \times
 {  | \cC_{ \lambda^{(1)} } |  \over | \cC_{\mu^{(1)}} |  }  \delta ( T_{\mu^{(1)}} T_{ \mu^{(1)}} \sigma_3^{*(1)} ) 
 {   | \cC_{ \lambda^{(2)} } |  \over | \cC_{\mu^{(2)}} |  } 
\delta ( T_{ \mu^{(2)}} T_{ \mu^{(2)}  } \sigma_3^{*(2)}  ) \cr 
&&   = {  | \Aut^{ (n_1) } ( \lambda^{(1)}  ) |  | \Aut^{ (n_2) } ( \lambda^{(2)}  ) |  \over 
  | \Aut^{ (n) } ( \lambda^{(1)}  \circ \lambda^{(2)} )  |} 
 \sum_{ \mu^{(1)} \vdash n_1 } \sum_{ \mu^{(2)} \vdash n_2 }  
 { | \Aut^{ (n) } ( \mu^{(1)}  \circ \mu^{(2)}  )  | \over  | \Aut^{ (n_1) } ( \mu^{(1)}  ) |  | \Aut^{ (n_2) } ( \mu^{(2)}  ) |  }  \cr \cr
 && 
 \qquad \qquad  \times {  | \cC_{ \lambda^{(1)} } |  \over | \cC_{\mu^{(1)}} |  }  \delta ( T_{\mu^{(1)}} T_{ \mu^{(1)}} \sigma_3^{*(1)} ) 
 {  | \cC_{ \lambda^{(2)} } |  \over | \cC_{\mu^{(2)}} |  } 
\delta ( T_{ \mu^{(2)}} T_{ \mu^{(2)}  } \sigma_3^{*(2)}  )
\eea
{ where we used the orbit stabilizer theorem to express 
the sizes of the conjugacy classes in terms of the 
order of the automorphism groups. In particular, 
we introduce the notation: 
\bea
&&
 | \cC_{ \rho^{(i)} } | = n! /  | \Aut^{ (n_i) } ( \rho^{(i)}  ) |\;, 
 \qquad \rho = \lambda, \mu\;, i=1,2 \crcr
 &&
 | \cC_{ \rho^{(1)} \circ   \rho^{(2)} } | = n! /  | \Aut^{ (n) } (  \rho^{(1)} \circ   \rho^{(2)}   ) | \;, \quad \rho = \lambda , \mu
\eea
} 
Now observe that 
\bea 
 { | \Aut^{ (n) } ( \mu^{(1)}  \circ \mu^{(2)}  )  | \over  | \Aut^{ (n_1) } ( \mu^{(1)}  ) |  | \Aut^{ (n_2) } ( \mu^{(2)}  ) |  }  \ge 1 
\eea
This means 
\bea 
G_{ \fac } &\ge& 
 {  | \Aut^{ (n_1) } ( \lambda^{(1)}  ) |  | \Aut^{ (n_2) } ( \lambda^{(2)}  ) |  \over 
  | \Aut^{ (n) } ( \lambda^{(1)}  \circ \lambda^{(2)} )  |}  \cr\cr
  &\times&  \sum_{ \mu^{(1)} \vdash n_1 } \sum_{ \mu^{(2)} \vdash n_2 }  
 {   | \cC_{ \lambda^{(1)} } |  \over | \cC_{\mu^{(1)}} |  }  \delta ( T_{\mu^{(1)}} T_{ \mu^{(1)}} \sigma_3^{*(1)} ) 
 {   | \cC_{ \lambda^{(2)} } |  \over | \cC_{\mu^{(2)}} |  } 
\delta ( T_{ \mu^{(2)}} T_{ \mu^{(2)}  } \sigma_3^{*(2)}  )
\eea
Equivalently 
\bea 
G_{ \fac }  \ge {  | \Aut^{ (n_1) } ( \lambda^{(1)}  ) |  | \Aut^{ (n_2) } ( \lambda^{(2)}  ) |  \over 
  | \Aut^{ (n) } ( \lambda^{(1)}  \circ \lambda^{(2)} )  |}  
  G_{ \lambda^{(1)} } G_{ \lambda^{(2)} } 
\eea
and we conclude 
\bea 
\label{eq:additive}
G_{ \lambda } \ge {  | \Aut^{ (n_1) } ( \lambda^{(1)}  ) |  | \Aut^{ (n_2) } ( \lambda^{(2)}  ) |  \over 
  | \Aut^{ (n) } ( \lambda^{(1)}  \circ \lambda^{(2)} )  |}  
  G_{ \lambda^{(1)} } G_{ \lambda^{(2)} } 
\eea 

So we have a   lower bound on the 
column sum of normalised characters for any conjugacy class $ \lambda $ associated with any disjoint decomposition of $ \lambda = \lambda^{(1)} \sqcup \lambda^{(2)} $. 

We define 
\bea 
F_{\lambda } = |\Aut ( \lambda )  | G_{ \lambda } 
\eea
and express the disjoint union as a sum of partitions, $ \lambda^{(1)}  \sqcup \lambda^{(2)} \equiv \lambda^{(1)}  + \lambda^{(2)} $.Note that 
\bea 
F_{ \lambda } =   |\Aut ( \lambda )  |  | \cC_{ \lambda } |   \sum_{ R } { \chi^R_{ \lambda } \over d_R }  = n! \sum_{ R } { \chi^R_{ \lambda } \over d_R }
\eea
We will refer to  $ F_{ \lambda } $ the column sum of normalised characters. 

The inequality \eqref{eq:additive} thus proves: 
\begin{theorem}\label{theo:prodpartition}
\bea 
F_{ \lambda^{(1)} + \lambda^{(2)}  } \ge  F_{ \lambda^{(1)} } F_{ \lambda^{(2)} } 
\eea 
\end{theorem}

As a direct corollary, we also obtain: 
\bea 
\log  ( F_{ \lambda^{(1)} + \lambda^{(2)}  } ) \ge  \log  ( F_{ \lambda^{(1)} }  ) + \log  ( F_{ \lambda^{(2)} }  ) 
\eea
In other words, the logarithm of the sum of normalised characters is super-additive. 
We are using the disjoint union to define an addition on partitions. Super-additivity and associated theorems are discussed, for example  in \cite{Horst2009}.

\section{ Discussion and outlook } 
\label{sec:towards}

In this paper, we have studied properties of the column sum of normalized central characters,  $ \sum_{ R \vdash n }  \chi^{R} ( T_\lam )/d_R$, where $T_\lam$ denotes the sum of elements in the conjugacy class $\cC_\lam$ of the symmetric group, and the sum is over irreducible representations $R$. 
A summary of our main results is included in the introduction. We will here make general remarks and discuss some future directions following from this work.

\vskip.2cm 

The starting point of our investigations has been the expression of the column sum of normalised central characters in terms of a sum over structure constants of multiplication in the centre of the group algebra of $S_n$ (Proposition \ref{columnsum}). This lends itself to efficient computational algorithms and combinatorial interpretations. A combinatorial interpretation of geometrical nature involves the expression of the column sum as a sum over bi-partite ribbon graphs with face structure determined by $ \lambda$. These graphs appear with a weight equal to an integer ratio of orders of automorphism groups, one group for the symmetry of some vertices and another group for the symmetry of the graph (Theorem \ref{theo:RGexpans}). It is interesting to note that bi-partite ribbon graphs appear summed, with slightly different weights, in other physical applications. In the computation of matrix model correlators, they appear weighted by the inverse of the automorphism group  of the graph. In tensor models, they appear with weight one in the enumeration of tensor model observables. In the context of matrix model correlators, interesting generating functions for correlators can sometimes be written explicitly, form special functions of interest in string theory, and relate to the phenomenon of super-integrability \cite{MorozovExact}\cite{AMM03}\cite{MiMoSint}. It will be interesting to consider generating functions for bi-partite graph counting, with  weights of the kind that have appeared here.

\vskip.2cm

In  deriving the lower bounds for the  column sums \eqref{Cmu_ell}, we have obtained the counting of bi-partite ribbon graphs with face structure $ \lambda$, and vertex structures for black vertices and white vertices given by partition $ \mu$, where $ \mu = [ 2^* , 1^* ] $. 
The techniques we used should also allow results for the case where the black and white vertex types are given by partitions $ \mu , \nu $, not necessarily equal but both of type $ [ 2^* , 1^* ]$. Further generalisation to $ \mu , \nu $ belonging to conjugacy classes of type $ [s^{ k_s} ,\cdots , 3^{ k_3} , 2^* , 1^* ] $ for small $s$ and small  $k_3, \cdots , k_s $ should be possible, using perturbative techniques from matrix models.  These  bi-partite graph counting problems are  special cases of Hurwitz counting of branched covers \cite{Hurwitz1891}, which are of broad and current  interest, e.g.  \cite{Irving2004CombinatorialCF,Irving2006MinimalTF}, in combinatorics.

\vskip.2cm 

Having established that the function computing the column sum of central normalized characters belongs to \shP (as well as the stronger result that the decision problem is in class P),   the ensuing natural question is whether it is \shP-hard (and so \shP-complete).  Given that we have expressed this function in terms of  the enumeration of bi-partite ribbon graphs possessing a specific structure, it seems logical to continue within the realm of graph theory, which abounds with \shP-hard and \shP-complete problems. The primary distinction between ribbon graphs and ordinary graphs lies in the symmetry: the cyclic symmetry $\Z_k$ at a $k$-valent vertex in ribbon graphs is substituted by a local $S_k$ symmetry in ordinary graphs. Nevertheless, permutations and their associated double cosets offer a unified description for both graph types \cite{Feynstring}. This could be instrumental as part of the strategy to prove the \shP-hardness of problems related to both ordinary and ribbon graphs. The crucial step involves devising a polynomial reduction  of a recognized \shP-hard problem in ordinary graph theory to the enumeration of bi-partite ribbon graphs arising in the column sum problem.

\

\noindent{\bf Acknowledgements.} 
JBG thanks Christophe Tollu for insightful discussions
on complexity theory.  We also thank Anirban Chowdhury, Robert de Mello Koch,  Charles Nash, Denjoe O' Connor, Yang-Hui He for useful discussions. SR is supported by the Science and Technology Facilities Council (STFC) Consolidated
Grants ST/P000754/1 “String theory, gauge theory and  duality” and ST/T000686/1
“Amplitudes, strings and  duality” and a Visiting Professorship at the  Dublin Institute for Advanced Studies. 

\appendix 

\section*{Appendix}

\section{Proof of Proposition \ref{columnsum}}
\label{app:proofcolumnsum}

Our goal is to show that, for any $\lam \vdash n$, 
\bea \label{app:sumC}
\sum_{  \mu \vdash n}  C_{ \mu  \lambda }^{\;\;\; \mu} 
 = \sum_{ R  \vdash n} {\chi^R ( T_{ \lambda } )  \over d_R }  
\eea
Start by  \eqref{tmutnu} and multiply it by $T_\lam$: 
\bea
T_\mu T_\nu T_\lam =   
\sum_{ \rho  \vdash n} C_{ \mu \nu }^{ \;\;\;\rho } T_{ \rho }  T_\lam 
\eea
We define the Kronecker delta function on the symmetric group as 
$\delta: S_n \to \R$, $\delta(\s)=1$, if $\s=\id$ and 0 otherwise, 
and extend $\delta$ by linearity over the group algebra $\C(S_n)$ of $S_n$. 
We evaluate the previous equation: 
\bea
\delta(T_\mu T_\nu T_\lam) =  
\sum_{ \rho  \vdash n} C_{ \mu \nu }^{\;\;\; \rho } \del(T_{ \rho }  T_\lam ) 
= \sum_{ \rho  \vdash n} C_{ \mu \nu }^{\;\;\;  \rho }\sum_{\s\in \cC_\rho, \tau \in \cC_\lam} \del(\s\tau) 
\eea
Using the fact that a permutation and its inverse belong to the same
conjugacy class, we infer $[\s]= \rho = [\tau^{-1}] = \lam$ and deduce: 
\bea
\label{deltaTTT}
\delta(T_\mu T_\nu T_\lam) 
= \sum_{ \rho  \vdash n} C_{ \mu \nu }^{ \;\;\;\rho } |\cC_\lam| \delta_{\rho \lam}  = |\cC_\lam| C_{ \mu \nu }^{\;\;\; \lam }
\eea
where in the last expression, $\delta_{\rho \lam} $ is a Kronecker delta on partitions. 
It gives 1 if two partitions are equal i.e. they have the same 
parts (each part in same number), and gives 0 otherwise. 
Thus, we have an expression of the structure constant 
\bea
\label{cmunulamabove}
  C_{ \mu \nu }^{ \;\;\; \lam } = \delta(T_\mu T_\nu T_\lam)  / |\cC_\lam|.
  \eea 
  Defining  $C_{ \mu \nu  \lam }  :=  \delta(T_\mu T_\nu T_\lam)$, we have the following identities 
\bea 
\label{craising}
&&
C_{ \mu \nu  \lam }  =  C_{ \mu \nu }^{\;\;\; \lam }   |\cC_\lam| \, ,
\qquad \qquad 
C_{ \mu \nu  \lam }  = C_{ \nu  \lam  \mu} =  C_{   \lam  \mu\nu } \cr\cr
&& 
C_{ \nu  \lam  \mu} =  C_{ \nu  \lam  }^{\;\;\; \mu }   |\cC_\mu|  
= C_{ \mu \nu  \lam } =  C_{ \mu \nu  }^{\;\;\;  \lam}   |\cC_\lam| 
\cr\cr
&& 
    C_{ \nu  \lam  }^{\;\;\; \mu } = C_{ \mu \nu  }^{\;\;\;  \lam}
    \frac{ |\cC_\lam| }{   |\cC_\nu| }
\eea 
This shows that  $C_{ \nu  \lam  }^{\;\;\; \mu }$ is proportional 
to $C_{ \mu \nu  }^{\;\;\;  \lam}$. 

Using the expansion in representation of the Kronecker delta function 
in terms of characters:  
\beq
\delta(\s) = \sum_{R \vdash n} \frac{d_R}{n!} \, \chi^R(\s) 
\eeq
we  expand the left-hand side of \eqref{app:sumC}: 
\bea 
\label{app:sumC2}
\sum_{ \mu \vdash n }  C_{ \mu \lambda }^{ \;\;\; \mu} 
 = 
 \sum_{ \mu \vdash n } { 1 \over | \cC_{ \mu} | }  
   \delta(T_\mu  T_\lam T_\mu ) 
  = 
\sum_{ \mu \vdash n } { 1 \over | \cC_{ \mu} | }  
 \sum_{R \vdash n} \frac{d_R}{n!} \, \chi^R( T_\mu  T_\lam T_\mu ) 
\eea
There is a crucial property of the characters: 
If $B$ is a central element, then  for any $A$
 \be \label{chiAB0}
 \chi^R ( A B ) = { 1 \over d_R } \chi^R ( A ) \chi^R ( B )  
 \ee
 Then, another expression of \eqref{app:sumC2} is
 \bea 
\label{app:sumC3}
&&
\sum_{ \mu \vdash n }  C_{ \mu   \lambda } ^{ \;\;\;  \mu} 
 =
\sum_{ \mu \vdash n } { 1 \over | \cC_{ \mu} | }  
 \sum_{R \vdash n} \frac{d_R}{n!} \,    { 1 \over d_R } \chi^R (  T_\mu T_\mu )  \chi^R( T_\lam ) 
 \crcr
 && 
  = \sum_{R \vdash n}  
  \left[\frac{d_R }{n!} \, 
  \sum_{ \mu \vdash n } { 1 \over | \cC_{ \mu} | }     \chi^R (  T_\mu T_\mu ) 
  \right]
   \widehat \chi^R( T_\lam ) 
\eea
The following lemma  ends the proof: 
\begin{lemma}
\bea
 \frac{ d_R }{n!}  
\sum_{ \mu \vdash n } { 1 \over | \cC_{ \mu} | }  
 \chi^R (  T_\mu T_\mu )  =1
 \eea
 \end{lemma}
 \proof 
 We use again the relation \eqref{chiAB0} for writing: 
 \bea
  \frac{ d_R }{n!}  
\sum_{ \mu \vdash n } { 1 \over | \cC_{ \mu} | }  
 \chi^R (  T_\mu T_\mu )  
 = 
   \frac{ 1 }{n!}  
   \sum_{ \mu \vdash n } { 1 \over | \cC_{ \mu} | }  
 \chi^R (  T_\mu )  \chi^R (   T_\mu )  = 
  \frac{ 1 }{n!}  
   \sum_{ \mu \vdash n } { | \cC_{ \mu} || \cC_{ \mu} | \over | \cC_{ \mu} | }  
 \chi^R (  g_\mu )  \chi^R (   g_\mu ) 
 \eea
 where $g_\mu$ is a representative of $ \cC_{ \mu}$, which 
 implies $ \chi^R (   T_\mu )  =  | \cC_{ \mu} | \chi^R (   g_\mu ) $ . We 
 then obtain
  \be
  \frac{ d_R }{n!}  
\sum_{ \mu \vdash n } { 1 \over | \cC_{ \mu} | }  
 \chi^R (  T_\mu T_\mu )  
   = 
  \frac{ 1 }{n!}  
   \sum_{ \mu \vdash n }  \sum_{g_\mu  \in \cC_{ \mu}} 
 \chi^R (  g_\mu )  \chi^R (   g_\mu ) 
  = 
   \frac{ 1 }{n!}  
   \sum_{g \in S_n } \chi^R (  g )  \chi^R (   g ) = 1  
 \ee
  where we use the orthogonality relation $ \sum_{g \in S_n } \chi^S (  g )  \chi^R (   g ) = n! \delta_{RS}$. 
 
 \qed

\section{Codes}

This appendix contains GAP and Sage codes for different countings discussed in the manuscript. We use a SageMath interface to run GAP, so all codes start with
 \verb|%%gap|.

\subsection{Counting using conjugacy classes}
\label{app:GapTTT} 

We deliver a code computing the integer
$C_{\mu \mu \lam}/ |\cC_\mu|$ 
\eqref{cmunulam}. 
Comments accompany each function and explain its calculation purpose.

\begin{verbatim}
%%gap
##  Produces the list of all terms in T_{ mu } T_{ mu }/|C_{ mu }|  
# n a positive integer
# mu a number between 1 and the number of partitions of n 
# mu labels a conjugacy class  

partsq := function (mu , n) 
    
    local LL, cc, Listmu, l , Grp ;
    
    Grp := SymmetricGroup(n) ;   
    cc := ConjugacyClasses (Grp) ; 
    Listmu := AsList(cc[mu]) ;   # conjugacy classes as lists 
    l := Length(Listmu);        # length of Listmu
    LL := [];   
        
    for k in [ 1 .. l ] do 
        Add ( LL , Listmu[1]*Listmu[k] ) ;  
        # register the product of  Listmu[1] and Listmu[k]
        # Listmu[1] is the 1st representative of T_{mu}
        # Listmu[1]* Listmu[k] computes the k-th element Listmu[1]*T_{mu}
    od; 
    return LL ; 
end; 
\end{verbatim}

\begin{verbatim}
%%gap
# Enumerates the number of elements of the conjugacy class 
# labeled by lam in the product of T_{ mu } T_{ mu } / |C_{ mu }|  
# uses partsq ( mu , n )
# Applies a filter to target the terms that are in the conjugacy class lam 
# n a positive integer
# lam, mu integers labeling some conjugacy classes
# lam, mu integers between 1 and the number of partitions of n 
  
Countmumulam := function ( mu , lam , n  )
    local cc , Grp ; 
    Grp  := SymmetricGroup (n) ; 
    cc := ConjugacyClasses (Grp) ; 
    return Length ( Filtered ( partsq (mu,n) , x -> x in  AsList (cc[lam ]) ) ) ;
end ; 
\end{verbatim}

Finally the following function achieves the enumeration 
of the number of  elements    $T_{ \mu }* T_{ \mu } / |\cC_{ \mu }|$
that belong to $\cC_\lam$:
\begin{verbatim}
%%gap 
# n a positive integer
# lam integer labeling a conjugacy class 

Tablam := function ( lam , n ) 
    
    local i, L, s , l ; 
    L := [ ] ; 
    s := 0; 
    l := NrPartitions ( n ); 
        
    Print("nb part = ", l, "\n" );
    for i in [1 ..l] do 
        Add ( L , [ cc [i]  , Countmumulam ( i , lam , n )] )   ; 
        if Countmumulam ( i , lam , n ) > 0 then 
            s := s + Countmumulam ( i , lam , n ); 
        fi; 
    od ; 
    
    Print("total s = ", s, "\n");  
    return L ; 
end ; 

%%gap 
cc[1] ; 
Tablam ( 1 , 3);

[out]
()^G
nb part = 3
total s = 3
[ [ ()^G, 1 ], [ (1,2)^G, 1 ], [ (1,2,3)^G, 1 ] ]
\end{verbatim}

\subsection{Basic functions on symmetric group 
computations and their complexity}
\label{app:basicfunctions}

\noindent{\bf Computing the orbits of a permutation:}
This is a simple Python code for computing the list of  orbits of a given 
permutation. 

\begin{verbatim}
# Function to find a single orbit
# visited is an array recording the visited indices 
# permut is a permutation 
# start is an index from which we start the construction of the orbit
# Assuming: start <= degree of permut 

def find_orbit(visited, permut, start):
    orbit = []
    while not visited[start]:
        visited[start] = True
        orbit.append(start)
        start = permut[start-1]
    return orbit

# Function which computes the set of orbits 
# of a permutation permut of n objects
def compute_orbits(n, permut):
    
    # Initialize all elements as not visited
    visited = [False] * (n + 1)
    orbits = []

    # Find all orbits
    for i in range(1, n + 1):
        if not visited[i]:
            orbits.append(find_orbit(visited,permut, i))

    return orbits

# Example:
# Permutation must be provided as a list, e.g., [2, 3, 1, 5, 4] for n=5
n = 5
perm = [2, 3, 1, 5, 4]
orbits = compute_orbits(n, perm)
print "The orbits are:", orbits

(out)
The orbits are: [[1, 2, 3], [4, 5]]
\end{verbatim}

Our task is to  evaluate the complexity of the function 
\verb|compute_orbits(n, permut)|.

First, 
 the complexity of  \verb|find_orbit(visited, permut, i)|
is evaluated as follows: 
we read, append, assign new values to $start$ until $visited[start]$ becomes $True$. 
To become $True$, we must reach  the size $c_i$ of the cycle containing   $start=i$. Thus running  \verb|find_orbit| costs  
 $\cO(|c_i|) \subset \cO(n)$, for a given orbit of size $c_i $
 and ignoring the size of each input key $ k_{i_j}$
 (adding them in the end, will simply cost another
 factor of $n$).

The permutation $permut$ is  stored as a list of the $n$ first integers.

-  Creating the list $visited$  is an allocation 
followed by an append of  $n$ strings $False$: this costs $\cO(n)$.  

- In the loop, we append $orbits$ only if $visited[i]$  is $False$; 
this occurs at most $n$ times and when a new cycle 
of $permut$ starts; 
the test `` if $visited[i]$  is $False$'' cost as much as
reading $i$ which is $\cO(1)$ (ignoring the size of $i$). 
Then follows the call of \verb|find_orbit(visited, permut, i)|
that we have already computed the complexity. 

In \verb|compute_orbits(n, permut)|, 
 we append to the list $orbits$, the orbit returned by 
 \verb|find_orbit(visited, permut, i)|. 
 This costs the length of the orbit appended. 
 We finally obtain, after summing over all  orbit sizes, 
 exactly $n$. Thus 
 the cost of the algorithm is polynomial in $n$, 
 taking into account the size of each key $i$.

 \

\noindent{\bf Computing the cycle structure of a permutation:}
This is a simple modification of the previous functions. 

\begin{verbatim}
# Function to find the size of an orbit
# visited is an array recording the visited indices 
# permut is a permutation 
# start an index from which we start constructing the orbit
# Assuming: start <= degree of permut 

def find_orbit_size(visited, permut, start):
    s = 0 
    while not visited[start]:
        visited[start] = True
        s=s+1
        start = permut[start-1]
    return s
\end{verbatim}

This new function should be inserted in the 
remaining program. Then the list obtained
 in \verb|orbits.append| is the cycle structure
 of  $permut$. The complexity remains unchanged.

\subsection{Sage code for lower bounds on the column sum}
\label{app:Bounds}

In this appendix, we provide a Sage implementation of the bounds on the column sum as elaborated in the text. 
We start  from  formulas \eqref{BSOddCyc} and \eqref{BPowOddCyc},
 for a given  partition $\lam$. Then, we extend the code 
to the most generic situation delivered in formula \eqref{eq:MostGenBound}. 
By restriction, we can recover \eqref{eq:GenBoundOdd} and \eqref{eq:GenBoundEven} when the partition $\lam$ consists solely of odd or even cycles. 

To obtain the sum of normalized central characters
in Sage, we use symmetric functions 
provided by the command \verb|SymmetricFunctions(QQ)|. 
 
 \

\noindent{\bf Sage code for the column sum of normalized central characters -} 

\begin{verbatim}
# For extra comments consult: 
# https://doc.sagemath.org/html/en/reference/combinat/sage/
combinat/sf/sfa.html
#
s = SymmetricFunctions(QQ).s()
# Schur symmetric function 
pp = SymmetricFunctions(QQ).power()
# power symmetric function 

# Ccomputes the sum of normalized central characters 
# for conjugacy class labeled by Lam in the permutation group Sn 
# n a positive integer 
# Lam a partition of n  

def NormCharSum (n , Lam) : 
  l = len ( Partitions(n).list() )
  s = 0 
  P = Partitions(n).list()
  for i in range( l ):
       s = s +
    s(P[i]).scalar(pp(Lam))*factorial(n)/Lam.centralizer_size()/dimension(P[i])
       # P[i] determines the irrep 
       # s(P[i]).scalar(pp(Lam)) gives the caracter chi^(P[i])_Lam         
       # Lam.centralizer_size() gives the size of the centralizer
       # dimension (P[i]) is the dimension of the irrep P[i]
  return s 
\end{verbatim}
We test the function as: 
\begin{verbatim}
NormCharSum (2, Partition([1,1])) 
(out)  2 
NormCharSum (5, Partition([2,2,1])) 
(out) 31
\end{verbatim}

\noindent{\bf Sage code for data in Table \ref{tab:singlecycl}  -} 
The following code delivers the data of 
 Table \ref{tab:singlecycl} which lists
  the column sum  and its lower bound 
 in the simplest case when $\lam = [n]$. 
 We  implement directly the formula \eqref{BSOddCyc}
 as: 

\begin{verbatim}
# Return the lower bound for the column sum of the character table
# when lambda is the cycle [n] of odd length n 
# n an positive odd integer  

def func (n):
    return 2^((n-1)/2)* factorial((n-1)/2)
    
# Returns a list L of 4-tuples  
# n a odd positive integer 
# for q = 3,5, ...., 2k+1 < n 
# L[q] = 
# [ q the number which is partitioned, 
# the column sum for lambda = [q], 
# the lower bound computed by func(q), 
# relative error between the two results ]
# N is for getting a numerical evaluation 

def Bound_SingleOddCycle (n): 
    L = []
    for q in range (3,n+1,2): 
        No = NormCharSum (q, Partition([q]))
        f = func(q)
        L.append( [q, No, f, N((No - f)/No)])
    return L
\end{verbatim}
We run this code by calling: 

\begin{verbatim}
Bound_SingleOddCycle (21):

(out)
[[3, 3, 2, 0.333333333333333],
 [5, 40, 8, 0.800000000000000],
 [7, 1260, 48, 0.961904761904762],
 [9, 72576, 384, 0.994708994708995],
 [11, 6652800, 3840, 0.999422799422799],
 [13, 889574400, 46080, 0.999948199948200],
 [15, 163459296000, 645120, 0.999996053329387],
 [17, 39520825344000, 10321920, 0.999999738823268],
 [19, 12164510040883200, 185794560, 0.999999984726507],
 [21, 4644631106519040000, 3715891200, 0.999999999199960]]
\end{verbatim}

\noindent{\bf Sage code for the Table \ref{tab:singleOdd}  -} 
We deal with the data of  Table \ref{tab:singleOdd} 
that records the column sum and its lower bound when $\lam = [a^p]$, with  $a$ an odd positive integer. 
 
 We  implement directly the formula \eqref{BPowOddCyc}
 as: 

 \begin{verbatim}
# Returns the lower bound for the column sum of the character table
# when lambda = [a^p] 
# a an odd positive integer 
# p is an integer 

def funcSingleOdd ( a, p ):
    s = 0 
    for k in range (0,floor(p/2)+1):
        s = s + factorial( p*(a-1)/2+ k )/ ( a^k *factorial(k))      
    return 2^((a-1)*p/2)*s
 \end{verbatim}
 
 We tabulate this function and the column sum to 
 collect the data of Table \ref{tab:singleOdd}: 
 
 \begin{verbatim}
#  Return a list L of 4-tuples  
# A a positive integer
# n  a positive integer
# for a = 3,5, ...., 2k+1 < A
# for q = 1,2,..., n 
# the (a,q)-th element of the list L 
# L[a,q] = [a*q  , the column sum for lambda = [a^q], 
# the lower bound (a,q), relative error between the two results]
# N is for getting a numerical evaluation 

def NCS_SingleOdd (A, n): 
    L = []
    for a in range (3,A+1,2): 
        for q in range (2,n+1): 
            l =[]
            for m in range (1,q+1): 
                l.append(a)   
            No = NormCharSum (a*q, Partition(l))
            f = funcSingleOdd(a,q)
            L.append( [a*q, No, f, N((No-f)/No)])
    return L
 \end{verbatim}
 We run this code as: 
  \begin{verbatim}
 NCS_SingleOdd (7, 4)
 (out)
 [[6, 99, 16, 0.838383838383838],
 [9, 4520, 112, 0.975221238938053],
 [12, 504012, 1664, 0.996698491305763],
 [10, 139535, 768, 0.994496004586663],
 [15, 3276085815, 110592, 0.999966242642518],
 [20, 309182846345600, 47480832, 0.999999846431222],
 [14, 1669223276, 92160, 0.999944788692247],
 [21, 47396002215287089, 451215360, 0.999999990479886],
 [28, 10217174199249113268200217, 9249384038400, 0.999999999999095]]
  \end{verbatim}

\noindent{\bf Sage code for Table \ref{tab:gen}  -} 
The following gives the code for the data found 
in Table \ref{tab:gen}. Based on  \eqref{eq:MostGenBound},
we compute the general lower bound of the column sum.

\begin{verbatim}
# Returns a list [Cmumulam, su] 
# L = [p_1,p_2,p_3] takes only the number of occurrences
# of a partition of n of the form 3* p_1 + 2*p_2 + p_3  
# where p_1, p_2, p_3  are fixed and never vanishing  

# Cmumulam is a list of pairs [mu, C] 
# where the partitions mu contribute
# to the factorization of L in two permutations of type mu
# mu = [(n- P-Q)/2 + ell, P+Q-2*ell] 
# where P = p_1 + p_3 is the number of odd parts in L
# Q = p_2  the number of even parts in L
# p_2 is an even positive integer
# and C is the structure constant C^mu_{mu L} of the centre 

# su is the bound on the column sum 
 
def BoundMK321 (L): 
  if (L[1]%2) != 0 :  
     print "Err: Q must be even" 
     return 0 
  else:      
     n = 3*L[0] + 2*L[1] + L[2]  
     P = L[0]+ L[2]
     Q = L[1]
        
     tQ = L[1]/2
     tP = floor(L[0]/2) + floor(L[2]/2)
              
     su = 0
     su2 = 0
     Cmumulam = []
     for ell in range(tP+tQ+1): 
       for M in range(tP+1): 
         for K in range (tQ+1):  
            if (K+M== ell): 
              su1 = 0
              for k0 in range( floor(L[0]/2)+1 ):  
                 for k2 in range( floor(L[2]/2)+1 ): 
                    if (k0+k2 == M):
                       for k1 in range( floor(L[1]/2) +1 ):  
                          if k1 == K : 
                             su1 += 1/((3^(k0)*factorial(k0)*
                              factorial(L[0]-2*k0)) )*
                              1/((2^(k1)*factorial(k1)* factorial(L[1]-2*k1)))*
                              1/((1^(k2)*factorial(k2)* factorial(L[2]-2*k2)))
               su1 *=  binomial(Q-2*K,(Q-2*K)/2) /(2^(2*M)) 
                                   
               su += 2^((1/2)*(n - P - 3*Q)+2*ell)
               * factorial((1/2)*(n - P - Q) + ell)
               *factorial(P + Q- 2*ell) *su1   
       Cmumulam.append( [[ 1/2*(n - (P+Q)) + ell, P+Q-2*ell ],su-su2])
       su2 = su
         
    return [Cmumulam,su]    
 \end{verbatim}
 
 To test this function, we call  
\begin{verbatim}
 BoundMK321 ([0,2,2])
 (out) [[[[1, 4], 6], [[2, 2], 4], [[3, 0], 6]], 16]
  \end{verbatim}
  
  Now, we define a function that tabulates the
  bounds on the column sum for various values of $n$. 
This gives the data of Table \ref{tab:gen}. 
   
\begin{verbatim}
# Computes a list tot_List of 4-tuples [L[i], no, nu[1], N((no-nu[1])/no)]
# where 
# given a positive integer M, for all 1 <= n <= M
# L the list of partitions of n with maximal part 3, 
# thus all partitions of the form [3^*,2^*,1^*] 
# 
# For all partitions L[i] elements of L 
# no is the sum of central normalized characters in conjugacy class L[i]
# (uses function NormCharSum (n, L[i]))
# nu[1] is the lower bound obtained for L[i]
# (uses BoundMK321 on the multiplicity list of L[i])
# and the relative error N((no-nu[1])/no)

def funBound321 (M): 
    
    for n in range(1,M+1):         
        L = Partitions(n, max_part=3).list()      
        l = len(L)
        tot_List =[]
        
        for i in range(l):
            
            li = list(L[i])
            if li.count(2)%2 ==0 :
                     
                no = NormCharSum (n, L[i])             
                Lcount = [li.count(3), li.count(2), li.count(1)]
                nu = BoundMK321(Lcount)   
                #nu[1] contains the bound, nu[0] is a list of the Cmumulam 
                   
                tot_List.append( [L[i],  no, nu[1], N((no-nu[1])/no)  ] ) 
                    
        print n, tot_List
        print "\n"
    return 0

\end{verbatim}
We run the following code to produce a table 
extending Table \ref{tab:gen}. 
\begin{verbatim}
funBound321 (10)

(out)
1 [[[1], 1, 1, 0.000000000000000]]
2 [[[1, 1], 2, 2, 0.000000000000000]]
3 [[[3], 3, 2, 0.333333333333333], 
[[1, 1, 1], 3, 2, 0.333333333333333]]

4 [[[3, 1], 12, 4, 0.666666666666667], 
[[2, 2], 7, 3, 0.571428571428571],
 [[1, 1, 1, 1], 5, 3, 0.400000000000000]]

5 [[[3, 1, 1], 42, 10, 0.761904761904762],
 [[2, 2, 1], 31, 5, 0.838709677419355], 
 [[1, 1, 1, 1, 1], 7, 3, 0.571428571428571]]

6 [[[3, 3], 99, 16, 0.838383838383838],
 [[3, 1, 1, 1], 99, 16, 0.838383838383838], 
[[2, 2, 1, 1], 118, 16, 0.864406779661017],
 [[1, 1, 1, 1, 1, 1], 11, 4, 0.636363636363636]]

7 [[[3, 3, 1], 612, 32, 0.947712418300654], 
[[3, 2, 2], 381, 24, 0.937007874015748], 
[[3, 1, 1, 1, 1], 231, 28, 0.878787878787879], 
[[2, 2, 1, 1, 1], 309, 24, 0.922330097087379], 
[[1, 1, 1, 1, 1, 1, 1], 15, 4, 0.733333333333333]]

8 [[[3, 3, 1, 1], 2898, 112, 0.961352657004831], 
[[3, 2, 2, 1], 3216, 72, 0.977611940298508], 
[[3, 1, 1, 1, 1, 1], 462, 40, 0.913419913419913], 
[[2, 2, 2, 2], 272, 21, 0.922794117647059], 
[[2, 2, 1, 1, 1, 1], 772, 50, 0.935233160621762], 
[[1, 1, 1, 1, 1, 1, 1, 1], 22, 5, 0.772727272727273]]

9 [[[3, 3, 3], 4520, 112, 0.975221238938053], 
[[3, 3, 1, 1, 1], 9432, 192, 0.979643765903308], 
[[3, 2, 2, 1, 1], 17022, 240, 0.985900599224533], 
[[3, 1, 1, 1, 1, 1, 1], 882, 60, 0.931972789115646], 
[[2, 2, 2, 2, 1], 2000, 45, 0.977500000000000], 
[[2, 2, 1, 1, 1, 1, 1], 1642, 70, 0.957369062119367], 
[[1, 1, 1, 1, 1, 1, 1, 1, 1], 30, 5, 0.833333333333333]]

10 [[[3, 3, 3, 1], 44093, 320, 0.992742612206019], 
[[3, 3, 2, 2], 50157, 432, 0.991387044679706], 
[[3, 3, 1, 1, 1, 1], 27639, 432, 0.984369912080755], 
[[3, 2, 2, 1, 1, 1], 63981, 528, 0.991747550053922],
[[3, 1, 1, 1, 1, 1, 1, 1], 1596, 80, 0.949874686716792],
[[2, 2, 2, 2, 1, 1], 11460, 186, 0.983769633507853], 
[[2, 2, 1, 1, 1, 1, 1, 1], 3391, 120, 0.964612208787968], 
[[1, 1, 1, 1, 1, 1, 1, 1, 1, 1], 42, 6, 0.857142857142857]]


0
\end{verbatim}

\begin{table}[ht]
\centering
\begin{tabular}{l|l|l|l|l}
$n$ &$\lambda $ & $\sum_{ R \vdash n } {  \chi^R ( T_{\lambda   } ) \over d_R }  $  & Lower bound & Relative error \\
\hline\hline
1 & $[1]$ & 1& 1& 0 \\ \hline
2 &$[1^2]$ & 2& 2& 0 \\ \hline
 3 &$[3]$ & 3& 2& 0.33 \\
 &$[1^3]$ & 3& 2& 0.33 \\ \hline
4 & $[3,1]$ & 12& 4& 0.66 \\
& $[2^2]$ & 7& 3& 0.57 \\
& $[1^4]$ & 5& 3& 0.40 \\ \hline
 5& $[3,1^2]$ & 42& 10& 0.76 \\
& $[2^2 , 1]$ &  31 & 5 & 0.83    \\
& $[1^5]$ & 7 & 3 & 0.57   \\ \hline
6& $[3^2]$ & 99& 16& 0.83 \\
& $[3 , 1^3]$ & 99 & 16 & 0.83   \\
& $[2^2 , 1^2]$ & 118 & {16} & 0.88   \\
& $[1^6]$ & 11 & 4 & 0.63 \\ \hline
7 &$[3^2 , 1]$ & 612 & 32 & 0.94  \\
&$[3 , 2^ 2]$ & 381 & 24 & 0.93 \\
&$[3 , 1^4 ]$ & 231 & 28 & 0.87 \\
&$[2 ^2 , 1^3]$ & 309 & {24} & 0.92 \\
&$[1 ^7]$ & 15 & 4 & 0.73 \\ \hline
8 & $[3^2,  1^2 ]$ & 2898 & 112 & 0.96 \\
 &$[3 ,2^2 , 1 ]$ & 3216 & 72 & 0.97    \\
& $[3 ,1^5 ]$ & 462 & 40 & 0.91  \\  
 &$[2^4 ]$ & 272 & 21 & 0.92   \\
 &$[2^2, 1^4 ]$ & 772 &  {50} & 0.94 \\
 &$[1^8 ]$ & 22 & 5 & 0.77 \\  \hline
9 &$[3^3]$ & 4520& 112& 0.97 \\
&$[3^2, 1^3]$ & 9432& 192& 0.97 \\
&$[3, 2^2, 1^2]$ & 17022&  {240} & 0.98 \\
&$[3, 1^6]$ & 882& 60& 0.93 \\
&$[2^4, 1]$ & 2000& 45& 0.97 \\
&$[2^2, 1^5]$ & 1642&  {70}  & 0.96  \\
&$[1^9]$ & 30& 5& 0.83 \\
\hline
\end{tabular}
\caption{The column sum evaluation and its lower bound for  
$\lambda= [3^*,2^*,1^*] \vdash n$, $n\in  \{1,2,\dots,9 \}$. Only non vanishing values
are shown.}
\label{tab:gen}
\end{table}

\subsection{GAP code for lower bounds on column sum}\label{sec:AppGapBds}  

We provide below a GAP code that produces the lower bounds
on the column sum problem. 
Some functions have been introduced in
Appendix \ref{app:GapTTT}, we recall them for pedagogical 
purpose. 

\begin{verbatim}
%%gap
partsq := function (mu , m ) 
        local LL, cc, Listmu , Grp ;
        LL := []  ; 
        Grp := SymmetricGroup(m) ;
        cc := ConjugacyClasses (Grp) ; 
        Listmu := AsList(cc[mu]) ; 
        for k in [ 1 .. Length(Listmu) ] do 
           Add ( LL , Listmu[1]*Listmu[k] ) ; 
        od; 
return LL ; 
end; 

# Returns the structure constant C_{mu lam}^{mu}
# Takes as input indices mu, lam of partitions of m 

Countmumulam := function ( mu , lam , m  )
    local cc , Grp ; 
    Grp  := SymmetricGroup (m ) ; 
    cc := ConjugacyClasses ( Grp ) ; 
    return 
    Length ( Filtered ( partsq ( mu , m  ) , x -> x in  AsList (cc[lam ]) ) ) ;
end ;

## Produces for lambda, a list of mu's , 
## along with the C_{ mu , mu , lambda }

Tablam := function ( lam , m ) 
    local L , cc  ;
    cc := ConjugacyClasses ( SymmetricGroup (m )) ;
    L := [ ] ; 
    for i in [ 1 .. NrPartitions ( m ) ] do 
       Add ( L , [ cc [i]  , Countmumulam ( i , lam , m )] )   ; 
    od ;  
return L ; 
end ; 

## Computes  the indices of [2*,1*] partitions 
# mm1 a positive integer 

Partitions2st1stind := function ( mm1) 
        local pp , Countones, Counttwos  , LL1 ; 
        LL1 := [] ; 
        pp := Partitions(mm1) ;  
        for i in [1 .. NrPartitions (mm1 )]  do
            Countones := Length( Positions (  pp[i] , 1 ) ) ; 
            Counttwos := Length( Positions (  pp[i] , 2 ) ) ;
            if 
               2*Counttwos + Countones  = mm1  then Add ( LL1 ,   i  ) ;
            fi; 
        od ; 
return LL1;
end;

partmunu := function (mu , nu ,  m ) 
        local LL, cc, Listmu , Listnu,  Grp ;
        LL := []  ; 
        Grp := SymmetricGroup(m) ;
        cc := ConjugacyClasses (Grp) ; 
        Listmu := AsList(cc[mu]) ; 
        Listnu := AsList(cc[nu]) ;  
        for k in [ 1 .. Length(Listmu) ] do 
           for l in [ 1 .. Length(Listnu) ] do 
             Add ( LL , Listmu[k]*Listnu[l] ) ; 
            od; 
        od; 
return LL ; 
end;


## We can do the same selection of 2st1st at the level of partitions ; 
## this function lists the partitions 
Partitions2st1st := function ( mm1) 
        local pp , Countones, Counttwos  , LL1 ; 
        LL1 := [] ; 
        pp := Partitions(mm1) ;  
        for i in [1 .. NrPartitions (mm1 )]  do
            Countones := Length( Positions (  pp[i] , 1 ) ) ; 
            Counttwos := Length( Positions (  pp[i] , 2 ) ) ;
            if 
               2*Counttwos + Countones  = mm1  then Add ( LL1 ,   pp[i]  ) ;
            fi; 
        od ; 
return LL1;
end;

## This is a variation that produces the indices 
Partitions2st1stind := function ( mm1) 
        local pp , Countones, Counttwos  , LL1 ; 
        LL1 := [] ; 
        pp := Partitions(mm1) ;  
        for i in [1 .. NrPartitions (mm1 )]  do
            Countones := Length( Positions (  pp[i] , 1 ) ) ; 
            Counttwos := Length( Positions (  pp[i] , 2 ) ) ;
            if 
               2*Counttwos + Countones  = mm1  then Add ( LL1 ,   i  ) ;
            fi; 
        od ; 
return LL1;
end;


## This computes  the bound of the colum sum 
## Lam is the partition 
## m is the integer which is partitioned 

Sumlam2st1stPartInp  := function ( Lam , m ) 
    local  S , i, lam  ;
    #pp2s1s := Partitions2st1st (m) ;
    S := 0 ; 
    lam := Position ( Partitions (m) , Lam );
    for i in Partitions2st1stind(m)  do 
        S := S + Countmumulam ( i , lam , m ) ; 
    od ;  
return  S  ; 
end ; 

\end{verbatim}
We run it as 
\begin{verbatim}
%%gap 
# For all lam at fixed n, compute the lower bound coming from 
# partition of the type [2*,1*]  
for  p in Partitions (5) do 
 Print (   Sumlam2st1stPartInp ( p  , 5 ) , " " ,p ,  " \n " )  ; 
od ; 

(out)
3 [ 1, 1, 1, 1, 1 ] 
0 [ 2, 1, 1, 1 ] 
5 [ 2, 2, 1 ] 
10 [ 3, 1, 1 ] 
0 [ 3, 2 ] 
0 [ 4, 1 ] 
8 [ 5 ]

%%gap
for  p in Partitions (8) do 
 Print (   Sumlam2st1stPartInp ( p  , 8 ) , " " ,p ,  " \n " )  ; 
od ; 

(out)
5 [ 1, 1, 1, 1, 1, 1, 1, 1 ] 
0 [ 2, 1, 1, 1, 1, 1, 1 ] 
50 [ 2, 2, 1, 1, 1, 1 ] 
0 [ 2, 2, 2, 1, 1 ] 
21 [ 2, 2, 2, 2 ] 
40 [ 3, 1, 1, 1, 1, 1 ] 
0 [ 3, 2, 1, 1, 1 ] 
72 [ 3, 2, 2, 1 ] 
112 [ 3, 3, 1, 1 ] 
0 [ 3, 3, 2 ] 
0 [ 4, 1, 1, 1, 1 ] 
72 [ 4, 2, 1, 1 ] 
0 [ 4, 2, 2 ] 
0 [ 4, 3, 1 ] 
72 [ 4, 4 ] 
80 [ 5, 1, 1, 1 ] 
0 [ 5, 2, 1 ] 
96 [ 5, 3 ] 
0 [ 6, 1, 1 ] 
48 [ 6, 2 ] 
96 [ 7, 1 ] 
0 [ 8 ]


\end{verbatim}
\bibliography{mybib}

\end{document}